\documentclass{jpp}
\usepackage{graphicx,psfrag}
\usepackage{epstopdf, epsfig,subfig}
\usepackage{graphicx, amsfonts, amsmath, amssymb}
\usepackage{color}

\shorttitle{Magnetic helicity in solar flux emergence}
\shortauthor{C. Prior \& D. MacTaggart}

\title{Interpreting magnetic helicity flux in solar flux emergence}

\author{C. Prior\aff{1}
 \corresp{\email{christopher.prior@durham.ac.uk}}
D. MacTaggart\aff{2}
}

\affiliation{\aff{1}Department of Mathematical Sciences, University of Durham, UK
\aff{2}School of Mathematics \& Statistics, University of Glasgow, UK}

\newcommand{\be}{\begin{equation}}
\newcommand{\en}{\end{equation}}
\newcommand{\norm}[1]{\left\vert#1\right\vert}
\def\pt{{\partial}}
\newcommand{\pder}[2]{\frac{\pt #1}{\pt #2}}

\def\d{{\rm d}}
\def\uv{{\boldsymbol u}}
\def\av{{\boldsymbol a}}

\def\nv{{\boldsymbol n}}

\def\Bv{{\boldsymbol B}}

\def\Av{{\boldsymbol A}}

\def\cl{{\mathcal{L}}}

\def\wdth{4cm}

\begin{document}

\maketitle

\begin{abstract}
Magnetic helicity flux gives information about the topology of a magnetic field passing through a boundary. In solar physics applications, this boundary is the photosphere and magnetic helicity flux has become an important quantity in analysing magnetic fields emerging into the solar atmosphere. In this work we investigate the evolution of magnetic helicity flux in magnetohydrodynamic (MHD) simulations of solar flux emergence. We consider emerging magnetic fields with different topologies and investigate how the magnetic helicity flux patterns corresponds to the dynamics of emergence. To investigate how the helicity input is connected to the emergence process, we consider two forms of the helicity flux. The first is the standard form giving topological information weighted by magnetic flux. The second form represents the net winding and can be interpreted as the standard helicity flux less the magnetic flux. Both quantities provide important and distinct information about the structure of the emerging field and these quantities differ significantly for mixed sign helicity fields. A novel aspect of this study is that we account for the varying morphology of the photosphere due to the motion of the dense plasma lifted into the chromosphere.  Our results will prove useful for the interpretation of magnetic helicity flux maps in solar observations.
\end{abstract}

\section{Introduction}
The year 2018 marks the 60th anniversary of Woltjer's observation \citep{woltjer1958theorem} that magnetic helicity in a {tangent magnetic field} is an invariant of ideal magnetohydrodynamics (MHD). That is, for a simply connected domain,
\be
\frac{{\rm d}}{{\rm d}t}\int_{\Omega}\Av\cdot\Bv\,{\rm d}V =0, \quad \Bv\cdot\nv=0\ {\rm on}\ \partial\Omega,
\en
where $\Bv$ is the magnetic field, $\Av$ is the vector potential of the magnetic field and $\nv$ is the surface normal. This form of helicity  is also gauge invariant in the sense that the transformation $\Av\rightarrow\Av+\nabla\chi$, for some scalar function $\chi$, does not change the value of helicity. A few years later, Moffatt \citep{moffatt1969degree} discovered the topological connection of magnetic helicity to Gauss' linking number. Thus, the subject of magnetic topology was born and has been influential in many areas of MHD ever since \citep[e.g.][]{taylor1974relaxation,frisch1975possibility,berger1984rigorous,schindler1988general,pevtsov1995latitudinal,devore2000magnetic,russell2015evolution}. Some important examples of applications include the prediction of relaxed magnetic states in reverse pinch experiments \citep{taylor1974relaxation} and the analysis of emerging solar active regions \citep[e.g.][]{pevtsov1995latitudinal,leka2005availability,demoulin2002magnetic,labonte2007survey}. The first example we have given is an instance of magnetic topology applied to magnetic relaxation. \cite{taylor1974relaxation} conjectured that (total) magnetic helicity is approximately conserved in a system with very small resistivity.  Minimizing the magnetic energy whilst preserving the magnetic helicity, the relaxed state is predicted to be a linear force-free field. This simple and elegant theory has had good experimental corroboration and is one of the successes of the application of magnetic helicity. 

The second application we gave, magnetic flux emergence, is the subject of this paper. `Bundles' of magnetic field rise up through the solar convection zone until they reach the Sun's surface, the photosphere. Here, the magnetic bundles emerge into the solar atmosphere where they can produce interesting phenomena such as flares and coronal mass ejections \citep[e.g.][]{hood20123d,cheung2014flux}. Since we cannot observe magnetic fields in the solar atmosphere directly, we rely on indirect information about their structure. This is where magnetic helicity input can play an important role.

When observers study the magnetic fields of solar active regions, the photosphere acts as a lower boundary for the magnetic field.  This fact introduces a complication in the original definition of magnetic helicity. It means that active region magnetic fields have a non-trivial normal boundary component at the photosphere and, hence,  magnetic field lines which leave the domain. In such circumstances the helicity is no longer gauge invariant. We can bypass the above complication, however, by considering \emph{relative magnetic helicity} \citep{berger1984topological} which provides a topological invariant for fields connected to a boundary. Magnetic helicity can, therefore, be used in the analysis of solar active regions, in particular as an indicator of eruptive activity \citep[e.g.][]{yeates2016global,pariat2017relative,guo2017magnetic}.

Although some researchers study relative helicity in solar atmosphere \citep[e.g.][]{valori2012relative,yeates2016global,pariat2017relative,guo2017magnetic}, this quantity requires knowledge of the magnetic field's full structure which is not observationally avaliable. Observational studies instead measure the relative magnetic helicity flux through the photosphere  \citep[e.g.][]{chae2001observational,kusano2002measurement,pevtsov2003helicity,yokoyama2003relation,pariat2006spatial,labonte2007survey,schuck2008track,vemareddy2015investigation}.  Photospheric maps of helicity flux during flux emergence can reveal interesting and complex patterns. Often, however, these results are interpreted based on a `standard model' of flux emergence where the emerging magnetic field is a twisted flux tube just below the photosphere \citep[e.g.][]{archontis2008eruption,mactaggart2009emergence,moreno2013plasma}. Twist {is an important part of helicity that, in flux emergence studies, leads to} eruptive behaviour in the atmosphere. Observations of active region helcity input, however, imply that even bipolar regions, associated with single sign helicity flux rope/sheared arcade formation, can often input both signs of helicity \citep[e.g.][]{leka1999value,pevtsov2008solar,vemareddy2015investigation,vemareddy2017successive,bi2018survey}. 
 Modelling and interpreting the emergence of mixed helicity regions should be a priority in order to provide an understanding of their behaviour.

The purpose of this article is to show how (relative) magnetic helicity flux can be interpreted in solar flux emergence with a variety of emerging  magnetic topologies. We first examine the `topological meaning' of helicity flux in terms of winding (field line entanglement). We then study the magnetic helicity and winding fluxes in flux emergence simulations with different initial profiles of the magnetic field. The paper ends with a summary and a discussion.

\section{Magnetic helicity flux}
Consider an emerging magnetic field subject to ideal motion and two points $\av_1(t)$ and $\av_2(t)$ which represent the intersection of two field lines and the photospheric plane $P$. The motion of these points in time follows the motion of the point of intersection of these field lines in $P$ as the field emerges. The net rotation of these points within $P$ as a function of time is  
\be\label{netr}
c({\boldsymbol a}_1,{\boldsymbol a}_2,t) = \frac{1}{2\pi}\int_{t_0}^{t_1}\frac{{\rm d}\theta_{12}({\boldsymbol a}_1,{\boldsymbol a}_2,t)}{{\rm d}t}\,{\rm d}t,
\en
where $\theta_{12}({\boldsymbol a}_1,{\boldsymbol a}_2,t)$ is the angle of the line segment linking $\av_1$ and $\av_2$. Equation (\ref{netr}) represents the entanglement of a pair of field lines due to emergence or in plane motions of the field lines, hence ${\rm d}c/{\rm d}t$ represents the rate of input  of entanglement through the photosphere and into solar atmosphere due to these fieldlines. 

{In order to determine the time derivative of the angle $\theta_{12}$, first note that,
{\be\label{theta}
\theta_{12} = 
\mathrm{arctan}\displaystyle\left(\frac{(\av_2-\av_1)\cdot\hat{{\boldsymbol  y}}}{(\av_2-\av_1)\cdot\hat{{\boldsymbol  x}}}\right).
\en
Differentiating gives,
\be
\label{dthe}
\frac{{\rm d}\theta_{12}}{{\rm d}t} =\frac{1}{(\av_2-\av_1)\cdot(\av_2-\av_1)} \hat{\boldsymbol  z}\cdot (\av_2-\av_1)\times \left(\frac{\d\av_2}{\d t}-\frac{\d\av_1}{\d t}\right).
\en
}} The (relative) magnetic helicity $H$ represents the total winding $c(\av_1,\av_2,t)$ over all pairs of paths weighted by the photospheric flux over a given period of time \citep{berger1988energyff,demoulin2003magnetic}. Thus, the rate of input of magnetic helicity into the solar atmosphere, through the {photosphere} $P$, is
\be\label{helflux}
\frac{{\rm d}H}{{\rm d}t} = {-}\frac{1}{2\pi}\int_{P}\int_{P} B_z(\av_1)B_z(\av_2)\frac{{\rm d}\theta_{12}({\boldsymbol a}_1,{\boldsymbol a}_2,t)}{{\rm d}t}\,{\rm d}x_1{\rm d}y_1\,{\rm d}x_2{\rm d}y_2.
\en
In order to use equation (\ref{helflux}), we need to know how to track points $\av_1$ and $\av_2$. This motion is given by both in-plane motions and the projection due to motion out of the plane (flux emergence or submergence). If $\uv$ is the velocity field and $\uv_{\parallel}$ is the projection of $\uv$ onto $P$, it can be shown that
\be\label{ub}
\frac{{\rm d}\av}{{\rm d}t} = \uv_{\parallel}-\frac{u_z}{B_z}\Bv_{\parallel} = \uv_{\parallel}-{u_z\left(\frac{{\rm d}\av}{{\rm d}z}\right)_{\parallel}},
\en
\citep[e.g.][]{berger1988energyff,demoulin2003magnetic}. {The term $\uv_{\parallel}$ accounts for the motions of the field lines within the plane. The term $u_z (\d \boldsymbol{a}/\d z)_{\parallel}$ accounts for the apparent motion within the plane due to the emergence (or submergence if $u_z<0$) of field lines. The total derivative ${\rm d}\av/{\rm d}t$ assumes that the field lines are advected with the plasma (ideal motion).}

 The helicity flux through $P$ represents the transmission of topological information into the atmosphere and can change in time. In magnetic relaxation, although magnetic helicity controls the final equilibrium, the path to that equilibrium can involve very complicated dynamics. Similarly in flux emergence, although magnetic helicity affects the overall evolution of the emerging field, the magnetic helicity flux is associated closely with the dynamics of emergence - the interplay of both plasma and magnetic field. For this reason, it is important to interpret carefully what maps of magnetic helicity flux mean.

A related quantity is the field line helcity input \citep[e.g.][]{berger1988energyff,vemareddy2015investigation,vemareddy2017successive} which we define here as 
\be\label{flhel}
\frac{{\rm d}{{\cal H}}}{{\rm d}t}({\boldsymbol a}_0) = {-\frac{1}{2\pi}}B_z({\boldsymbol a}_0)\int_{P}B_z({\boldsymbol a})\frac{{\rm d}\theta({\boldsymbol a}_0,{\boldsymbol a})}{{\rm d}t}\,{\rm d}x{\rm d}y.
\en
Equation (\ref{flhel}) describes the contribution to the helicity flux ${\rm d}H/{\rm d}t$ from a single point $\av_0\in P$ (thus the contribution to helicity flux from a single field line). {Although the integrand of (\ref{flhel}) has a singularity at  ${\boldsymbol  a}_0$, the integral still converges as the pole is of a lower order than the domain of integration.} In the solar physics literature, ${\rm d}{\cal H}(\av_0)/{\rm d}t$ is often labelled `$G_{\theta}$' and is used as a proxy for helicity flux density \citep[e.g.][]{pariat2005photospheric}.

  
The contribution from a neighbourhood around the pole (call it $a_0^{\epsilon})$ can be shown to be decomposed into the local twisting of the field around this point (which relates to the local electric current)  and a contribution called the writhe, due to the geometry of the field line passing through $\av_0$ \citep{calugareanu1959integrale,calugareanu1961classes,white1969self,berger2006writhe}.
Considering the contribution of the integral (\ref{flhel}) from  $a_0^{\epsilon}$, we can relate this to \emph{self helicity} as it measures the winding of the field in the locality of the field line passing through $\av_0$. The contribution due to the rest of the field, on the set $P-a_0^{\epsilon}$, with this local neighbourhood is related to \emph{mutual helicity} since the contribution represents the winding of the rest of the field with the small `flux tube' in $a_0^{\epsilon}$.  This idea of decomposing the helicity into mutual and self components has been used in numerous studies. Often it is the case that the field will have finite volume flux ropes, as in \cite{pariat2006spatial} and \cite{guo2017magnetic}, be the result of clear large scale shearing motion, as in \cite{demoulin2002magnetic}, or have regions of similar squashing factor, such as in \cite{guo2017magnetic}.  In these cases it is possible to extend the notion of self helicity to these finite bundles  of field lines and gain further insight into the field's morphology by tracking these self/mutual interactions.  In this study, we will consider fields that exhibit highly complex local current structure and topology. Hence, a decomposition of the helicity into self and mutual components is not a simple task or necessarily well-defined. Therefore, in this study, we do not pursue the self/mutual approach to describing helicity.

 As a final note, before presenting our analysis, we add that \cite{pariat2006spatial} showed that it is possible to get spurious input values from (\ref{flhel}) due to field structures with no helicity. This is not the case for the emergence/submergence events analysed here but is a possibility that should always be considered when looking at the distributions of this quantity (such spurious contirbutions vanish for the net input (\ref{helflux})).


\subsection{Separating field strength and topological information}
To disentangle the field strength (flux) and topological information associated with the helicity input, we can also consider the net winding input
 \be\label{windingflux}
\frac{{\rm d} L}{{\rm d}t} ={-}\frac{1}{2\pi}\int_{P}\int_{P} \sigma_z(\av_1)\sigma_z(\av_2)\frac{{\rm d}\theta_{12}({\boldsymbol a}_1,{\boldsymbol a}_2,t)}{{\rm d}t}\,{\rm d}x_1{\rm d}y_1\,{\rm d}x_2{\rm d}y_2,
\en
where{ 
\begin{equation}\label{sigma1}
\sigma_z({\boldsymbol  x}) = \left\{
\begin{array}{cc}
1 & \mbox{ if $B_z({\boldsymbol  x})>0$},\\
-1  & \mbox{if $B_z({\boldsymbol  x})<0$},\\
0 & \mbox{if $B_z({\boldsymbol  x})=0$},
\end{array}
\right.
\end{equation}
{\citep[e.g.][]{prior2014helicity}}.
The rate ${\rm d}{L}/{\rm d}t$} measures the input of field line entanglement through the photosphere and ignores the extra flux information provided by the helicity. {A similar suggestion made in the literature is to divide the helicty input by the square of the flux \citep[e.g][]{yamamoto2005helicity,labonte2007survey,vemareddy2017successive}}. We propose that ${{\rm d} L}/{{\rm d}t}$ is more meaningful in this context as it is based on a quantity, the net winding, which is also an ideal invariant \cite[]{prior2014helicity}. Indeed, it  was shown recently to be more topologically meaningful than helicity as it can be used, under certain circumstances, to classify field topologies and hence precisely measure changing field connectivity \cite[]{prior2018quantifying}.

{\subsection{Evolving photosphere}
{The standard practice in both theoretical and observational studies of helicity is to treat the lower (photospheric) boundary as fixed. In almost all studies, the photosphere is a plane \citep[e.g][]{pariat2005photospheric,pariat2006spatial,jeong2007magnetic,sturrock2015sunspot,vemareddy2015investigation,vemareddy2017successive}, although calculations in spherical geometry have also been performed \citep{berger1985spherical,mactaggart2016spherical,moratis2018spherical}. In our calculations, which we will describe shortly, our fixed photospheric plane $P$ will represent the fixed height $z=0$ in a Cartesian domain. 

In addition to this approach, we also consider the effect of an \emph{evolving photosphere}. Including this effect means that the lower boundary is no longer fixed but is a surface that deforms in space and time. In our calculations, we will define the photospheric boundary as the surface where the plasma density $\rho$ has the non-dimensional value of 1. In order to evaluate how a changing photosphere will affect the helicity and winding fluxes, we use the following procedure. First, we calculate the varying surface on which $\rho=1$, i.e. the set
\begin{equation}
\left\{ P_v({\boldsymbol  a}, t)\,\vert\,{\boldsymbol  a} \in P,\,\rho(P_v)=1\right\}.
\end{equation}
$P$ will always represent the horizontal plane at $z=0$ in our model (more details will be given later). At $t=0$ in the simulations, $P_v\equiv P$. As indicated, $P_v$ is a function of the coordinates $(x,y)$ of $P$.

 We then evaluate $\d\av/\d t$ at $P_v(\av,t)$ rather than at $P(\av)$. We can use this quantity to calculate an effective rotational derivative $\d{\theta_{12}}/\d{t}$ using (\ref{dthe}). To account for the surface geometry, we replace the area element $\d x_1 \d y_1$ with $J({\boldsymbol  a}_1)\d x_1 \d y_1$, where $J$ the Jacobian of the surface $P_v$ at the coordinates of ${\boldsymbol  a}_1$ (and similarly for  ${\boldsymbol  a}_2$). For example, the calculation (\ref{windingflux}) will become
\begin{align}\label{windingfluxmov}
\frac{{\rm d} L}{{\rm d}t} ={-}\frac{1}{2\pi}\int_{P}\int_{P}\left[ \sigma_z( P_v(\av_1,t))\sigma_z( P_v(\av_2,t))\frac{{\rm d}\theta_{12}( P_v({\boldsymbol a}_1,t), P_v({\boldsymbol a}_2,t))}{{\rm d}t}\right.\\\nonumber \times J( P_v({\boldsymbol a}_1,t))J( P_v({\boldsymbol a}_2,t)) \Big]{\rm d}x_1{\rm d}y_1\,{\rm d}x_2{\rm d}y_2.
\end{align}
Equation (\ref{windingfluxmov}) \emph{does not} represent the \emph{exact} winding flux through a spatially non-uniform surface. We could, for example, have made use of the differential form definition of relative helicity given in Chapter 3 of \cite{arnold1999topological} in order to calculate the actual flux. However, we argue that what the above procedure represents is more akin  to the effective projection of information onto the plane $P$ which occurs in the creation  of magnetogram and vector magnetogram data  \citep[e.g.][]{scherrer2012helioseismic,demoulin2009modelling}. These magnetogram data are obtained from line-of-sight information, Zeeman splitting and normal to line-of-sight linear polarization information. These data are effectively projected onto a plane $P$. Since we are trying to gauge the consequences of an effect which is already uncertain (the interpretation of optical information) we feel the above approach is sensible first step. To the best of our knowledge, this is the first theoretical helicity study which has attempted to take account of a moving photosphere.

}

}


\section{Simulation setup and initial conditions}
In this study, we will consider small active regions of photospheric area $\sim75$ Mm$^2$. Studying regions of this size is common in the flux emergence literature \citep{hood20123d,cheung2014flux} and is justified on the grounds of achieving suitable spatial and temporal resolution. {Of course, care must be taken when comparing the results of such studies to observations of regions of different sizes. Although qualitative behaviour may be found across spatial and temporal scales, quantitative information will be different.}

  As our focus is on the dynamics of the magnetic field, we will consider an idealized description of the solar atmosphere. The bulk properties of the plasma and magnetic field dynamics are described by compressible MHD. The 3D resistive and compressible MHD equations are solved using a Lagrangian remap scheme \citep{arber2001staggered}. In dimensionless form, the MHD equations are
\begin{equation}\label{mass_con}
\frac{{\rm D}\rho}{{\rm D} t} = -\rho{{\boldsymbol \nabla\cdot  u}},
\end{equation}
\begin{equation}\label{mom_con}
\frac{{\rm D}{\boldsymbol u}}{{\rm D} t} = -\frac{1}{\rho}{\boldsymbol \nabla} p + \frac{1}{\rho}(\nabla\times{\boldsymbol B})\times{\boldsymbol B}+\frac{1}{\rho}{{\boldsymbol \nabla\cdot \boldsymbol\sigma}}+{\boldsymbol g},
\end{equation}
\begin{equation}\label{induction}
\frac{{\rm D}{\boldsymbol B}}{{\rm D} t} = ({\boldsymbol {B}\cdot\nabla}){\boldsymbol u} - ({\boldsymbol \nabla\cdot{\boldsymbol u}}){\boldsymbol B} +\eta\nabla^2{\boldsymbol B},
\end{equation}
\begin{equation}\label{energy_con}
\frac{{\rm D}\varepsilon}{{\rm D} t} = -\frac{p}{\rho}{\boldsymbol \nabla\cdot{u}} + \frac{1}{\rho}\eta|{\boldsymbol j}|^2 + \frac{1}{\rho}Q_{\rm visc},
\end{equation}
\begin{equation}\label{divB}
{\boldsymbol \nabla\cdot{B}} = 0,
\end{equation}
with specific energy density
\begin{equation}\label{energy_den}
\varepsilon = \frac{p}{(\gamma-1)\rho}.
\end{equation}
The basic variables are the density $\rho$, the pressure $p$, the magnetic induction ${\boldsymbol B}$ (referred to as the magnetic field) and the velocity ${\boldsymbol u}$. ${\boldsymbol j}$ is the {electric} current density, ${\boldsymbol g}$ is {the surface gravitational acceleration} (uniform in the $z$-direction) and $\gamma =5/3$ is the ratio of specific heats. The dimensionless temperature $T$ can be found from
\begin{equation}\label{temp}
T = (\gamma-1)\varepsilon.
\end{equation}
We make the variables dimensionless against photospheric values, {a standard procedure in flux emergence studies \citep[e.g.][and references therein]{hood20123d}}. We have pressure $p_{\rm ph} = 1.4\times 10^4$ Pa; density $\rho_{\rm ph} = 2\times 10^{-4}$ kg~m$^{-3}$; scale height $H_{\rm ph}=170$ km; {surface gravitational acceleration} $g_{\rm ph} = 2.7\times 10^2$ m~s$^{-2}$; speed $u_{\rm ph} = 6.8$ km~s$^{-1}$; time $t_{\rm ph} = 25$ s; magnetic field strength $B_{\rm ph} = 1.3\times 10^3$ G and temperature $T_{\rm ph} = 5.6\times 10^3$ K. In the non-dimensionalization of the temperature we use a gas constant $\mathcal{R}=8.3\times 10^{3}$ m$^2$~s$^{-2}$~K$^{-1}$ and a mean molecular weight $\tilde{\mu}=1$. $\eta$ is the resistivity and we take its value to be $10^{-3}$. This value is close to the lowest physical resistivity that can be chosen before numerical resistivity dominates \citep[see][]{arber2007emergence,leake2013simulations}. Note that in the simulations we present, reconnection plays a minor role and motion, to a good approximation, can be considered ideal.  The fluid viscosity tensor and the viscous contribution to the energy equation are respectively
\begin{equation}\label{sigma}
\boldsymbol\sigma = 2\mu\left[\boldsymbol{D}-\frac13({\rm tr}\boldsymbol{D})\boldsymbol{I}\right]
\end{equation}
and
\begin{equation}\label{q_visc}
 Q_{\rm visc} = \boldsymbol\sigma:\nabla\boldsymbol{u},
\end{equation}
where
\begin{equation}
{\boldsymbol D} = \frac12\left(\nabla\boldsymbol{u} + \nabla\boldsymbol{u}^{\rm T}\right)
\end{equation}
is the symmetric part of the rate of strain tensor and $\boldsymbol{I}$ is the identity tensor. We take $\mu = 10^{-5}$ and use this form of viscosity primarily to aid stability. The code accurately
resolves shocks by using a combination of shock viscosity \citep{wilkins1980use} and
Van Leer flux limiters \citep{van1979towards}, which add heating terms to the
energy equation. Values will be expressed in non-dimensional form unless explicitly stated otherwise. 

The equations are solved in a Cartesian computational box of
(non-dimensional) sizes $[-45, 45]\times[-45, 45]\times{[-30, 65]}$ in the
$x$, $y$ and $z$ directions respectively. The boundary conditions are
closed on the top and base of the box and periodic on the sides.  Damping layers are included at the side and top boundaries to reduce the reflection/transmission of waves. The computational mesh contains 486$\times$486$\times$729 points.

\subsection{Initial background velocity perturbation}
The idealized initial equilibrium atmosphere is given by prescribing the temperature profile
{\begin{equation}\label{initial_temp}
{T(z) = \left\{\begin{array}{cc}
1-\frac{\gamma-1}{\gamma}z, & z < 0, \\
1, & 0 \le z \le 10,  \\
150^{[(z-10)/10]}, & 10 < z < 20,  \\
150, & z \ge 20.
\end{array}\right.}
\end{equation}
Starting from the top of (\ref{initial_temp}), the sections represent the solar interior, the photosphere/chromosphere, the transition region and the corona.  The above model temperature profile is standard in many works on flux emergence \citep[e.g.][]{murray2006emerge,fan2009emergence,sturrock2015sunspot}. The above temperature profile is also used in \cite{prior2016emergence}.

The other state variables, pressure and density, are found by solving the hydrostatic equation in conjunction with the ideal equation of state
\begin{equation}\label{EOS}
\frac{{\rm d}p}{{\rm d}z} = -\rho g, \quad p = \rho T.
\end{equation}

To study emergence, we must place a particular form for the magnetic field in the solar interior and apply a perturbation to allow it to emerge. {In the following simulations, we will apply an initial velocity perturbation of the form
\begin{equation}
{\boldsymbol  u}\cdot\hat{\boldsymbol  z} = u_0\exp\left(-\frac{x^2}{x_0^2}\right)\exp\left(-\frac{y^2}{y_0^2}\right)\exp\left(-\frac{(z+\epsilon_0-R)^2}{z_0^2}\right)\sin\left(\frac{t}{t_0}\pi\right),
\end{equation}
where we set the constants $u_0=0.05$, $x_0=5$, $y_0=3$, $z_0=5$, $\epsilon_0=2.5$ and $t_0=6$. $R$ is the major radius of the tube and is described below. After $t=6$ the perturbation is switched off.
}

We will consider two {magnetic field} models. The first model is a twisted toroidal tube that has been used in many other studies \citep[e.g.][]{mactaggart2009emergence,mactaggart2014emergence,mactaggart2015emergence,sturrock2015sunspot}. The second model is a `mixed helicity' field which has a complex topology but zero total helicity and will serve as a model for mixed helicity emergence. We now discuss the construction of such fields.

\begin{figure}
\centering
\subfloat[]{\includegraphics[width=8cm]{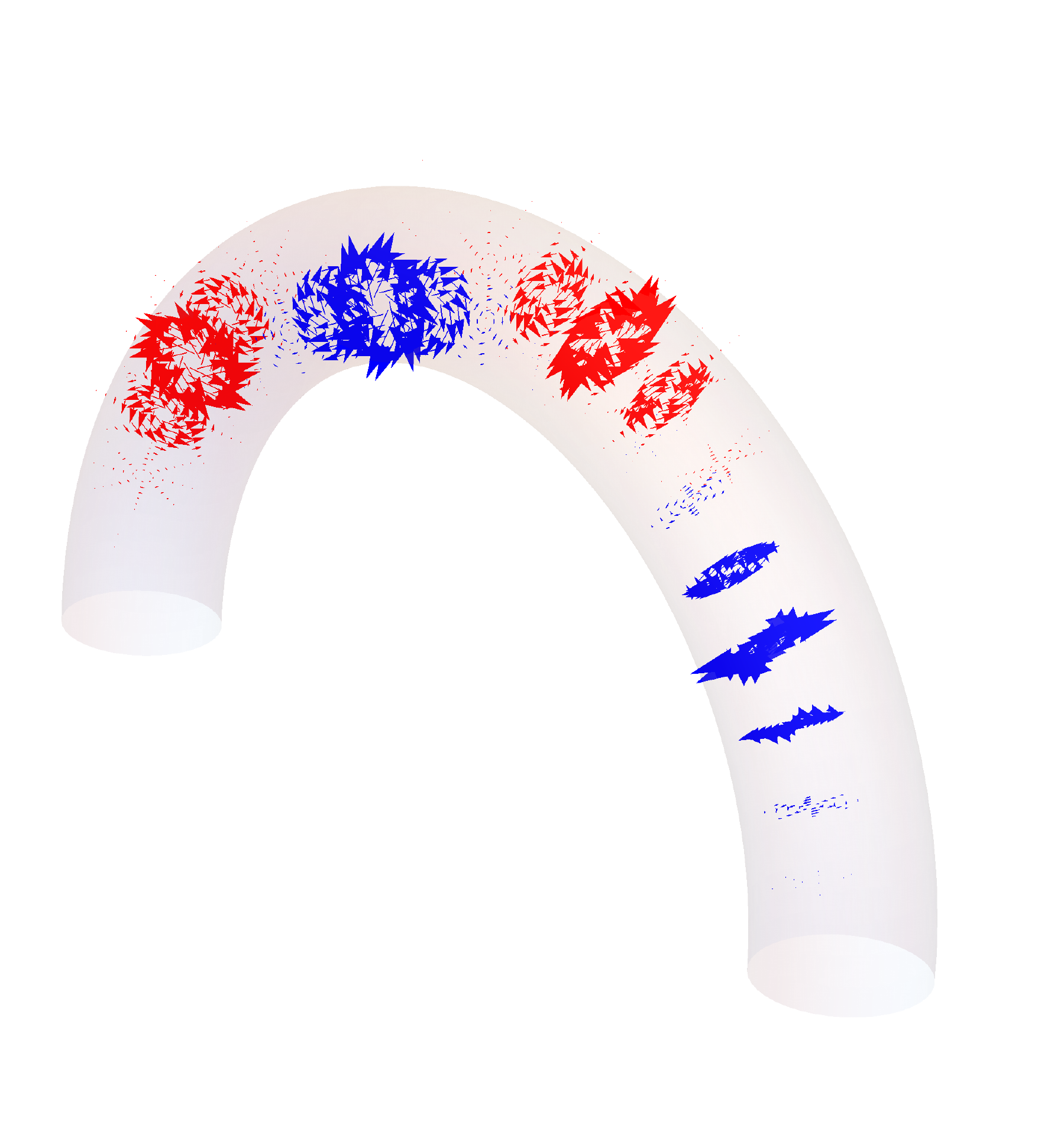}}\quad\subfloat[]{\includegraphics[width=8cm]{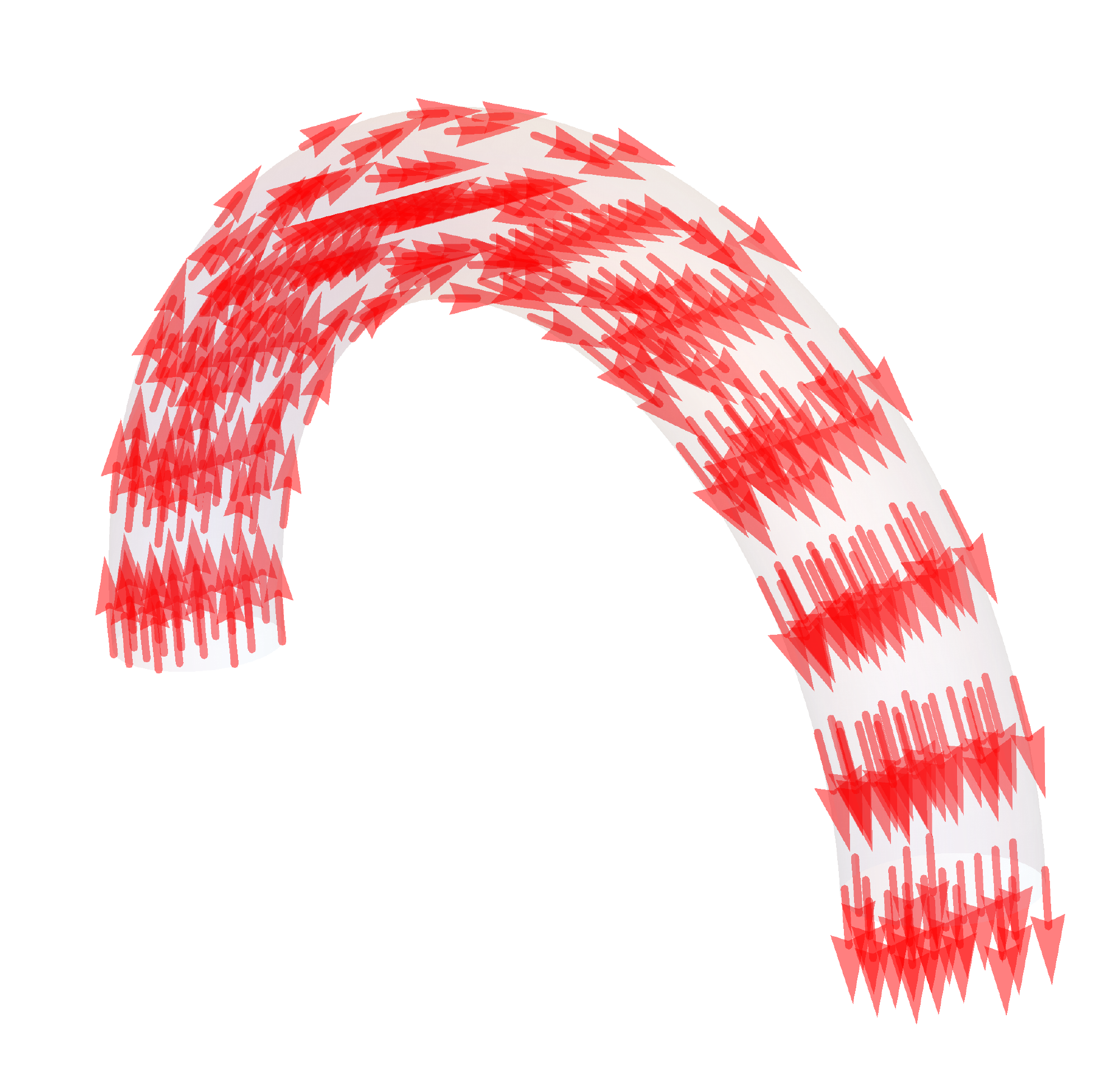}}
\caption{\label{braidfield} An example of the mixed helicity field we consider. (a)  a field with $n=2$, i.e. two right-handed twist elements (red) and two left-handed (blue).  (b) shows the background axial field which is added to (a) in equation (\ref{braidfieldtube}).}
\end{figure}

\subsection{How to construct mixed helicity fields}\label{sec:braid}
The following represents a specific case of the general mathematical form of magnetic flux ropes with general field line topology, introduced in \cite{prior2016twisted1} and first applied to flux emergence in \cite{prior2016emergence}. We assume the flux rope has a toroidal shape with an axis curve ${\boldsymbol r}(s)$ parameterised by its arclength $s$ as
\begin{equation}
{\boldsymbol r}(s) = \left(-R \cos(s/R),0,R\sin(s/R) + z_0\right),\,s\in[0,\pi R],
\end{equation}
where $R$ is the major radius of the torus {and $z_0$ is the height of the ropes's anchoring footpoints at the base of the computational domain}. This expression can be used to define a moving orthonormal frame $\left\{{\boldsymbol T},{\boldsymbol d}_1,{\boldsymbol d}_2\right\}$ for ${\boldsymbol r}(s)$, which can, in turn, be used to define a tubular coordinate system through the map ${\boldsymbol f}(s,x_1,x_2)$, where
\begin{align}
  &{\boldsymbol f}(s,x_1,x_2) = {\boldsymbol r}(s) + x_1{\boldsymbol d}_1 + x_2{\boldsymbol d}_2,\\
  &{\boldsymbol d}_1 = \left(\cos(s/R),0,-\sin(s/R)\right),{\boldsymbol d}_2 = (0,1,0). 
\end{align}
From this mapping, we can find a set of basis vector fields $\left\{\pder{{\boldsymbol f}}{s},\pder{{\boldsymbol f}}{x_1},\pder{{\boldsymbol f}}{x_2}\right\}$ which can be used to define general vector fields in this toroidal coordinate system as 
\be
{\boldsymbol B} = B_s\pder{{\boldsymbol f}}{s} + B_1\pder{{\boldsymbol f}}{x_1} + B_{2}\pder{{\boldsymbol f}}{x_2}. 
\en
For such fields to be divergence-free the following condition must hold
\begin{equation}
 \pder{JB_s}{s} +  \pder{JB_1}{x_1} +  \pder{JB_{2}}{x_2} =0,
\end{equation}
where $J= \frac{(R-x_1)}{R}$ is the Jacobian of the map $\boldsymbol{f}$. It follows that a magnetic field defined {in} a cylinder can be transferred to a torus simply by dividing the components by $J$. In particular, we take the mixed helicity (braided) field used in numerous studies  \citep[e.g][]{wilmot2010dynamics,wilmot2011heating,pontin2015structure}, which is composed of a series of overlapping counter-twists. The magnetic field has no helicity (the average total entanglement) but subsets of its field lines are braided. In our toroidal coordinate system this field is composed of exponential twists ${\boldsymbol B}_t(b_0,k,a,l,x_{1c},x_{2c},s_{c})$ given by
\begin{align}
  \label{bt}
  {\boldsymbol B}_t(b_0,k,a,l,x_{1c},x_{2c},s_{c}) &= \frac{2 b_0 k}{a J}\mathrm{exp}\left(-\frac{(x_1-x_{1c})^2+(x_2-x_{2c})^2}{a^2}-\frac{(s-s_c)^2}{l^2}\right){\boldsymbol R},\\
 \label{Rdef} {\boldsymbol R}&= -(x_2-x_{2c})\pder{{\boldsymbol f}}{x_1}+(x_1-x_{1c})\pder{{\boldsymbol f}}{x_2},
\end{align}
where the parameter $b_0$ determines the strength of the field, $a$ the horizontal width of the twist zones, $l$ their vertical extent and $k$ the handedness of the twist ($k=1$ is right handed). The centre of rotation is $(x_{1c},x_{2c},s_{c})$. The mixed helicity field is then defined as a superposition of $n$ pairs of positive and negative twists and an axial background field 
\begin{align}
  \label{braidfieldtube}
        {\boldsymbol B}_{br}(b_0,a,l,d,R,n) = &\sum_{i=1}^{n}\left[{\boldsymbol B}_t(b_0,1,a,l,0,-d,s_d i)\right. \nonumber\\
      &+ \left.{\boldsymbol B}_t(b_0,-1,a,l,0,d,s_d(i+1))\right] + \frac{b_0}{J}\pder{{\boldsymbol f}}{s},
\end{align}
where $s_d = \pi R/(2n+1)$ and $d$ is the axial offset of the two opposing twists (see Figure \ref{braidfield}). The final component is the axial field (see Figure \ref{braidfield}(b)). Finally, to express (\ref{braidfieldtube}) in the ambient Cartesian coordinate system $(x,y,z)$ we use the following maps between the tubular coordinates $(s,x_1,x_2)$ and the Cartesian coordinates
\begin{equation}
  \label{tubetocart}
  s = \arctan(z,x) +\frac{\pi}{2},\quad x_1 = (x+R\cos s)\cos s-(z-R \sin s)\sin s,\quad x_2 =y,
\end{equation}
where the  branch cut for the arctan function is at $\pi$. For a magnetic field confined to a tube of minor radius $r$, the field ${\boldsymbol B}_{b}(x,y,z)$ takes the following form
\begin{equation}
  \label{braidedfield}{\boldsymbol B}_b(x,y,z)=\left\{\begin{array}{cc}
  {\boldsymbol B}_{br}(b_0,a,l,d,R,n) & \mbox{ if } x_1^2(x,y,z)+x_2^2(x,y,z)\leq r,\\
  0 & \mbox{ if }x_1^2(x,y,z) +x_2^2(x,y,z) > r.
  \end{array}\right.
\end{equation}
For completeness, we can define a uniformly twisted field ${\boldsymbol B}_t$ as
\begin{equation}
  \label{twistedfield}{\boldsymbol B}_{tw}(x,y,z)=\left\{\begin{array}{cc}
  \frac{\phi}{J}{\boldsymbol R} & \mbox{ if } x_1^2(x,y,z) +x_2^2(x,y,z)\leq r,\\
  0 & \mbox{ if }x_1^2(x,y,z) +x_2^2(x,y,z) > r.
  \end{array}\right.
\end{equation}
with $\phi$ determining the rate of rotation of the field and ${\boldsymbol R}$ given by (\ref{Rdef}). The actual twisted model that we will use in the simulations is that in (\ref{twistedfield}) but weighted with an exponential term \citep[see][]{mactaggart2009emergence}. { In these studies, we set the tube parameters to be $z_0 = -30$, $R=17.5$  and $r=2.5$, so that the flux rope is anchored at the bottom boundary and its maximum initial height is $-10$. The mixed helicity parameters are} $a=\sqrt{0.04}$, $l=0.04 \pi R$, $d=2.5/3$, $b_0=5$ and we consider $n=2$.

Finally, we remark that the more general field specification in \cite{prior2016twisted1} allows for arbitrary tube shapes ${\boldsymbol r}(s)$ (as well as varying tube radius). The frame vectors ${\boldsymbol d}_1(s)$ and ${\boldsymbol d}_2(s)$ are defined by parallel transport, which requires an arclength integration. Thus, in general, the relationship between the tube coordinates $(s,x_1,x_2)$ and $(x,y,z)$ will require numerical integration. 

\section{Quantities analyzed}
We now list the quantities we will make use of in our analysis of the flux emergence simulations.

\subsection{Field line helicity input rate}
For completeness, we restate equation (\ref{flhel}),
\be
{\frac{{\rm d}{{\cal H}}}{{\rm d}t}({\boldsymbol a}_0) = {-\frac{1}{2\pi}}B_z({\boldsymbol a}_0)\int_{P}B_z({\boldsymbol a})\frac{{\rm d}\theta({\boldsymbol a}_0,{\boldsymbol a})}{{\rm d}t}\,{\rm d}x{\rm d}y,}
\en
which represents the contribution to the magnetic helicity input rate ${\rm d}H/{\rm d}t$ from a single point $\av_0\in P$. For the moving photosphere calculation we evaluate the field at the points of the surface $P_v({\boldsymbol  a},t)$ and account for the surface Jacobian, i.e. ${\rm d}x{\rm d}y \rightarrow J {\rm d}x{\rm d}y$. We label this quantity ${\rm d}{\cal H}_v/{\rm d}t$.

\subsubsection{Net helicity input}
The net helicity flux  is given by
\be
\frac{{\rm d}H}{{\rm d}t} = \int_P \frac{{\rm d}{\cal H}}{{\rm d}t}(\av_0)\,{\rm d}x{\rm d}y,
\en
where the integration is taken over all $\av_0\in P$. The total helicity input over a period $[t_0,t]$ through $P$ is then given by 
\be\label{nethel}
H(t) = \int_{t_0}^{t}\frac{{\rm d}H}{{\rm d}{t'}}\, {\rm d}{t'}.
\en
In this study we choose $t_0$ to be the time at which the emerging magnetic field first reaches $z=0$. The equivalent moving photosphere calculations will be labelled ${\rm d}H_v/{\rm  d}t$ and $H_v(t)$. We point out that whilst the calculation concerns a moving surface, the quantitiy ${\rm d}H/{\rm d} s$ is calculated on a fixed domain $P$ by projection (see equation \ref{windingfluxmov}). This applies to all calculations which consider the moving domain.
\subsection{{Winding input rate}}
The winding field line input is
\be
{\frac{{\rm d}{{\cal L}}}{{\rm d}t}({\boldsymbol a}_0) = {-\frac{1}{2\pi}}\sigma_z({\boldsymbol a}_0)\int_{P}\sigma_z({\boldsymbol a})\frac{{\rm d}\theta({\boldsymbol a}_0,{\boldsymbol a})}{{\rm d}t}\,{\rm d}x{\rm d}y,}
\en
and the integral of this over $P$ gives ${\rm d} L/{\rm d}t$. The net winding input over a time period $[t_0,t]$ is
\be\label{netwind}
L(t) = \int_{t_0}^{t}\frac{{\rm d} L }{{\rm d}{t'}}\, {\rm d}{t'}.
\en
The moving photosphere calculations will be labelled with a $v$ subscript as for the helicity inputs, with analogous changes to the calculations. 

\subsection{Weak field corrections}

In our simulations, there is no magnetic field outside the emerging regions. Therefore, there is a thin current sheet around the emerging field where the field strength rapidly descreases to zero. In this region the, field line topology is incoherent and not likely to be resolved numerically (typical field strengths in this layer are measured to be $<0.01$\% of the peak field strength). Such regions make a negligible contribution to the helicity, due to the field strength weighting, but can have a significant effect on the winding calculations. To remove such potentially spurious information we will, in some calculations, modify the definition of the indicator function $\sigma_z$ to be 
\begin{equation}
\label{modsig}
\sigma_z({\boldsymbol  x}) = \left\{
\begin{array}{cc}
1 &\mbox{if $B_z({\boldsymbol  x})>0$ and $\vert{\boldsymbol  B} \vert>\epsilon $},\\
-1 &\mbox{if $B_z({\boldsymbol  x})<0$  and $\vert{\boldsymbol  B} \vert>\epsilon $},\\
0 &\mbox{if $B_z({\boldsymbol  x})=0$ or $\vert{\boldsymbol  B} \vert\leq \epsilon $}.
\end{array}
\right.
\end{equation}
Where we use  this definition in the article, we will be explicit and specify the value of $\epsilon$. Otherwise it is to be assumed that $\sigma_z$ has its usual definition (\ref{sigma1}).

\subsection{Winding and helicity ratios}
Both the helicity and winding densities can be positive and negative corresponding to right and left handed field entanglement. For mixed helicity flux ropes there can be significant entanglement, {locally}, but little overall average. This is true of the mixed helicity fields ${\boldsymbol B}_b$ which have zero {total} helicity. To quantify whether the helcity input for a given simulation is biased to one sign or is mixed, we calculate the following ratios
{\begin{align}
H_t^r  = \frac{{\rm d}{H}}{{\rm d}{t}}\Big/\frac{{\rm d}H^{abs}}{{\rm d}t},\quad \frac{{\rm d}H^{abs}}{{\rm d}t} =  \frac{1}{2\pi}\int_{P}\int_{P}\left| B_z(\av_1)B_z(\av_2)\frac{{\rm d}\theta_{12}({\boldsymbol a}_1,{\boldsymbol a}_2,t)}{{\rm d}t}\right| {\rm d}x_1{\rm d}y_1\,{\rm d}x_2{\rm d}y_2,\\
L_t^r  =  \frac{{\rm d}L}{{\rm d}t}\Big/\frac{{\rm d}L^{abs}}{{\rm d}t},\quad \frac{{\rm d}L^{abs}}{{\rm d}t} =   \frac{1}{2\pi}\int_{P}\int_{P}\left| \sigma_z(\av_1)\sigma_z(\av_2)\frac{{\rm d}\theta_{12}({\boldsymbol a}_1,{\boldsymbol a}_2,t)}{{\rm d}t}\right| {\rm d}x_1{\rm d}y_1\,{\rm d}x_2{\rm d}y_2.
\end{align}}

\subsection{Weighted velocity flux ${\cal V}_z$}
The plasma flow across the photosphere will transpire to be an important quantity in what follows. We calculate the net rate of plasma flow through the photosphere weighted by the absolute plasma flow rate,
\begin{equation}
{\cal V}_z = \frac{\int_{P}u_z {\rm d}x{\rm d}y}{\int_{P}\vert u_z\vert {\rm d}x{\rm d}y}. 
\end{equation}

\section{Simulation analysis}
\subsection{Twisted flux emergence}
First, we analyze the helicity input of a twisted field with a {non-dimensional twist of $\phi=-0.4$ and a negative (left handed) chirality (the other parameters are as stated in Section \ref{sec:braid})}. This value of twist is a common choice for flux emergence simulations \citep[e.g.][and references therein]{hood20123d} and represents a level of twist slightly below what would be required for the onset of the kink instability.

\subsubsection{Physical characteristics of emergence}
\begin{figure}
 \subfloat[$t=35$]{\includegraphics[width=\wdth]{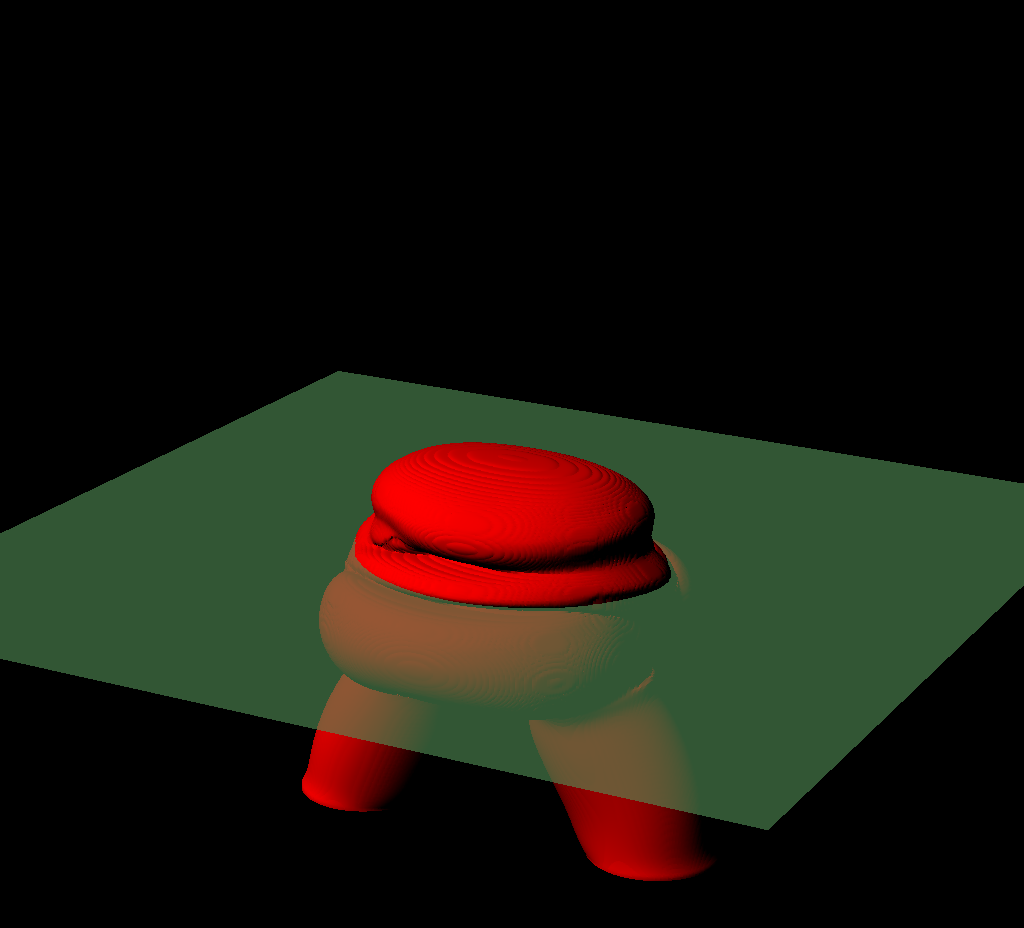}}\quad\subfloat[$t=35$]{\includegraphics[width=\wdth]{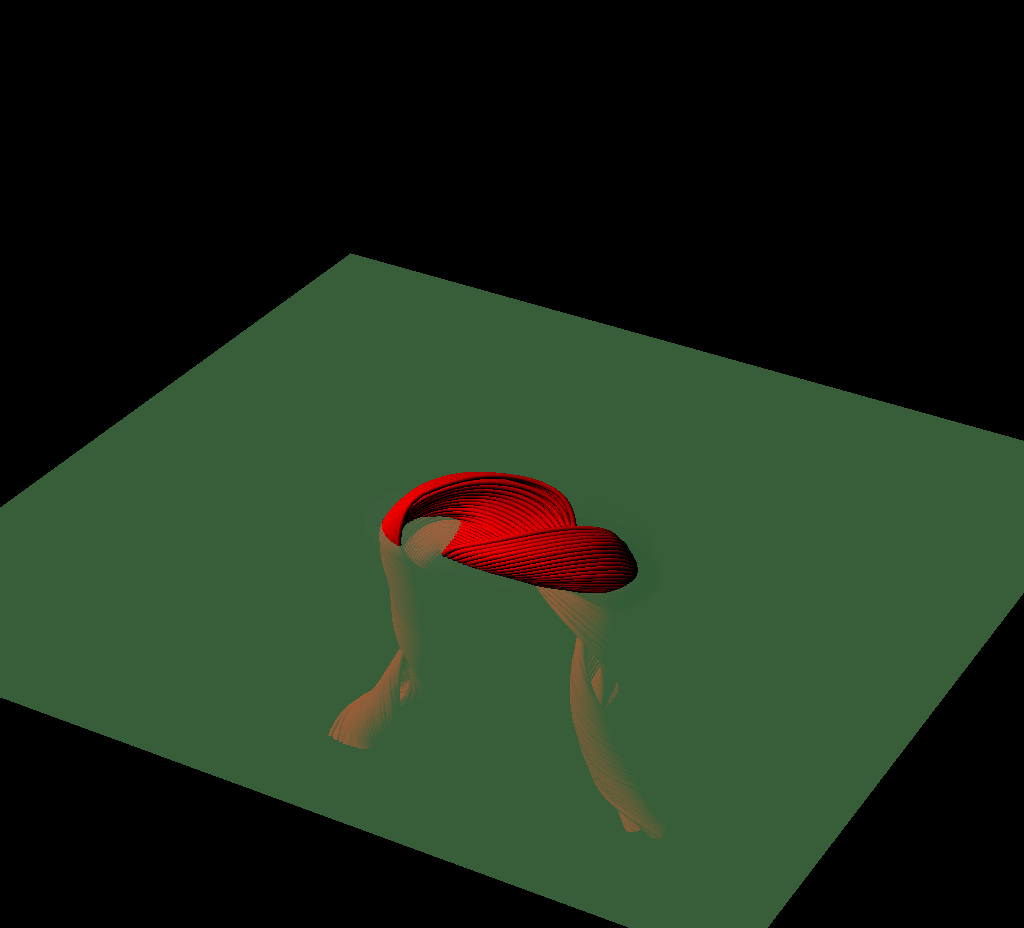}}\quad\subfloat[$t=35$]{\includegraphics[width=4.5cm]{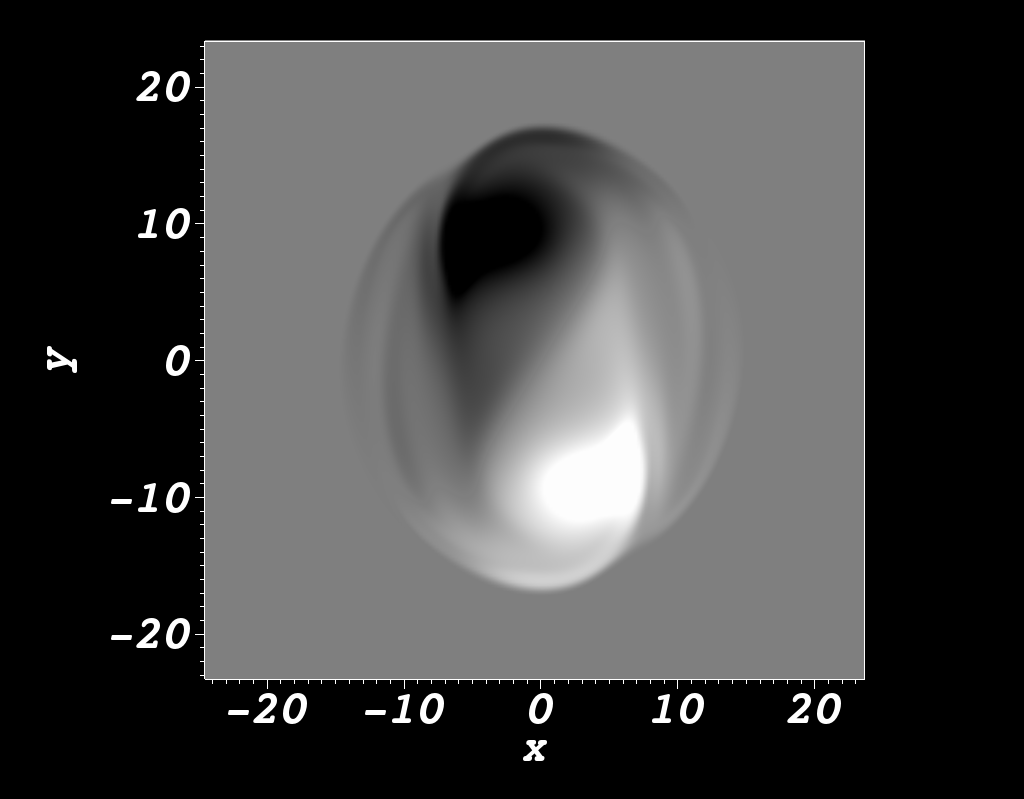}}\quad
   \subfloat[$t=55$]{\includegraphics[width=\wdth]{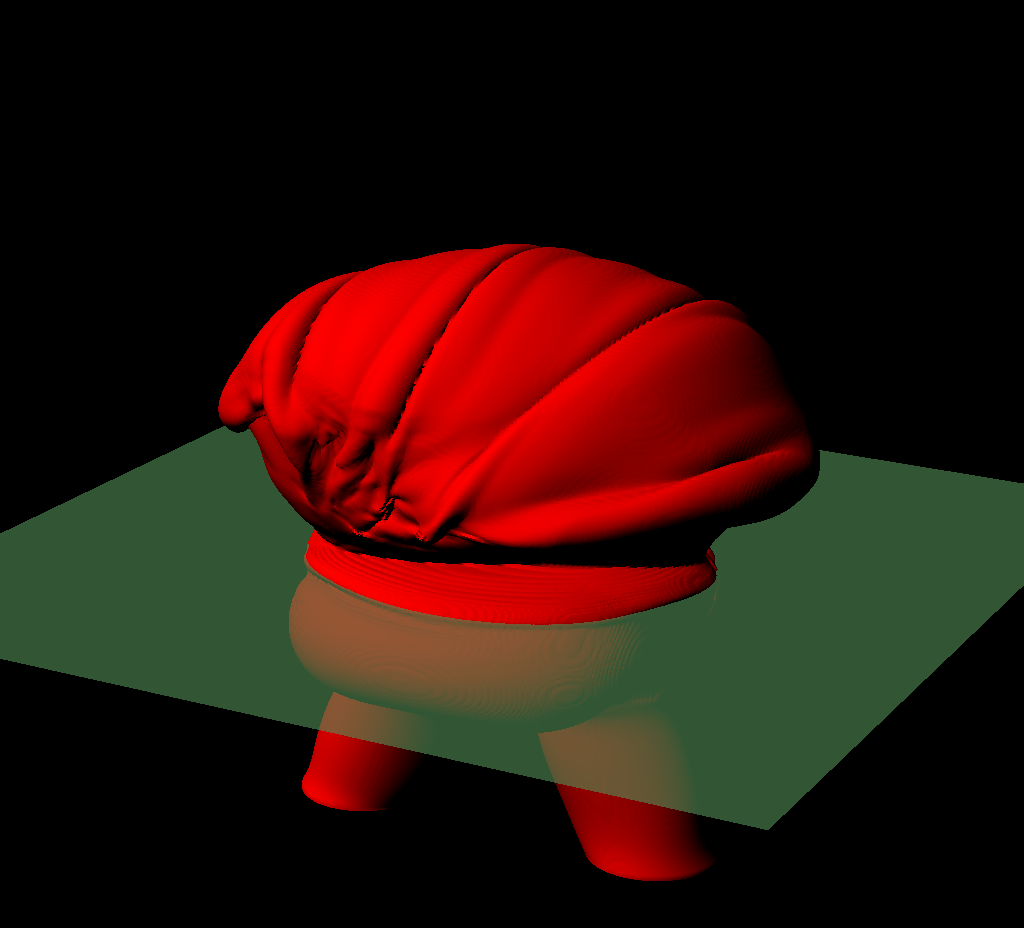}}\quad\subfloat[$t=55$]{\includegraphics[width=\wdth]{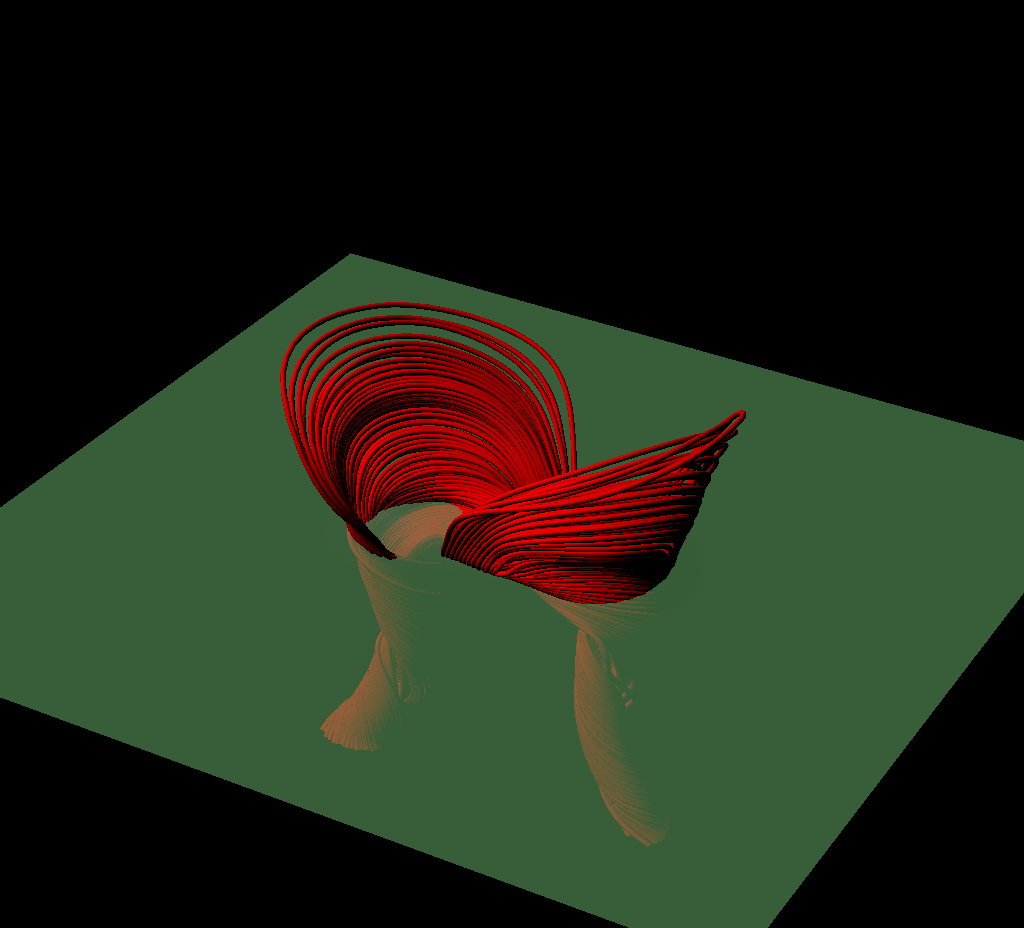}}\quad\subfloat[$t=55$]{\includegraphics[width=4.5cm]{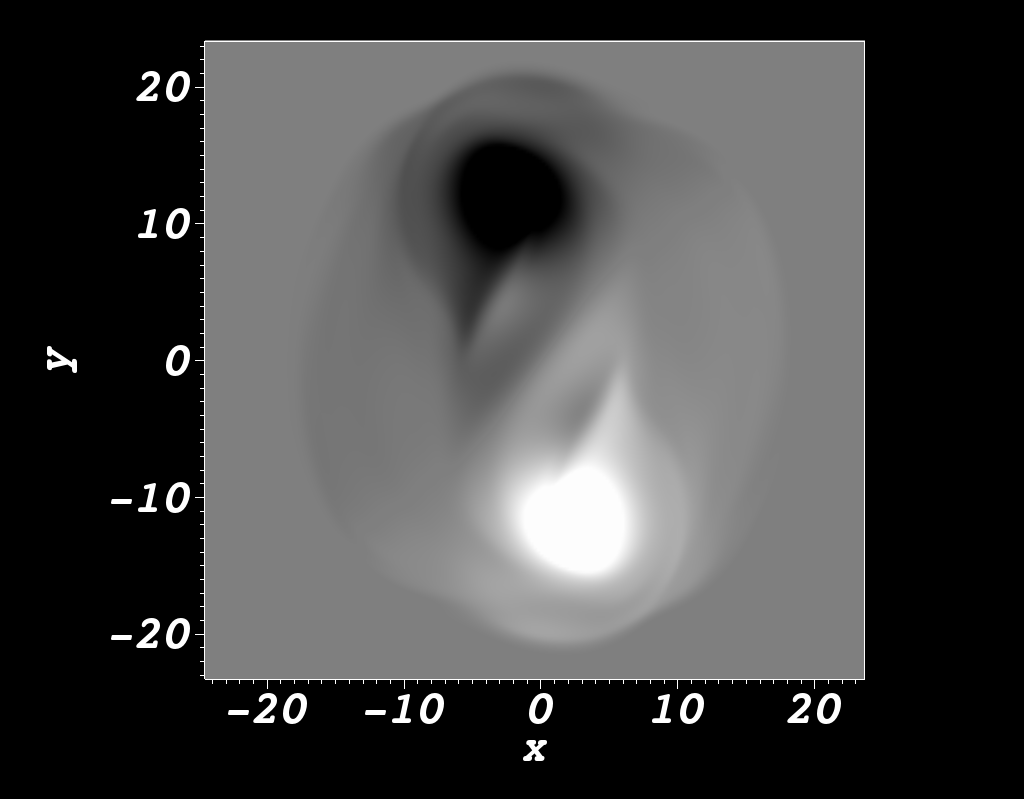}}\quad
   \subfloat[$t=71$]{\includegraphics[width=\wdth]{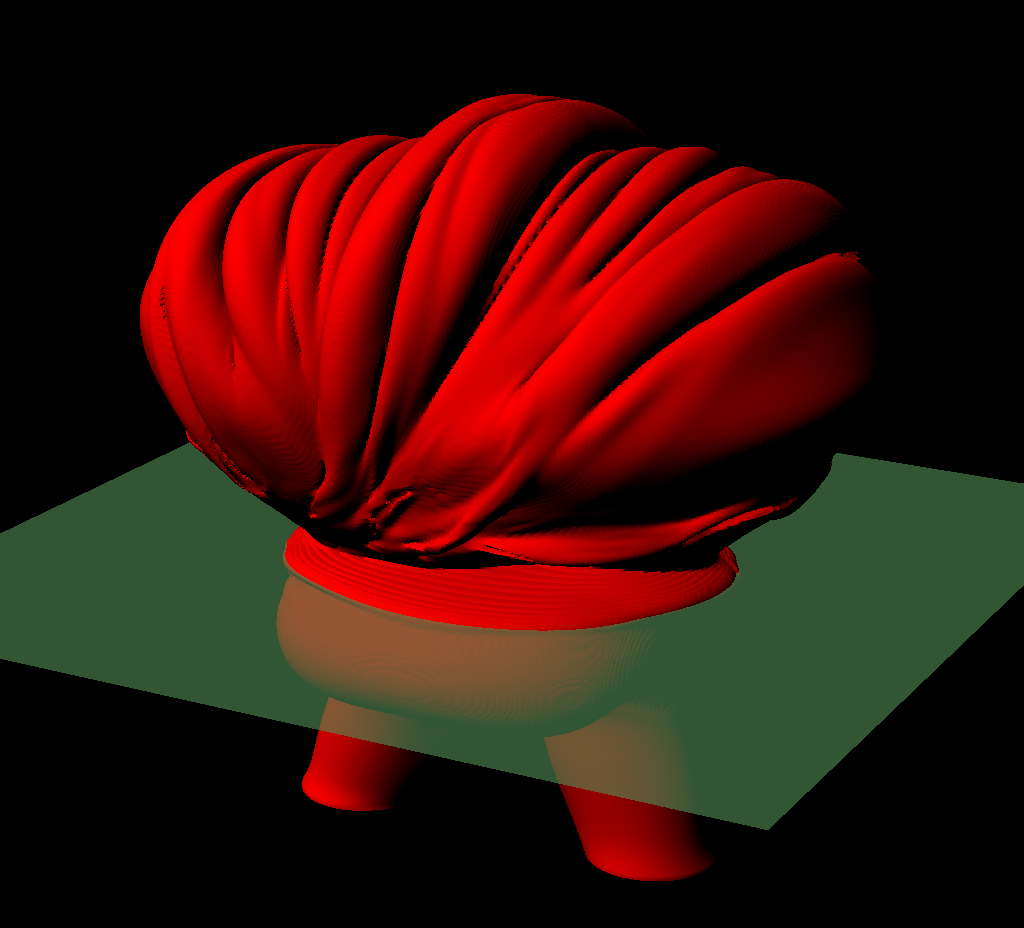}}\quad\subfloat[$t=71$]{\includegraphics[width=\wdth]{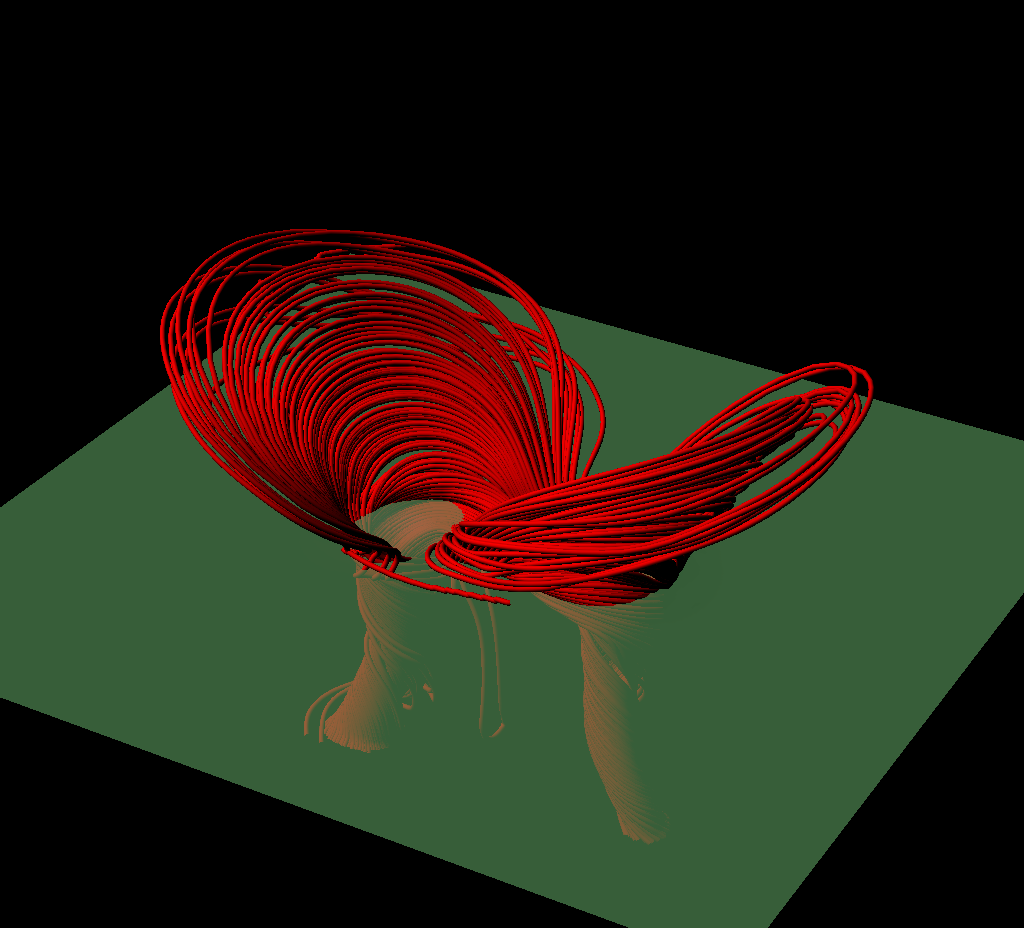}}\quad\subfloat[$t=71$]{\includegraphics[width=4.5cm]{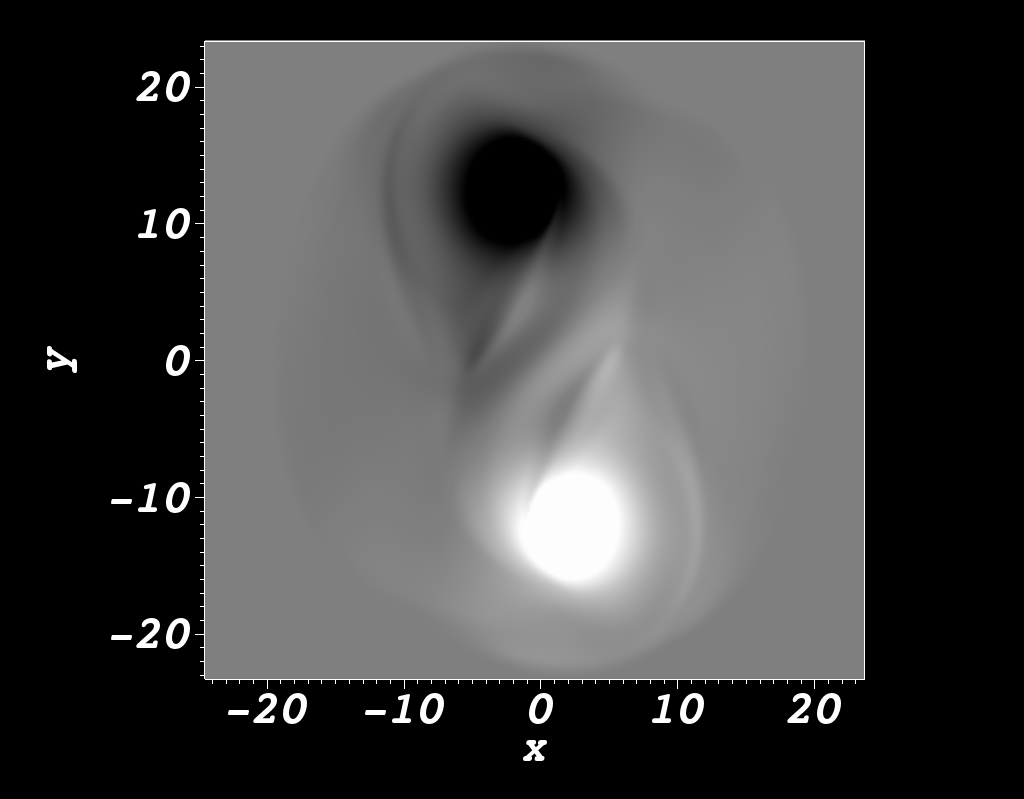}}
\caption{\label{twistemergence} Illustrative distributions characterizing the emergence of the twisted flux rope into the Sun's atmosphere. (a),(d), and (g)  are contour plots of the current density, showing the buoyancy instability-triggered rise and expansion into the corona. Also indicated is the plane $z=0$. (b), (e), and (h) represent subsets of the field lines at these times. (c), (f) and (i) are the corresponding magnetograms with a clear bipole structure.}   
\end{figure}
\begin{figure}
 \subfloat[]{\includegraphics[width=7cm]{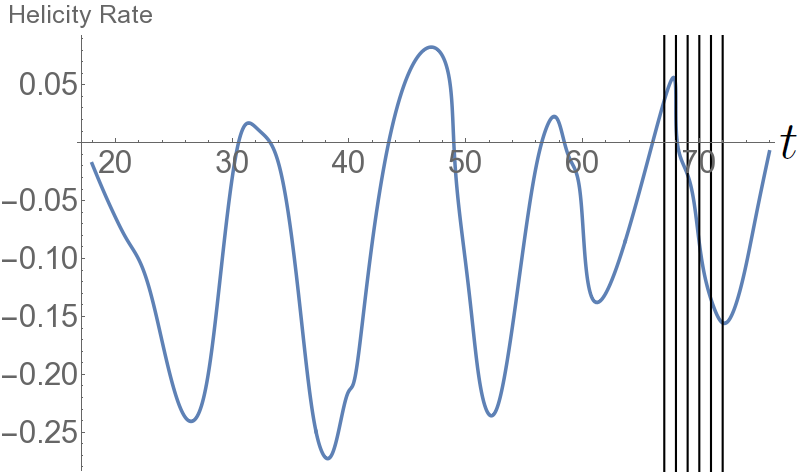}}\quad\subfloat[]{\includegraphics[width=7cm]{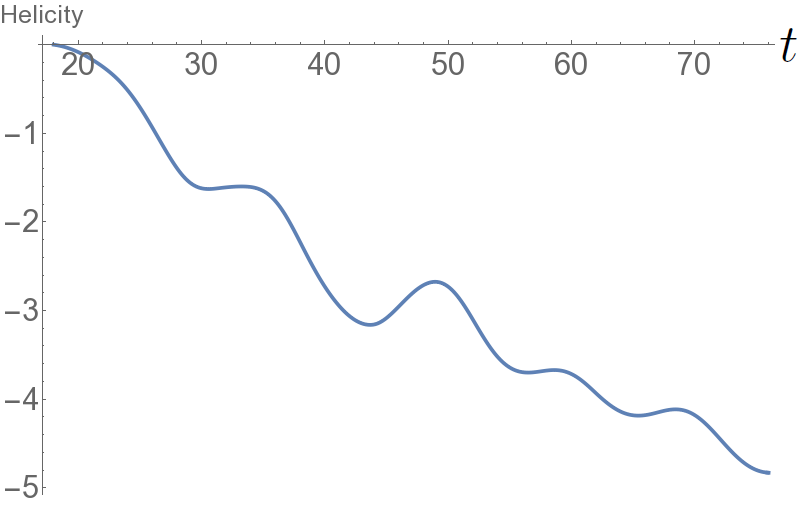}}\quad\subfloat[]{\includegraphics[width=7cm]{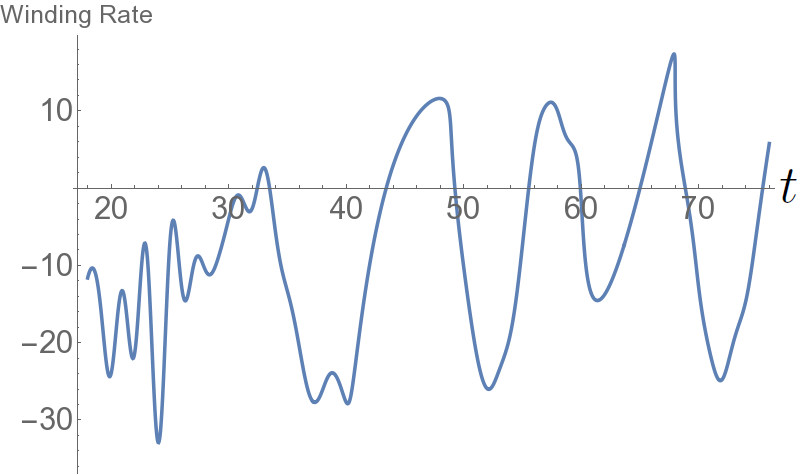}}\quad\subfloat[]{\includegraphics[width=7cm]{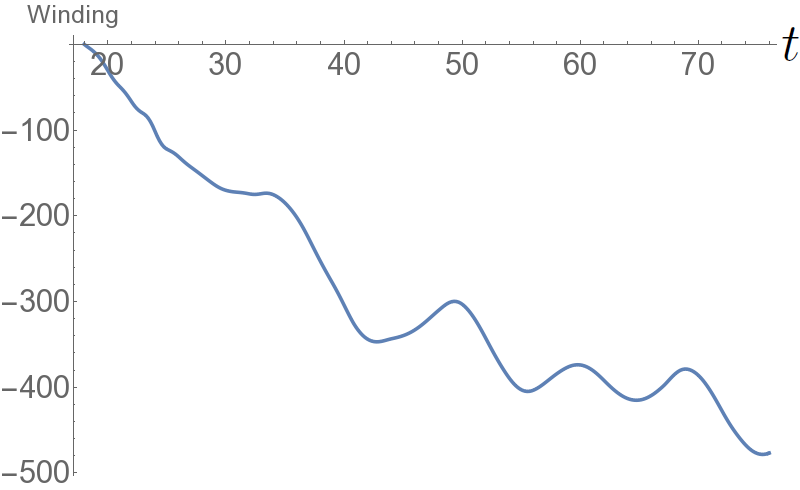}}
  \caption{\label{twistedhelcityinput}Photospheric helicity and winding input for the emergence of the twisted field. (a) the helicity input rate ${\rm d}{H}/{\rm d}{t}$. It is negative on average and has a consistent oscillation.  The vertical lines indicate the times $t=67$-$72$ at which the field's helicity distribution is analyzed in what follows. (b) the net helicty input $H(t)$ which is always negative and increases in magnitude over time. (c) the winding input rate ${\rm d} L/{\rm d}{t}$ and (d)  the total winding input $L(t)$.}
\end{figure}
A description of the emergence of twisted flux tubes has been described in detail in many other studies \citep[e.g.][]{mactaggart2009emergence,hood20123d}  so we only highlight aspects critical to the subsequent helicity input analysis. Illustrative figures are shown in Figure \ref{twistemergence} at times $t=35,55$ and $71$. {The magnetic field reaches the photosphere where it remains until the plasma beta drops to order of unity and the field becomes subject to the magnetic buoyancy instability, subsequently expanding into the corona \cite[]{hood20123d}}. The initial toroidal flux tube leads to a bipole emergence which is clearly visible in the magnetograms. 

\begin{figure}
\subfloat[]{\includegraphics[width=6.5cm]{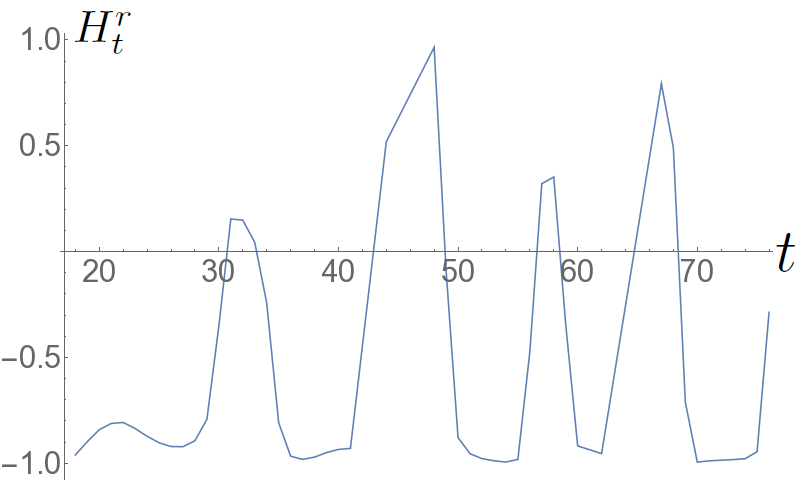}}\quad\subfloat[]{\includegraphics[width=6.5cm]{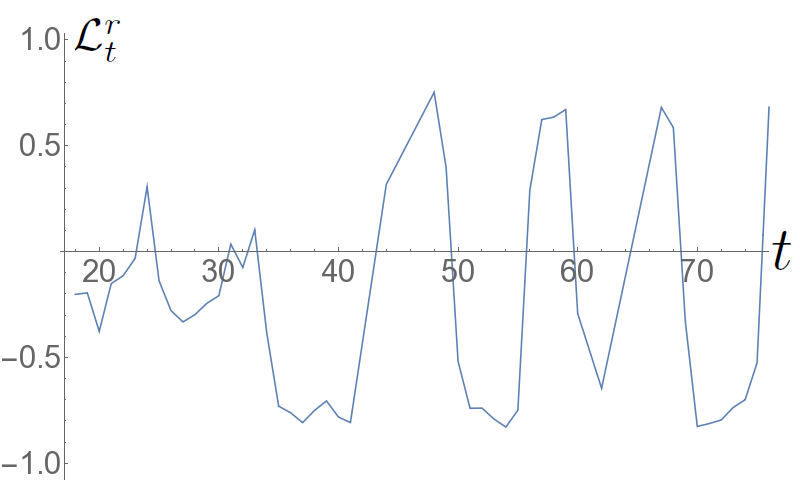}}\quad
\caption{\label{twistedratios}Time series of the ratios (a) $H_t^r$ and (b) $L_t^r$ for the twisted field emergence. } 
\end{figure}
\subsubsection{Helcity and winding input time series}
The temporal evolution of both ${\rm d}{H}/{\rm d}{t}$  and $H$ are shown in Figures \ref{twistedhelcityinput}(a) and (b) respectively. The tracking of the helicity input begins when the magnetic field reaches the photospheric boundary which, in this simulation, is almost coincident with the onset of the magnetic buoyancy instability ($t=18$). With the initial expansion of the field into the corona there is an input of negative helicity (Figure \ref{twistedhelcityinput}(a)), as is to be expected for this negative helicity flux rope \cite[]{mactaggart2009emergence}. Later, the helicity input settles  {into an oscillatory pattern with the input rate, occasionally, becoming net positive}. The net helicity input $H$, shown in Figure \ref{twistedhelcityinput}(b), exhibits, on average, an almost linear input increase of negative helcity (in line with the results of \cite{sturrock2015sunspot} accounting for the opposing sign of twist) with a smaller oscillation about this trend.  The time series ${\rm d} L/{\rm d}{t}$ and $L$, shown in Figures \ref{twistedhelcityinput}(c) and (d), are qualitatively similar to their helicity counterparts so we do not focus on them in what immediately follows. In contrast to these results, the oscillations in the ${\rm d}{H}/{\rm d}{t}$ time series calculated in \cite{sturrock2015sunspot} are relatively small (though similarly coherent) and do not cause the change in sign we find here. We will expand more on this comparison later. Finally, the ratios $H_t^r$ and ${L}_t^r$, whose time series are shown in Figure \ref{twistedratios},  indicate, as expected, a preference towards one sign of helicity  input, {i.e.} the quantity is often close to 1 in magnitude so  that most of the winding/helicity density is coherent in sign (the sign, of course, oscillates from positive to negative).  An important question to address is what is causing the oscillations and, further, why are they so significant in comparison to previous studies? 

\begin{figure}
\subfloat[]{\includegraphics[width=6cm]{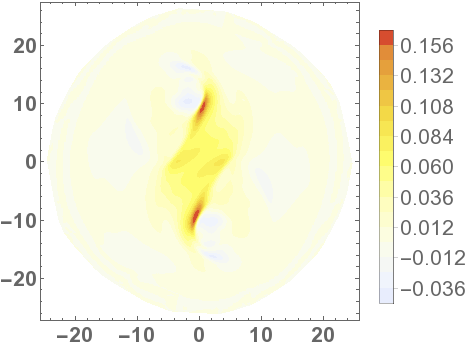}}\quad\subfloat[]{\includegraphics[width=6cm]{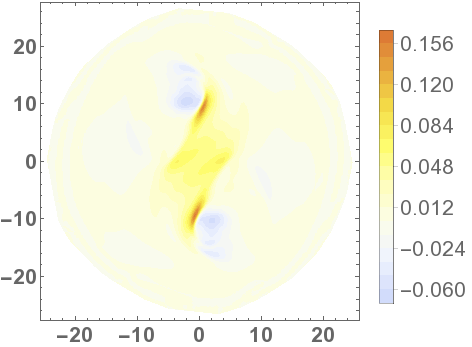}}\quad\subfloat[]{\includegraphics[width=6cm]{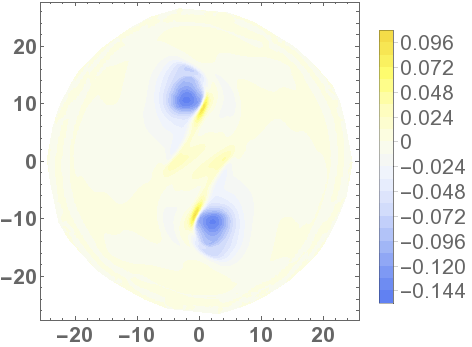}}\quad\subfloat[]{\includegraphics[width=6cm]{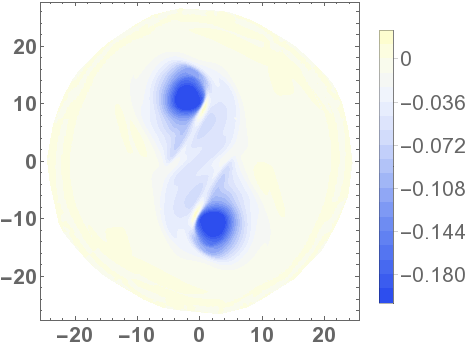}}\quad\subfloat[]
{\includegraphics[width=6cm]{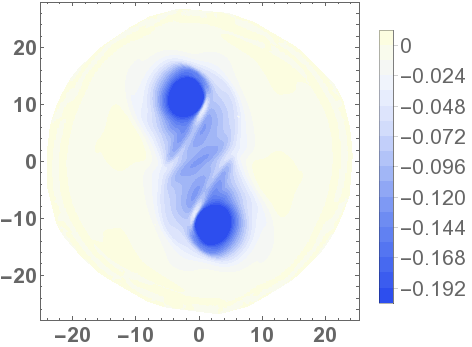}}\quad\subfloat[]
{\includegraphics[width=6cm]{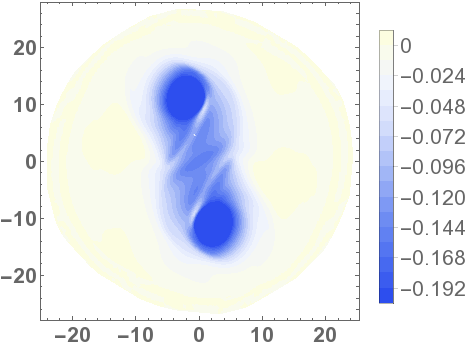}}
\caption{\label{twisthelinputvary}Helicity input rate density distributions $\d {\cal H}/\d t$ of the emerged region during the period $t\in[67,72]$ over which the sign of $\d H/\d t$ varies form positive to negative (net).  The plots shown in (a)-(f) correspond times $t=67$-$72$ indicated by the set of vertical lines on Figure \ref{twistedhelcityinput}(a). In panel (a) the density is dominantly positive around the PIL.  In (b)-(f), the helicity input at the PIL changes from dominantly positive to negative. The helicity input in the two flux poles becomes dominantly negative over the cycle.}
\end{figure}

\begin{figure}
\centering
\subfloat[]{\includegraphics[width=6.5cm]{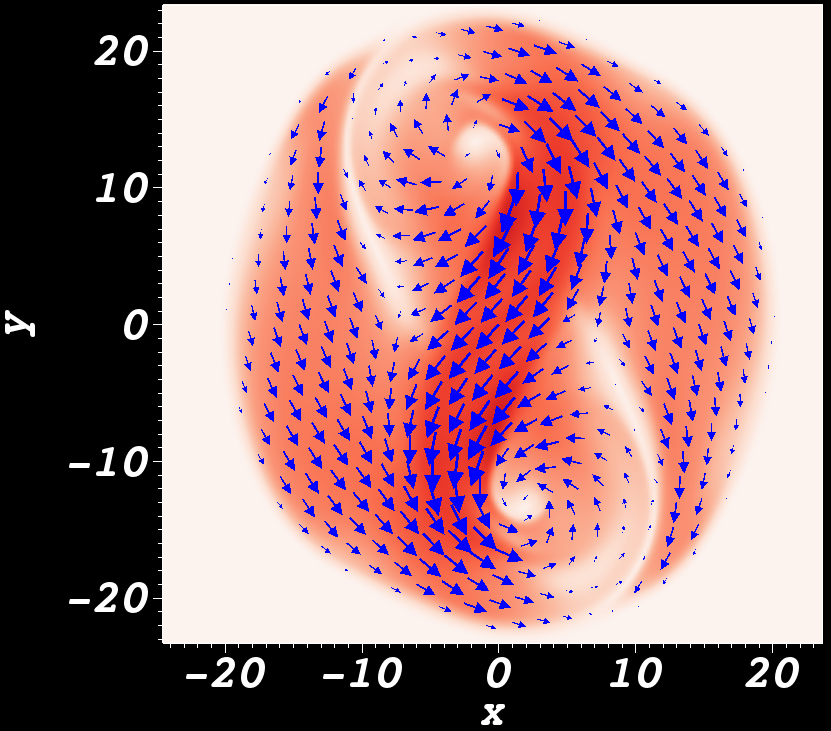}}\quad\subfloat[]{\includegraphics[width=6.5cm]{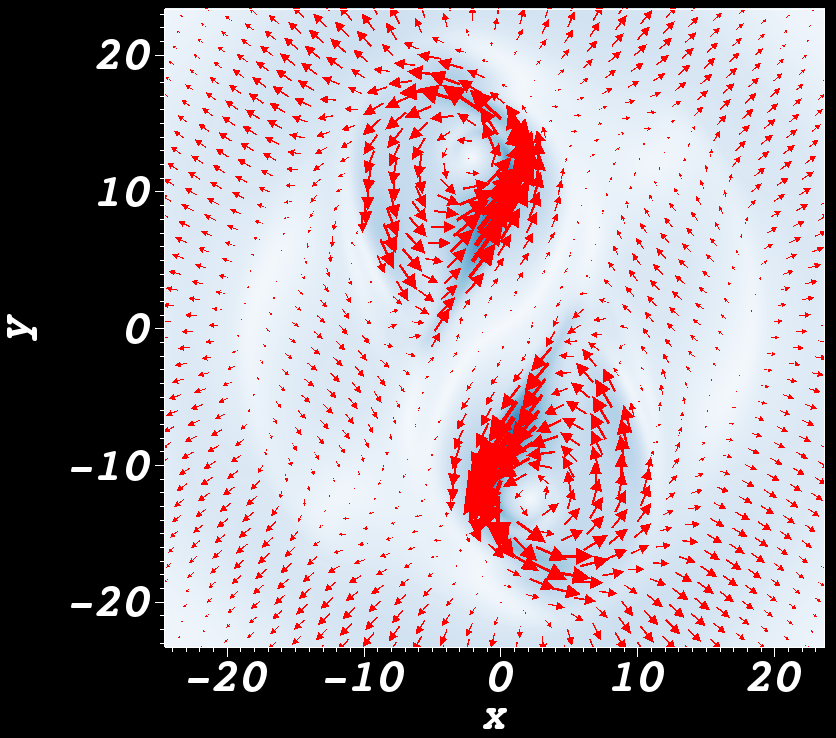}}
\caption{\label{twistedFieldDiagnostics}Magnetic and velocity field distributions in the photosphere at $t=67$. (a) the transverse field ${\boldsymbol  B}_{\parallel}$ superimposed on a scalar plot of its magnitude (stronger red colouring implies a stronger field). The twist at the positive pole is left-handed and the twist at the negative pole is right handed. (b)  the transverse velocity field ${\boldsymbol  u}_{\parallel}$ superimposed on a scalar plot of its magnitude (stronger blue coloring implies a stronger field). Both vorticies have right handed rotation. }
\end{figure}

\begin{figure}
\centering
\subfloat[]{\includegraphics[width=6.5cm]{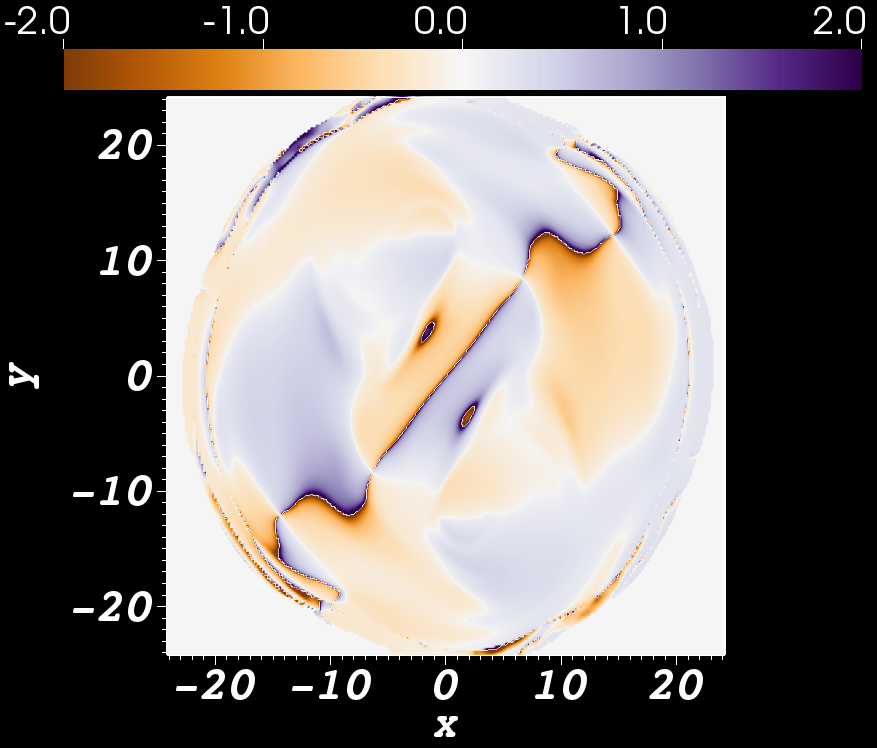}}\quad\subfloat[]{\includegraphics[width=6.5cm]{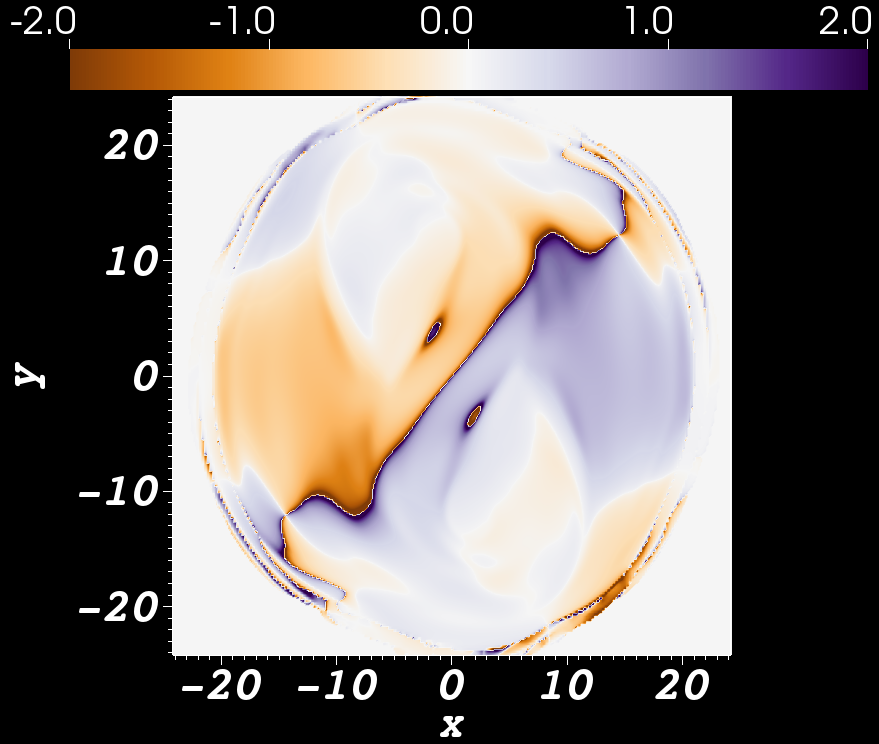}}\quad\subfloat[]{\includegraphics[width=6.5cm]{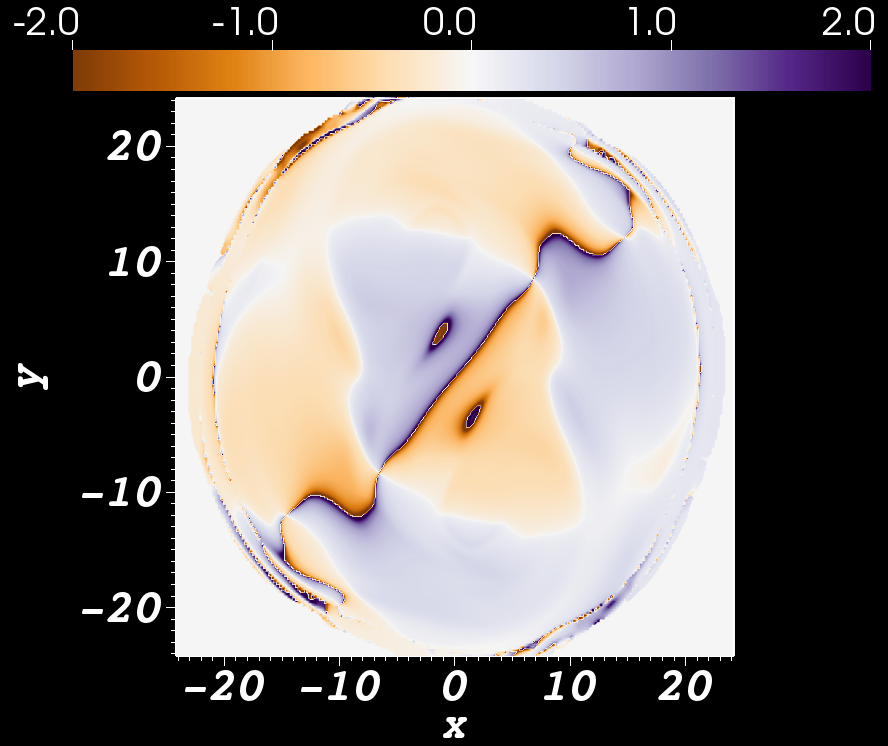}}\quad\subfloat[]{\includegraphics[width=6.5cm]{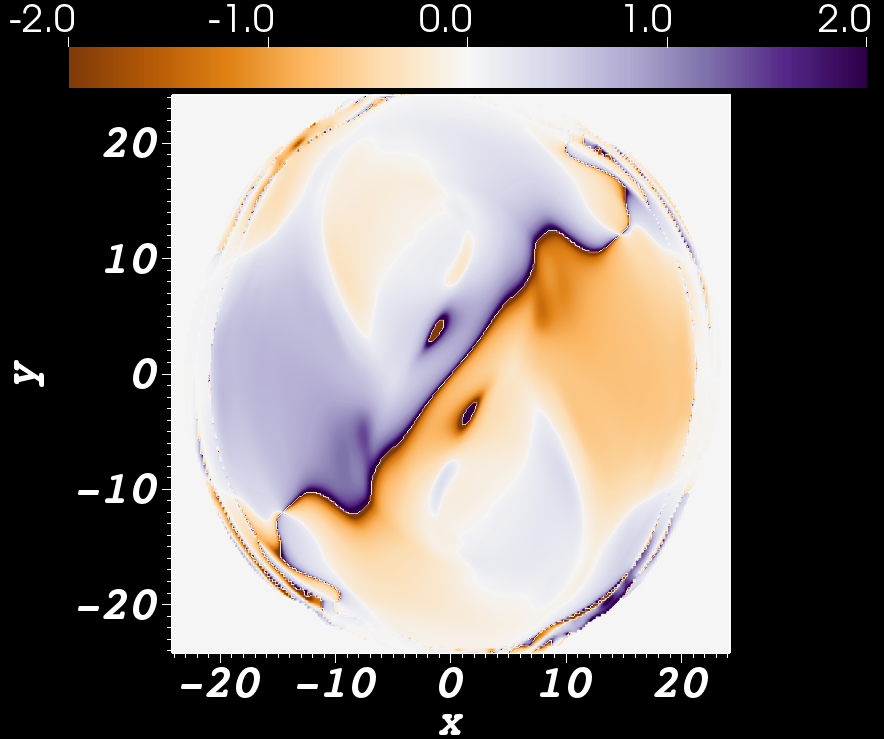}}
\caption{\label{twistucomps} Distributions of the components of the helicity producing field ${\boldsymbol w}$, at times $t=67$ and $72$. Panels (a) and (b) are the distributions of $\sqrt{w_x({\boldsymbol a}_0)}$ and $\sqrt{w_y({\boldsymbol a}_0)}$ respectively at $t=67$. Panels (c) and (d) are the same distributions but at $t=72$. For both  $w_x({\boldsymbol a}_0)$ and $w_y({\boldsymbol a}_0)$, the sign of the distribution either side of the PIL reverses.}
\end{figure}

We now focus on the time period $t\in[67,72]$, indicated in Figure \ref{twistedhelcityinput}(a) by a set of vertical lines. There is a variation between positive and negative input rates over this period. There is no particular reason why we choose to present the results of this period over the rest of the input cycle (except the rise stage). Indeed we checked that the following analysis would have led to similar conclusions for the behaviour of the system over the period $t\in[40,76]$.

{Distributions of ${\rm d}\mathcal{H}/{\rm d}t$, during the considered time interval}, are shown in Figure \ref{twisthelinputvary}. At $t=67$ (a) and $t=72$ (f) there are respectively strong positive and then negative field line helicity input densities around the polarity inversion line (PIL). For the intermediate times (b)-(d), the field line helicity density in this region shows a gradual variation from positive to negative. Patches of strong negative density at the two poles of flux distribution  also develop over the period. There are fluid vorticies, in the in-plane velocity field, which are centered on these poles. It is shown in Figure \ref{twistedFieldDiagnostics} that at the magnetic field's positive pole, the vortex opposes the direction of magnetic twist, whilst at the negative pole the signs of the vortical and magnetic twists agree. We might speculate that some kind of relative balance between the twisted field input and fluid rotation might lead to the oscillation in helicity input around the inversion line. However, we now show this oscillation occurs instead due to the cyclic submergence and re-emergence of the {flux}, {that is, the movement of part of the magnetic field below and above the photospheric plane}. In what follows, \emph{submergence} does not necessarily indicate that the field passes deep beneath the photosphere. As will be demonstrated, any part of the field that makes contact with the photospheric boundary, and perhaps passing only slightly beneath it, will register a signal in the magnetograms and in the helicity and winding inputs. Therefore, in this work, submergence is any previously emerged field moving down from the atmosphere or otherwise making contact with the photosphere from above.

\subsection{Helicity sign change around the PIL}
\begin{figure}
\subfloat[]{\includegraphics[width=6.5cm]{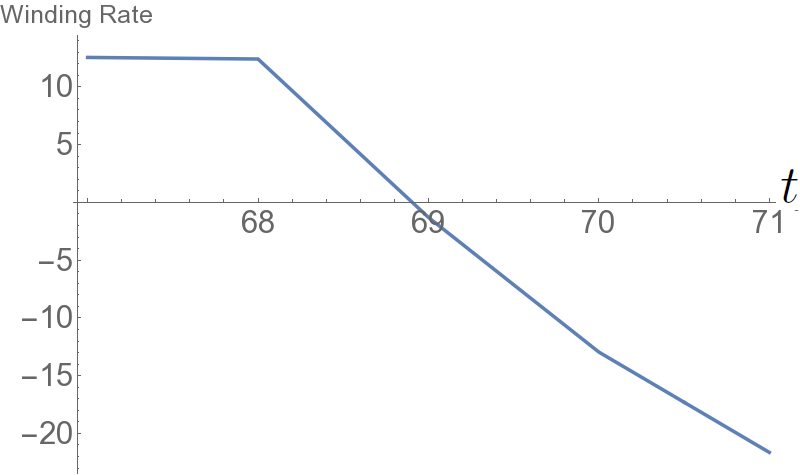}}\quad \subfloat[]{\includegraphics[width=6.5cm]{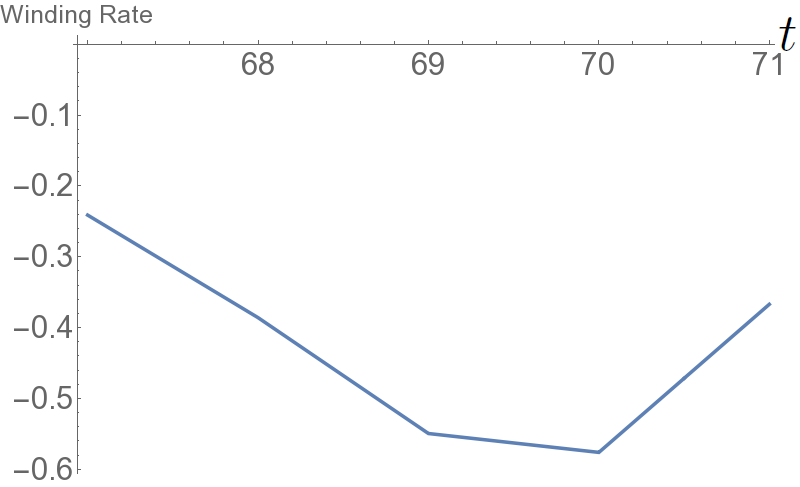}}
\caption{\label{cappedhel}Plots of the winding input rate over the period $t=67$ to $71$ for (a) the unrestricted calculation $\d L/ \d{t}$ and (b) the capped version $\d L^c/ \d{t}$ for which the threshold is $c=2$.}
\end{figure}
The distributions of the $x$ and $y$-components of ${\boldsymbol w}=\d {\av}/\d t$ at $t=67$ and $t=72$ are shown in Figure \ref{twistucomps}. The most significant helicity producing velocities arise in the neighbourhood of the PIL (the plots are of $\sqrt{w}_x$ and $\sqrt{w}_y$ {are shown} for clarity, thus relatively exaggerating the magnitudes in weaker regions). There is a clear switch in sign of both the $w_x$ and $w_y$ densities on both sides of the PIL over this cycle. {To test if it is the temporal change in the sign of the field line velocity field ${\boldsymbol w}$ close to the PIL which determines the change in sign of the helicity input}, we consider a modified vector field ${\boldsymbol w}^{c}$ as follows
\begin{equation}
  {\boldsymbol w}^c = \left\{
  \begin{array}{cc}
    {\boldsymbol w} & \mbox{ if } \norm{{\boldsymbol w}}\leq c,\\
    0 & \mbox{ if } \norm{{\boldsymbol w}}>c, 
  \end{array}
  \right.
\end{equation}
where $c$ is some chosen threshold. We choose $c=2$ here, which represents a cut-off speed much smaller than typical field line speeds at the PIL ($\approx10$). As long as the cut-off is small enough, the following results are robust.  The main behaviour of the following results was also obtained for $c=1$ and $c=5$.

In essence, we cut out the higher helicity producing velocities which reside primarily in the region of the PIL and {not} at the centres of the magnetic footpoints. We also calculate the modified field line winding input rate $\d  L^c/ \d t$ using ${\boldsymbol w}^c$. {We see in Figure \ref{cappedhel}(a) that the full winding input rate $\d L/\d t$ changes sign from positive to negative over the time period $t\in[67,72]$. By contrast, for the restricted input $\d  L^c/ \d t$ (b),  the sign of winding input is always negative over this period and two orders of magnitude smaller. Therefore, ignoring the winding input due to the field line motions ${\boldsymbol w}$ around the PIL leads to a time series which does not have a positive input rate over this period (the story is the same for the helicity input rate). It was confirmed that this is true throughout the simulation (for the time period when the oscillations in $\d H/ \d t$ occur).}

\begin{figure}
\centering
  \subfloat[]{\includegraphics[width=6cm]{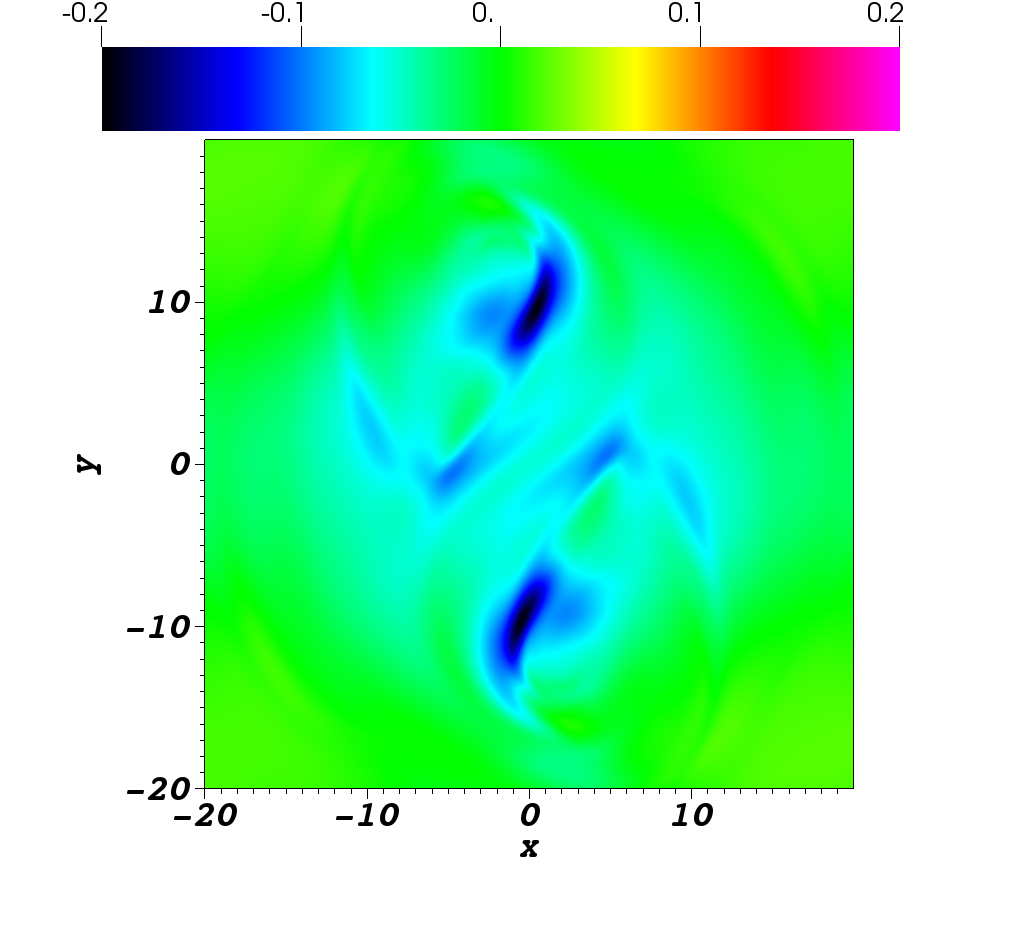}}\quad\subfloat[]{\includegraphics[width=6cm]{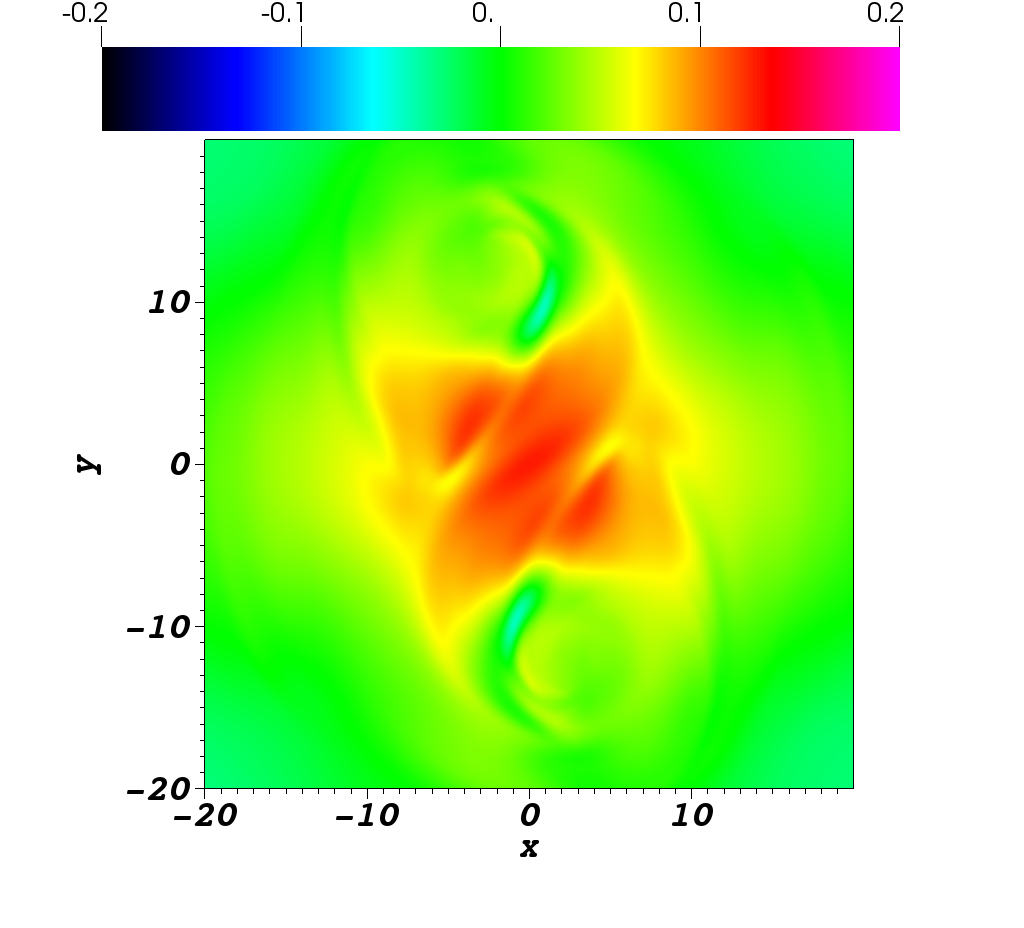}}\quad \subfloat[]{\includegraphics[width=6.5cm]{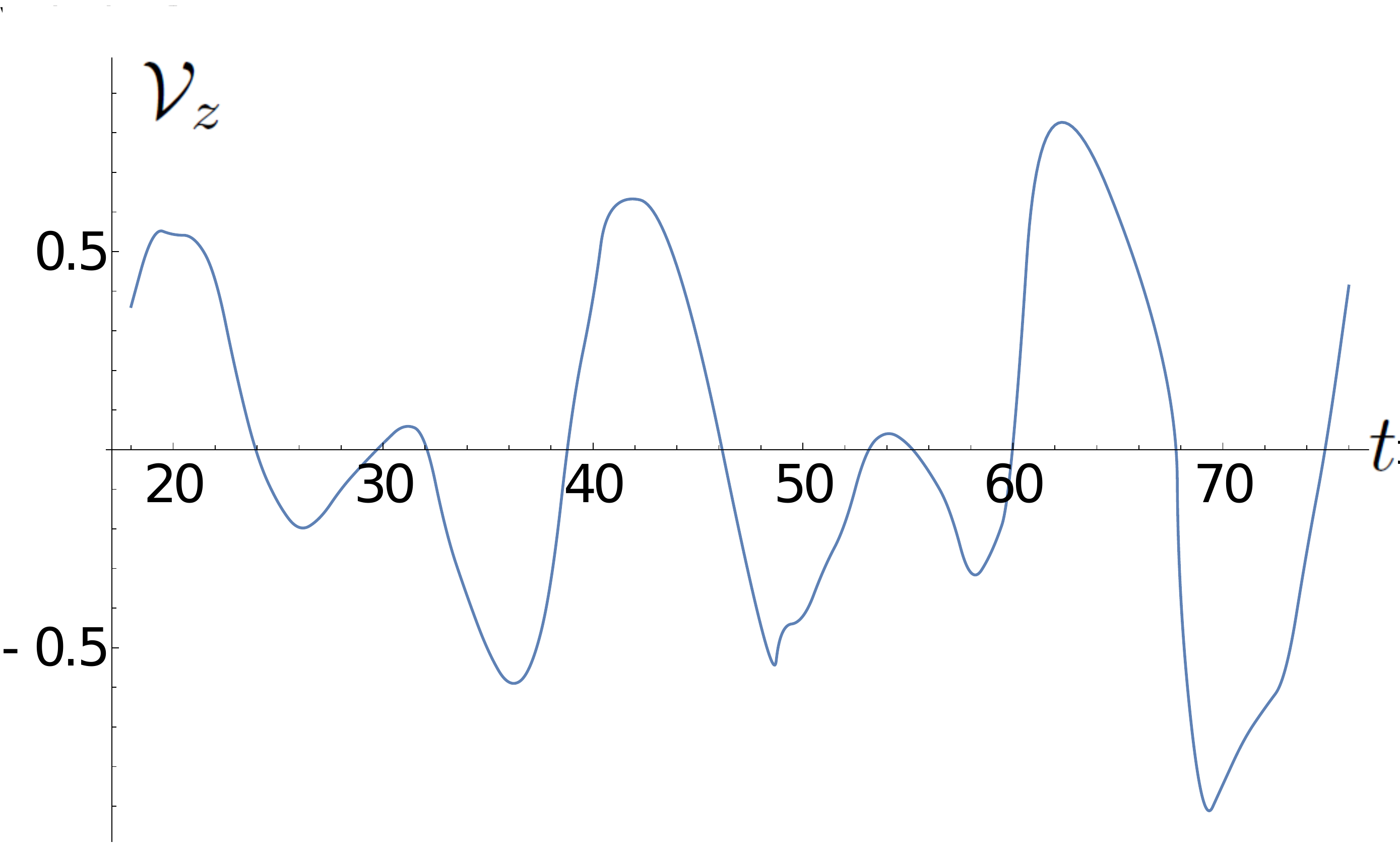}}\quad
\subfloat[]{\includegraphics[width=6.5cm]{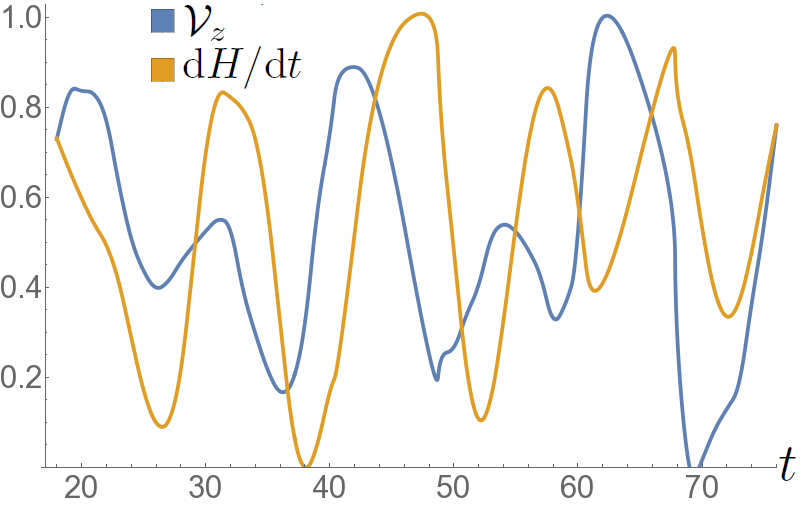}}\caption{\label{velocityvariation} Distributions of the vertical velocity $u_z({\boldsymbol a}_0)$ at times (a) $t=67$ and (b) $72$. (c) a plot of the net velocity flux ${\cal V}_z$ through $P$. (d)  plots of $\d H/ \d t$ and ${\cal V}_z$  scaled to have values between $0$ and $1$.}
\end{figure}

The cause of the helicity sign change is found to be {related to} a switch in the sign of $u_z$ in the region around the PIL. To establish this, we first evaluate the relative magnitudes of the terms ${\boldsymbol  u}_{\parallel}$ and ${\boldsymbol  B}_{\parallel}u_z/B_z$. The submergence/emergence term ${\boldsymbol  B}_{\parallel}u_z/B_z$ is typically found to be an order of magnitude higher in the regions where ${\boldsymbol w}$ is (relatively) large. 
The $u_z$ distributions at the start ($t=67$) and end ($t=72$) of the cycle are shown in Figures \ref{velocityvariation}(a) and (b). Initially, there is a net negative velocity with most of the flow surrounding the PIL. Then at the end of the period the velocity is net positive with most of the flow being at the PIL. A time series plot of the net velocity {flux} ${\cal V}_z$ is shown in Figure \ref{velocityvariation}(c). A scaled comparison of the maxima and minima of $\d H/ \d t$ and ${\cal V}_z$ is shown in Figure \ref{velocityvariation}(d) emphasizing that it is motion across the photosphere that is dominating the helicity in this phase of emergence.



\subsubsection{Flux rope centre at the photosphere.}
\begin{center}
\begin{figure}
\centering
\subfloat[]{\includegraphics[width=11cm]{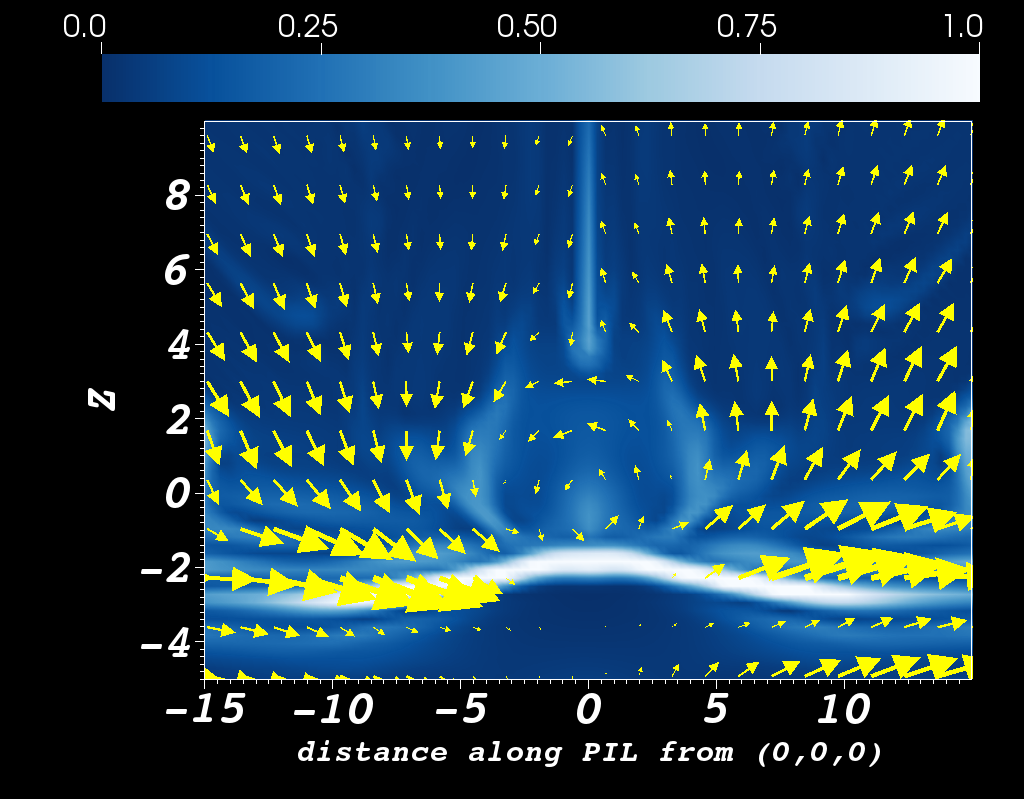}} 
\caption{\label{twistphoto}Arrows indicating the magnetic field in the plane orthogonal to the PIL at $(0,0,0)$ at $t=67$. This is superimposed on a plot of the current magnitude in the same plane. The core of the flux rope is clearly visible and is centered at the PIL.}
\end{figure}
\end{center}
The fact that the oscillations in helicity (and winding) are sufficient to lead to a sign of input which opposes the field's chirality is a result of the fact that the bulk of the field's initially twisted flux rope remains trapped at the photosphere, see Figure \ref{twistphoto}. The plasma $\beta$ $(=2p/|{\boldsymbol  B}|^2)$ has a value of about 5 at the photosphere in this simulation. This implies that the magnetic field is not dynamically dominant (as in the low-$\beta$ corona) and so can moved by the surrounding plasma. A combination of upward motion, from emergence, and downward motion, from draining plasma \citep[e.g.][]{hood20123d} causes the flux rope centre (axis) to oscillate about the $z=0$ plane. In this topologically simple field (relative to mixed helicity model that we will study shortly) the bulk of the topological information is concentrated at the flux rope centre. Therefore, if the centre crosses the photospheric plane, the response in the helicity and winding rates is large. As mentioned before, it is transport \emph{across} the photospheric plane and not (un)twisting motions \emph{on} the plane cause the largest changes in helicity and winding.

If the plasma $\beta$ were smaller, it would be expected that the oscillations would be less pronounced. In \cite{sturrock2015sunspot}, who perform a very similar simulation but with a flux rope of almost double the initial field strength to the one we consider, they still find oscillations in  $\d H/ \d t$ but much less pronounced than those found here (their oscillations do not change the sign of the helicity input rate). Two factors are important in contributing to this change in behaviour. The first is that the axes of flux tubes with higher field strengths can emerge further in the atmosphere compared to those with weaker field strengths \citep{mactaggart2009emergence}. This makes the flux rope centre further away from the $z=0$ plane and so there is less flux crossing the plane in any partial submergence event. Secondly, the plasma $\beta$ is smaller and the magnetic field is less susceptible to surrounding plasma motions. 



{
\subsection{Tracking the moving photosphere}
\begin{figure}
 \subfloat[]{\includegraphics[width=7cm]{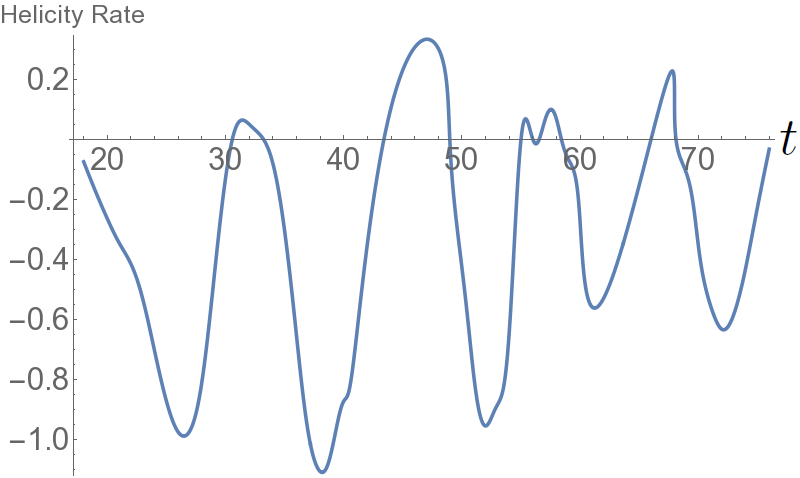}}\quad\subfloat[]{\includegraphics[width=7cm]{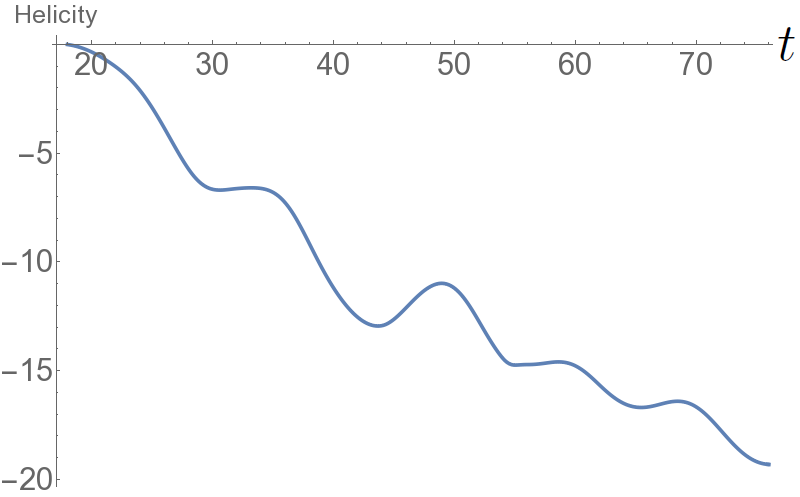}}\quad\subfloat[]{\includegraphics[width=7cm]{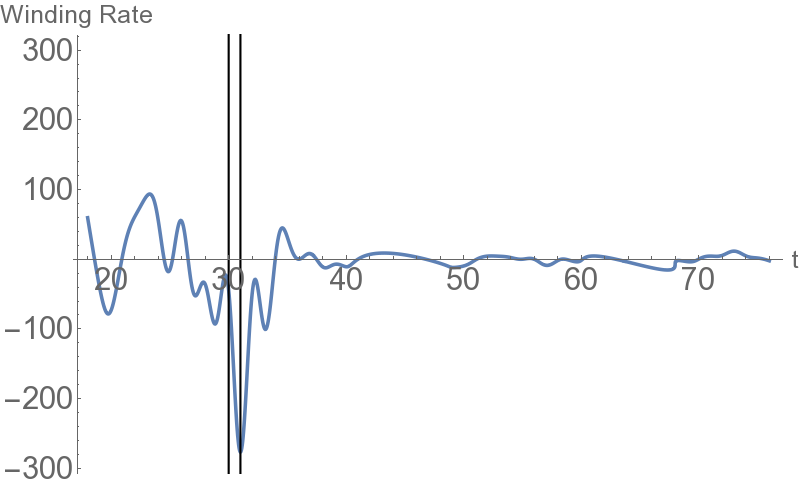}}\quad\subfloat[]{\includegraphics[width=7cm]{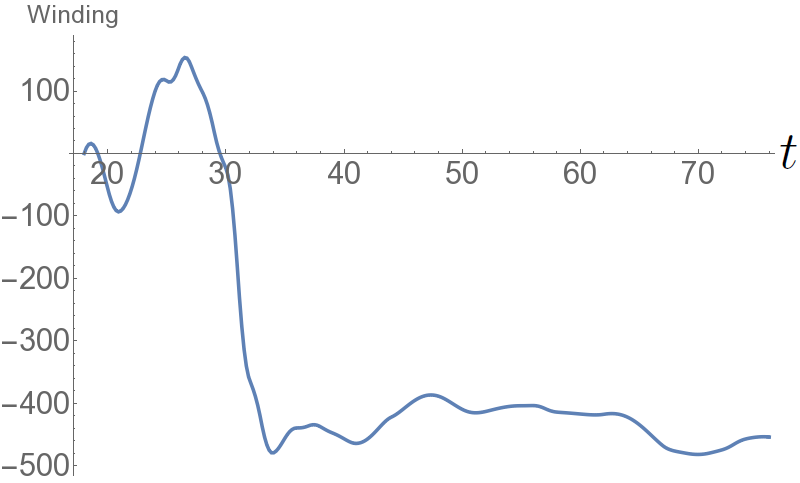}}
  \caption{\label{twistedhelcityinputnonEuc}Photospheric helicity and winding inputs for the emergence of the twisted field which account for the tracking of the changing photosphere geometry. (a) the helicity input rate ${\rm d}{H_{v}}/{\rm d}{t}$. (b) the net helicty input $H_{v}(t)$. (c) the winding input rate ${\rm d}L_{v}/{\rm d}{t}$.  The vertical lines are at times $t=30,31$ when the field structure is analyzed further in what follows. (d) the total winding input $ L_{v}(t)$.} 
\end{figure}
\begin{figure}
\subfloat[]{\includegraphics[width=7cm]{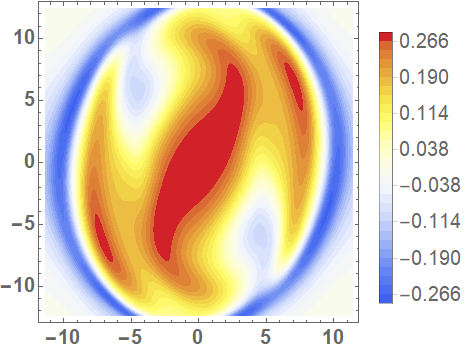}}\quad \subfloat[]{\includegraphics[width=7cm]{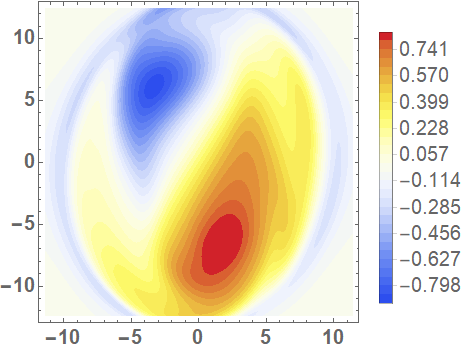}}
\caption{\label{velcompstwist}Vertical velocity distributions at $t=28$. (a) the velocity density $u_z(P_v(x,y))J$ at the moving photosphere line $\rho=1$. (b) the difference between the varying photosphere velocity density $u_z(P_v(x,y))J$  and the $z=0$ velocity density $u_z(x,y)$.}
\end{figure}
The velocity {flux} oscillations indicated in Figure \ref{velocityvariation} imply that the surface $\rho=1$ will be varying in space and time. As discussed in the introduction, we can track this variation and calculate modified quantities which account for this motion. The time series of the adjusted quantities ${\rm d}{H_{v}}/{\rm d}{t}$, $H_v$, ${\rm d}{L_{v}}/{\rm d}{t}$ and $L_v$ are shown in Figure \ref{twistedhelcityinputnonEuc}. On a qualitative level, the helicity time series (panels (a) and (b) respectively) are effectively the same as the $z=0$ calculations shown in Figure \ref{twistedhelcityinput}. The magnitudes, however, of the various peaks of the input rate ${\rm d}{H_{v}}/{\rm d}{t}$ are roughly four times larger than those of ${\rm d}{H}/{\rm d}{t}$. This magnification is due to increased \emph{variations} in velocities at the $\rho=1$ surface. As an example, Figure \ref{velcompstwist} displays the varying photosphere velocity  distribution $J u_z(P_v(x,y))$ in (a) and the difference $J u_z(P_v(x,y))-u_z(x,y)$ between the moving surface and static surface velocity distributions in (b). In (a), the strongest velocities are, as before, at the PIL. In (b), major differences are at the main footpoints. We note that the differences can be greater in magnitude than a typical velocity $J u_z(P_v(x,y))$.   Therefore, it is not simply a case of larger velocity magnitudes at the $\rho=1$ surface compared to the $z=0$ plane, rather a difference in the distribution of flows on these surfaces.

\begin{figure}
\centering
\subfloat[]{\includegraphics[width=13cm]{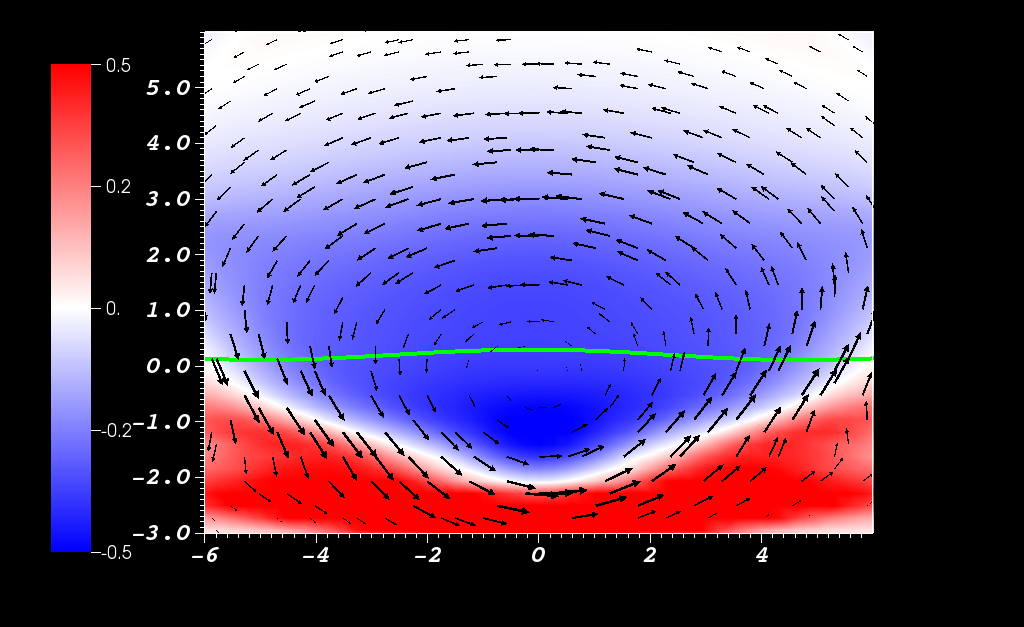}}
\caption{\label{o30}A slice of the vector field in the plane orthogonal to the PIL at $t=30$, just prior to the spike in the time series ${\rm d}{L_{v}}/{\rm d}{t}$ shown in Figure \ref{twistedhelcityinputnonEuc}(c). Also shown as a green curve is the intersection of the surface $P_v$ and this plane. The background distribution is the out of plane component of the current density.}
\end{figure}

\begin{figure}
\centering
\quad \subfloat[]{\includegraphics[width=13cm]{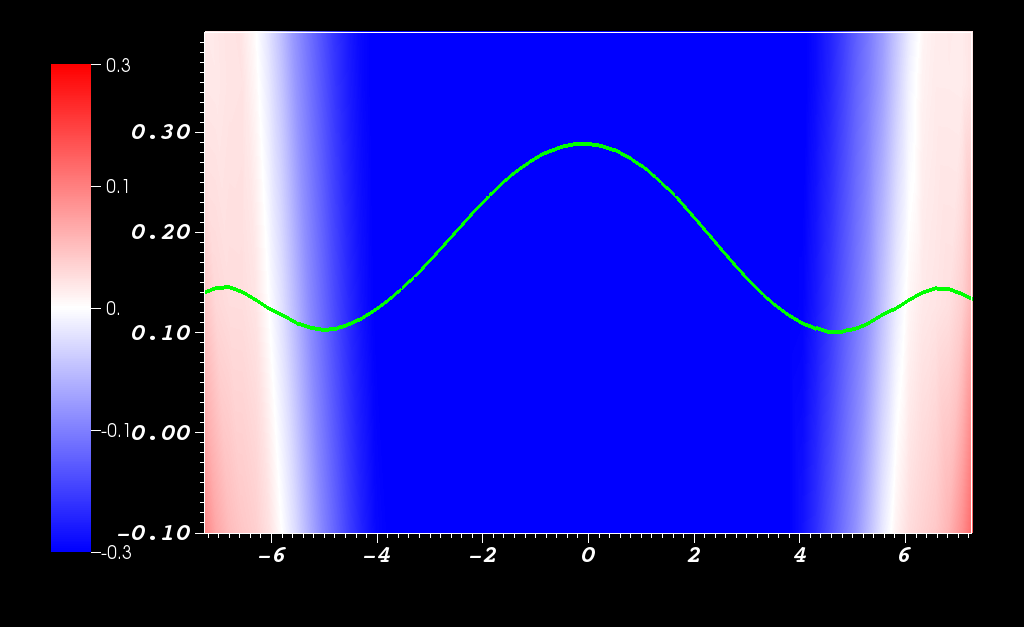}}\quad 
\subfloat[]{\includegraphics[width=13cm]{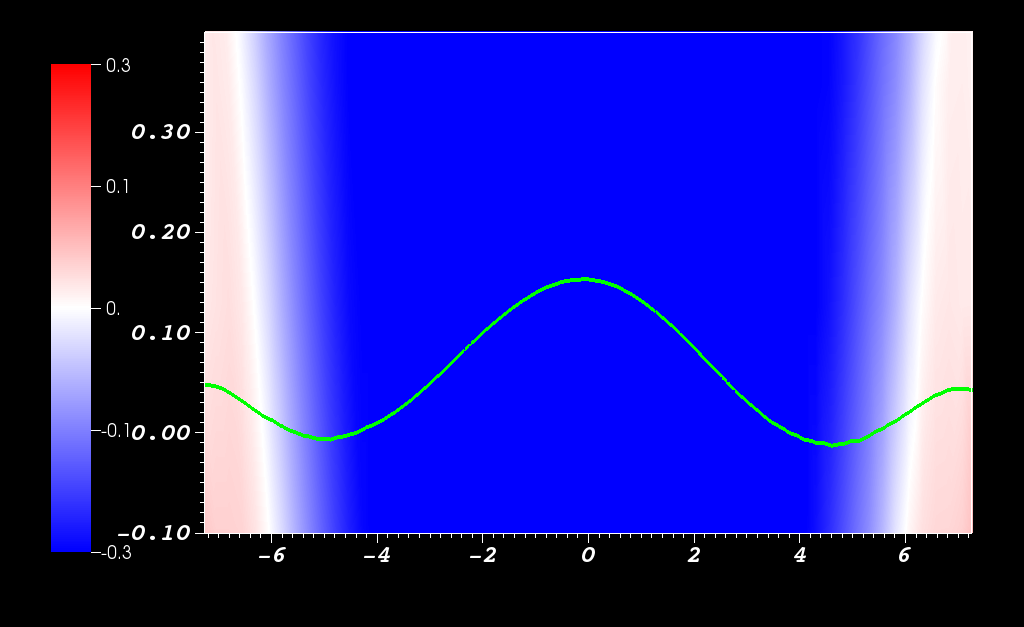}}
\caption{\label{twistedspikeslice}Magnified representations of a slice of the vector field in the plane orthogonal to the PIL at $t=30$ and  $t=31$, prior to and at the spike in the time series ${\rm d}{L_{v}}/{\rm d}{t}$ shown in Figure \ref{twistedhelcityinputnonEuc}(d). Shown as a green curve is the intersection of the surface $P_v$ and this plane. The background distribution is the out of plane component of the current density. (a) ($t=30$) a magnification of the region of Figure \ref{o30} which contains the curve $\rho=1$. (b) ($t=31$) the same distribution as in (a) but at the time of the spike. The $\rho=1$ curve has dropped vertically by a value of approximately $0.5$. }
\end{figure}

The winding time series, panels (c) and (d) of  Figure \ref{velocityvariation},  show significant  qualitative differences compared to the $z=0$ input calculations shown in panels (c) and (d) of Figure \ref{twistedhelcityinput}. There is one significant period of negative input rate ${\rm d}{L_{v}}/{\rm d}{t}$ at $t=31$. This is near the end of the period when the central part of the field (containing the original tube axis) emerges at the photosphere (see Figure \ref{o30}). The spike coincides with a relatively sharp (negative) change in the height of the $\rho=1$ surface from $t=30$ (Figure \ref{twistedspikeslice}(a)) to $t=31$ (Figure \ref{twistedspikeslice}(b)). This change results in a significant change in the negative helicity input as the flux rope core moves further beyond the photospheric boundary. It was confirmed that this jump in the moving photospheric surface was atypically large from $t=30$ to $t=31$ by comparison to typical changes in its morphology over single time step of size $1$. After this, the input rate ${\rm d}{L_{v}}/{\rm d}{t}$ falls to a relatively small rate. It is interesting to note that this occurs due to a relatively fast change in the geometry of the $\rho=1$ surface, rather than a sudden change in field topology, indicating that this may be an important factor in interpreting obsevational inputs of field topology. The oscillations shown in the helicty input rate ${\rm d}{H_{v}}/{\rm d}{t}$ are still present but, as shown in Figure \ref{twistedhelcityinputnonEuc}(d), these have a small effect on the net helicity injected into the photosphere, by comparison to the sharp negative input around $t=30$. 


It is intriguing that the (moving photosphere) winding input more clearly represents the transition between the emergence of the field's twisted core to the photosphere (the $\rho=1$ surface) followed by the lack of changing topological input due to the core getting stuck at this surface. The significantly reduced (at least on a relative scale) size of the oscillations in comparison to the helicity input results from the lack of field strength weighting in the winding input. The inclusion of field strength in the helicity input rate magnifies the oscillatory signal due to the central core of the field being  immediately surrounded by strong field - slight dips under the photospheric boundary produce high signals because of the strong field strengths. Another way to view this result is that the winding input treats the oscillation at the photosphere as not especially interesting at later times as it is not really indicating the input of new topological information into the atmosphere. 

}

\subsection{Mixed helicity emergence}
{We now} consider a mixed helicity field ${\boldsymbol  B}_b$, specified by (\ref{braidedfield}) with $n=2$. The emergence of {mixed helicity (braided) magnetic fields and their associated dynamics  was first described} in \cite{prior2016emergence}. 
\subsubsection{Physical characteristics of the emergence}
\begin{figure}
  \subfloat[$t=27$]{\includegraphics[width=\wdth]{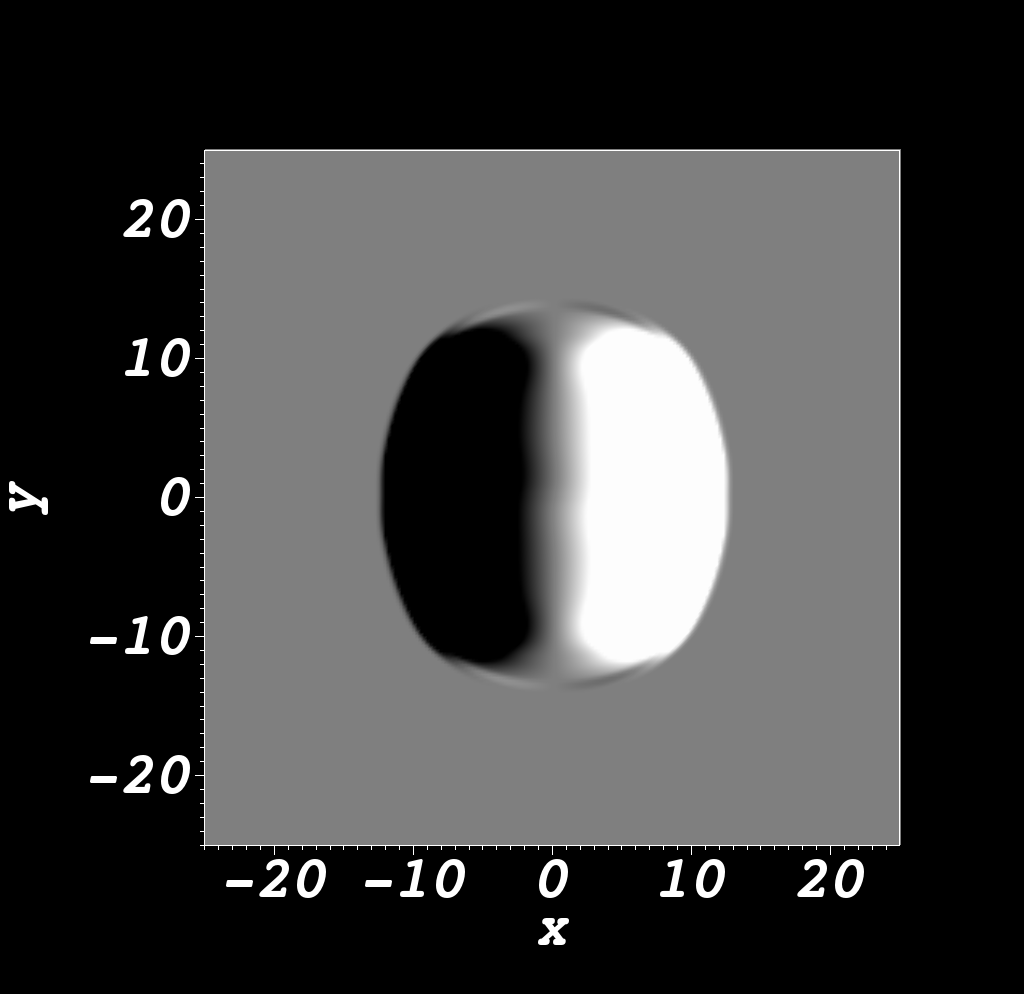}}\quad\subfloat[$t=27$]{\includegraphics[width=\wdth]{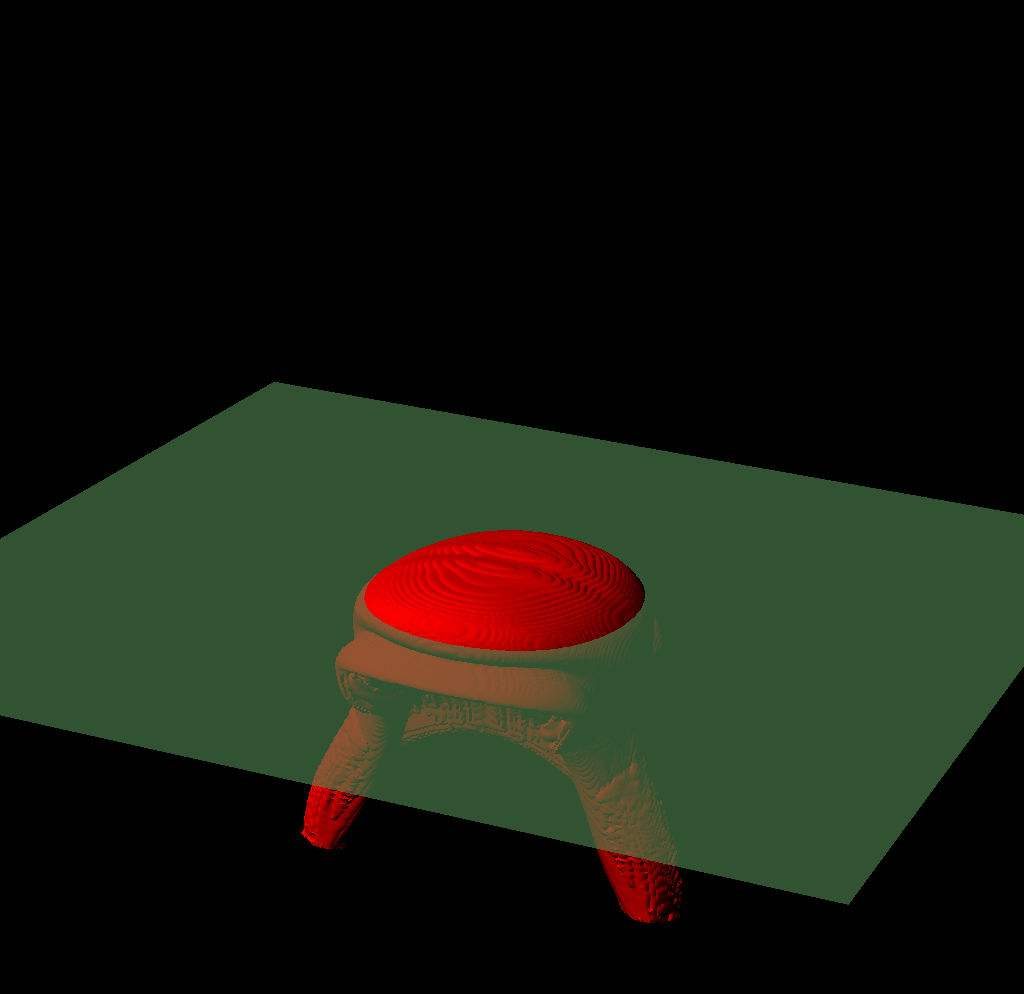}}\quad\subfloat[$t=27$]{\includegraphics[width=\wdth]{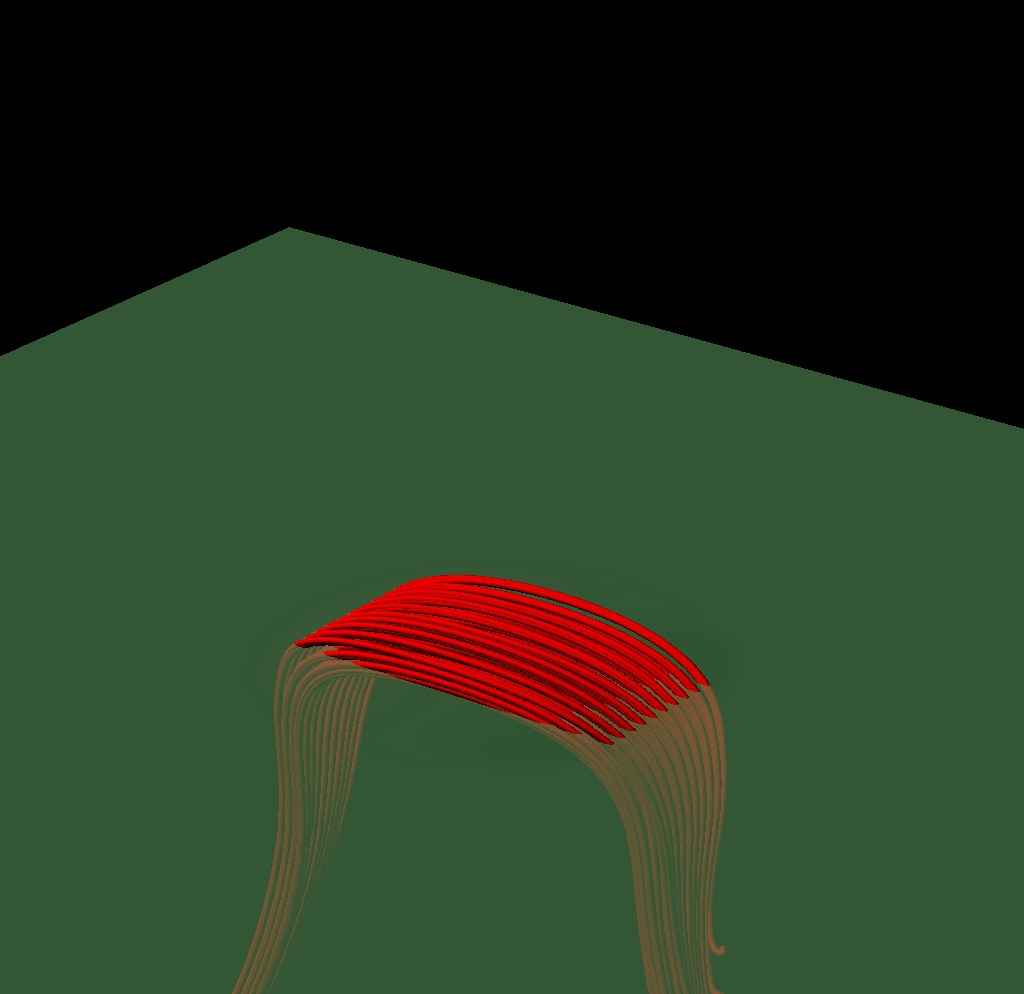}}\quad \subfloat[$t=38$]{\includegraphics[width=\wdth]{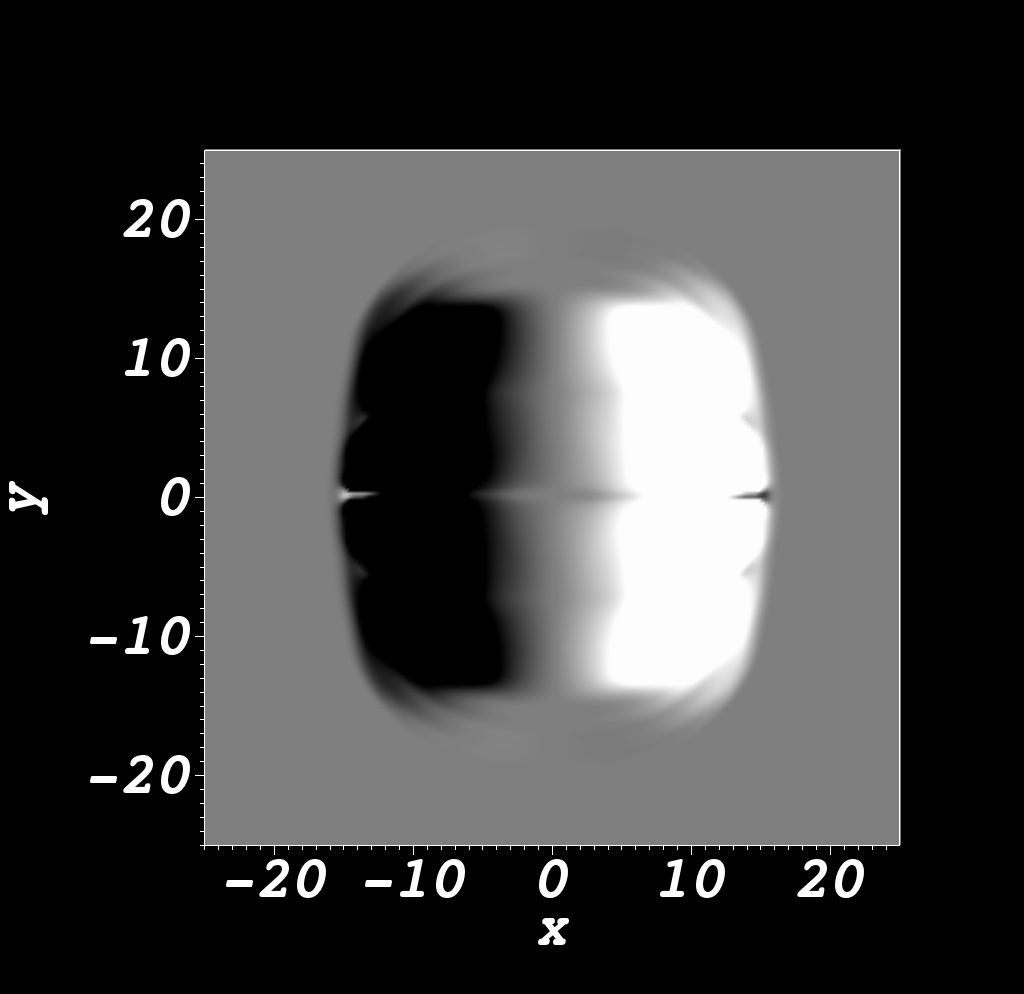}}\quad\subfloat[$t=38$]{\includegraphics[width=\wdth]{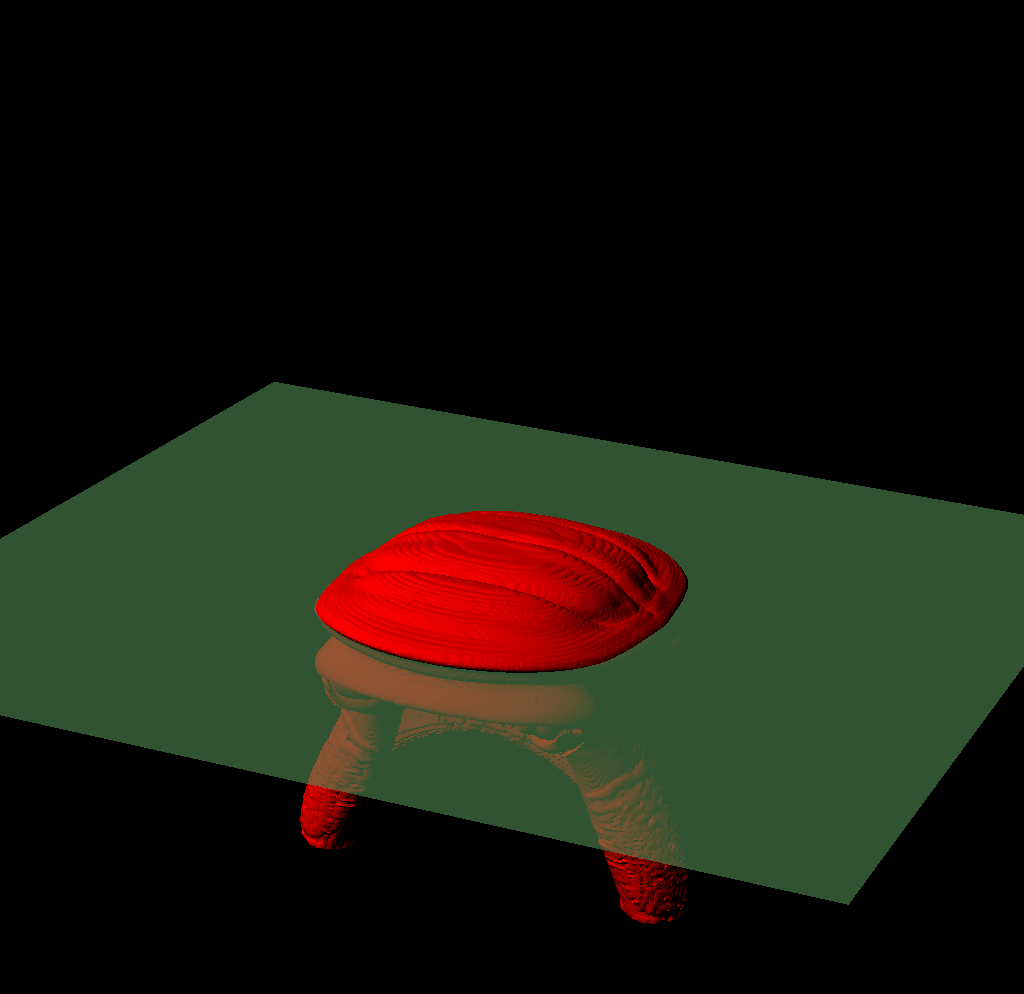}}\quad\subfloat[$t=38$]{\includegraphics[width=\wdth]{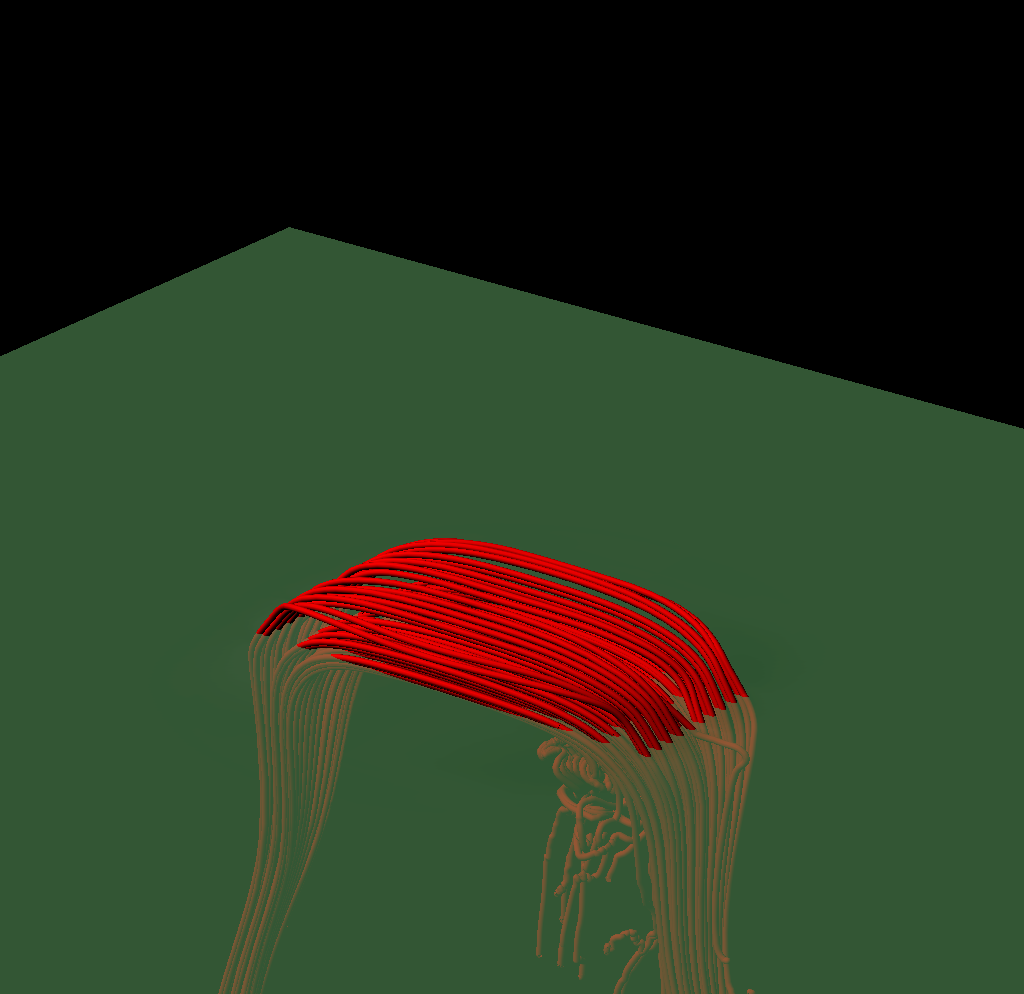}}\quad
   \subfloat[$t=62$]{\includegraphics[width=\wdth]{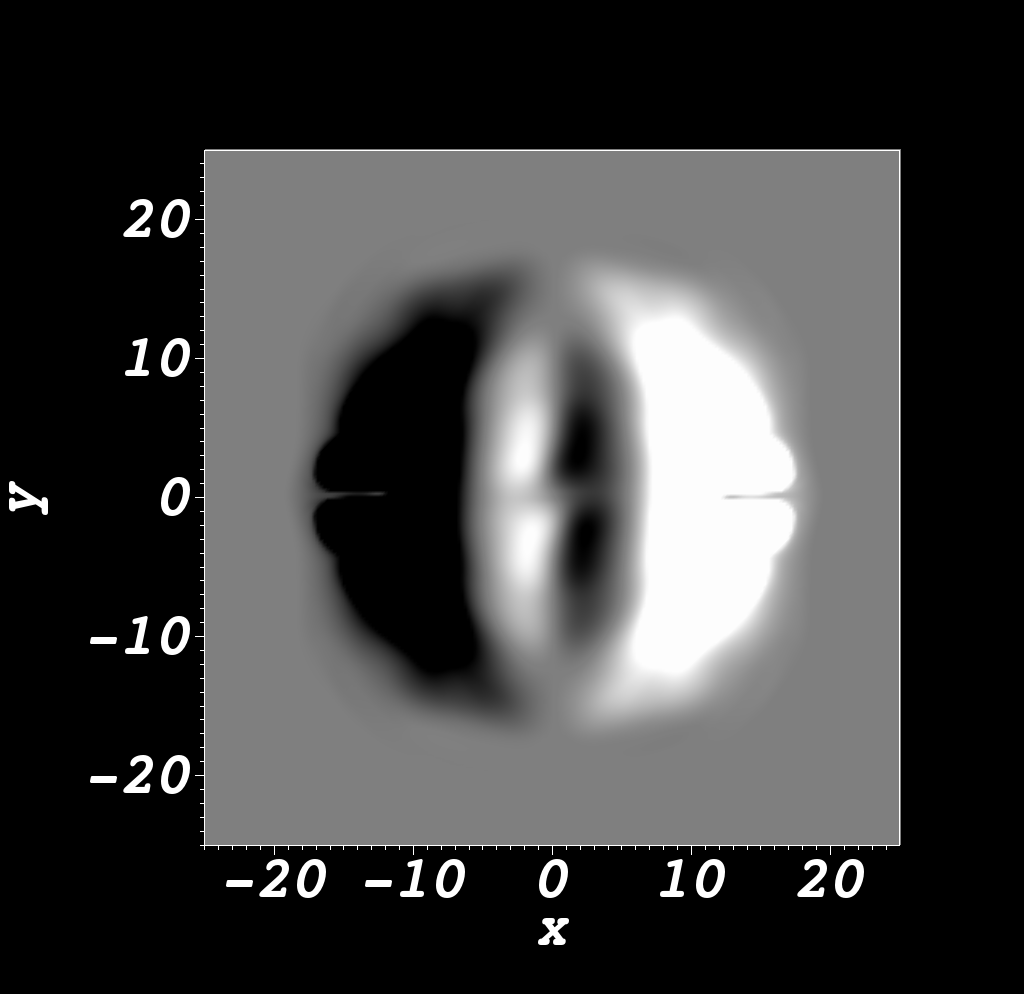}}\quad\subfloat[$t=62$]{\includegraphics[width=\wdth]{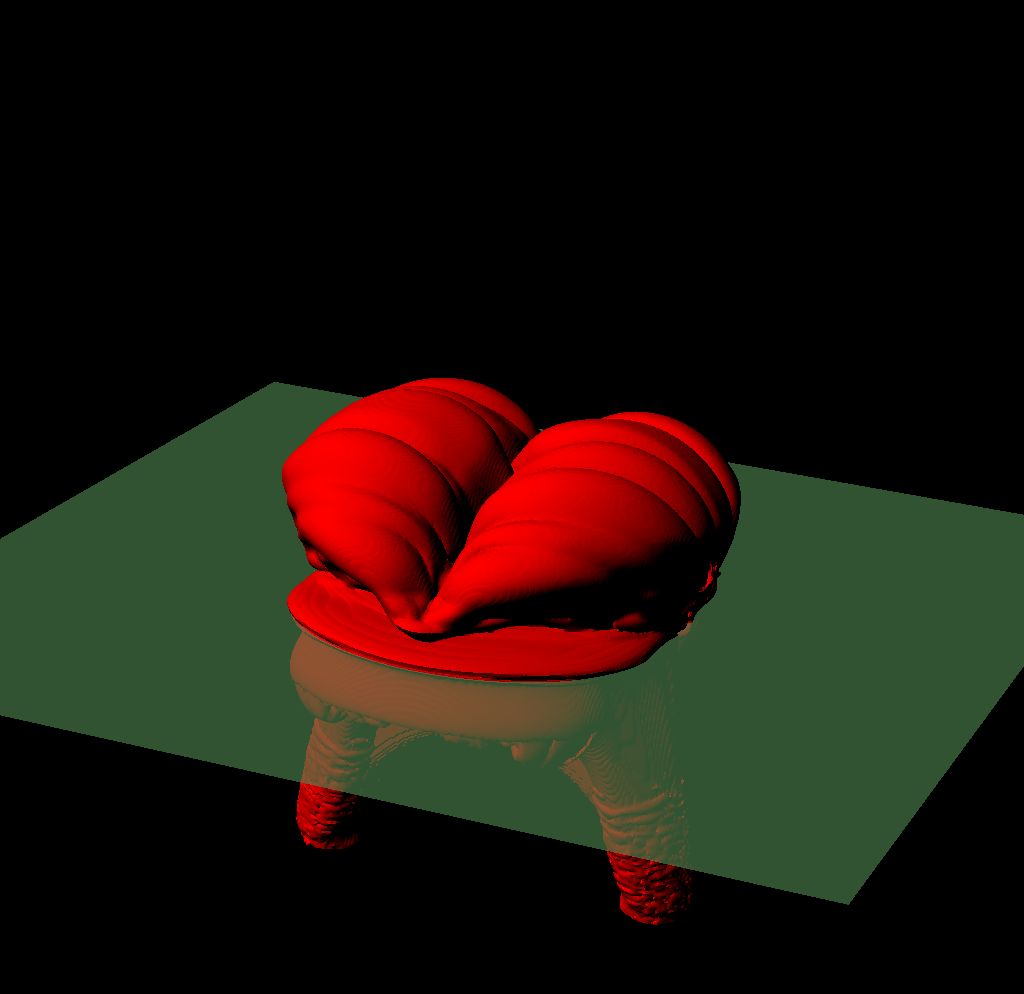}}\quad\subfloat[$t=62$]{\includegraphics[width=\wdth]{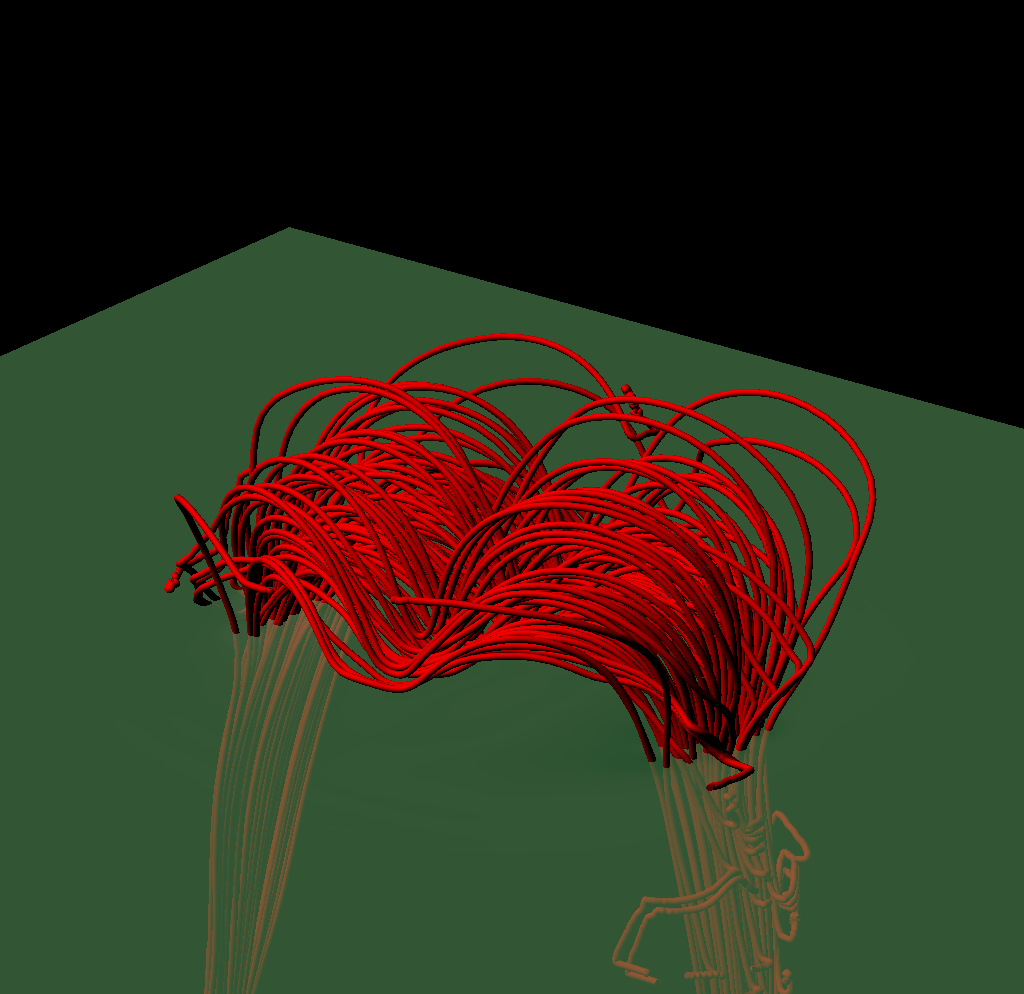}}
\caption{\label{braidemergence} Illustrative distributions characterizing the emergence of a mixed helicity flux rope with $n=2$ twist pairs. (a),(d), and (g) are magnetograms. (a) ($t=27$) depicts a {bipole} structure associated in the initial emergence phase. In (d) ($t=38$),  this distribution has separated slightly and there also appear to be some weak thin horizontal structures appearing at the centre of the domain. In (g) ($t=62$), these weak additional structures have developed into pairs of {bipole}s whose polarity oppose that of the larger initial {bipole}. (b), (e) and (h) are current contours ($\vert {\boldsymbol  j} \vert =0.01$) which indicate the field's expansion. Also indicated is the plane $z=0$. (c), (f) and (i) are representative field lines.}   
\end{figure}
Various illustrative visualizations of the emerging field's evolution are shown in Figure \ref{braidemergence}. As for the twisted tube, {the emerging field gets trapped below the photosphere before} becoming subject to the magnetic buoyancy instability and rising into the coronal region. The magnetograms (a), (d) and (g) show that, in addition to a larger bipole structure developing, there is also an additional formation of smaller bipole structures inbetween these larger poles with orientations in opposition to that of the large bipole. In (b), (e) and (h), contours of the {current density structure} show the field's expansion, which is restricted along the PIL of the main bipole. This behaviour is in contrast to the single structure found in the twisted case (Figure \ref{twistemergence}) which shows a more uniform expansion. The field lines, shown in Figures (c), (f) and (i) indicate a pair of intertwined field line arcades linked by a centrally dipped section (which partially submerges, as we will see later). There is no significant twisting in the field lines. It was established in \cite{prior2016emergence} that the dips in the current contours coincide with plasma draining and are the nonlinear manifestations of a Rayleigh-Taylor instability.

\subsubsection{Helicity and winding input: static photosphere}
\begin{figure}
 \subfloat[]{\includegraphics[width=7cm]{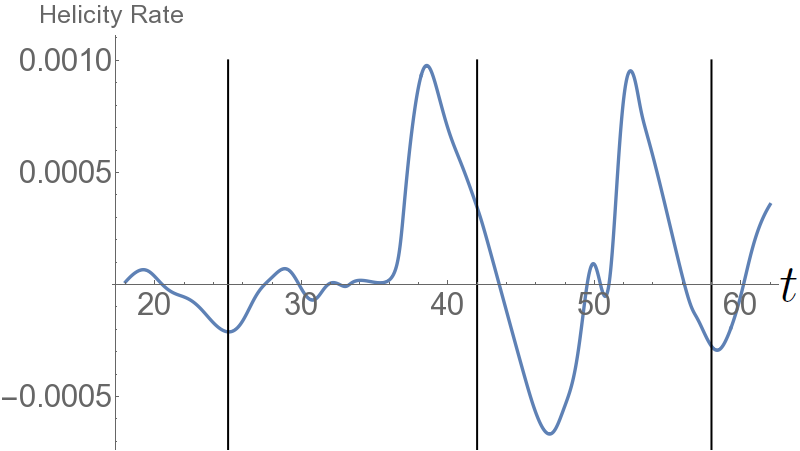}}\quad\subfloat[]{\includegraphics[width=7cm]{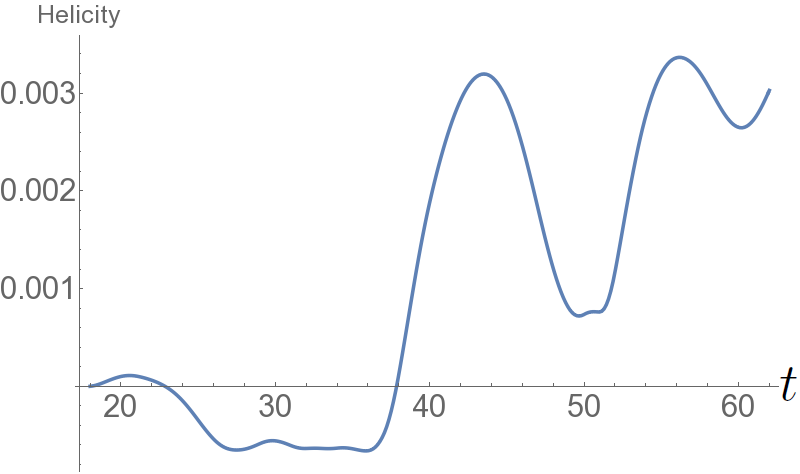}}\quad\subfloat[]{\includegraphics[width=7cm]{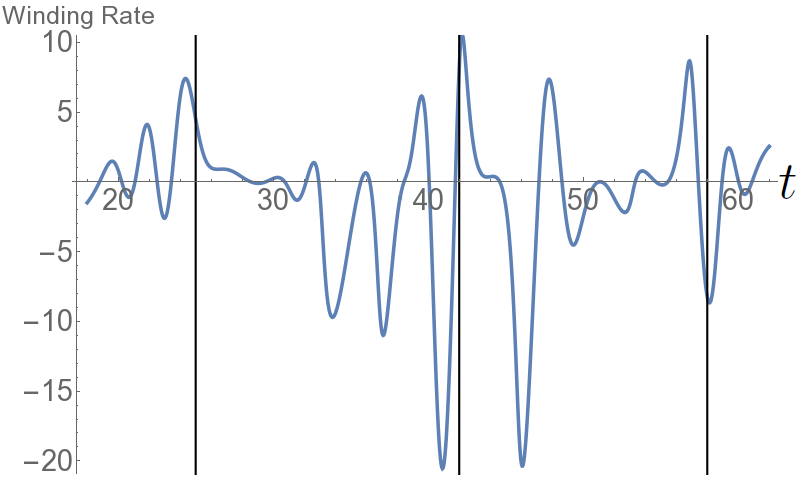}}\quad\subfloat[]{\includegraphics[width=7cm]{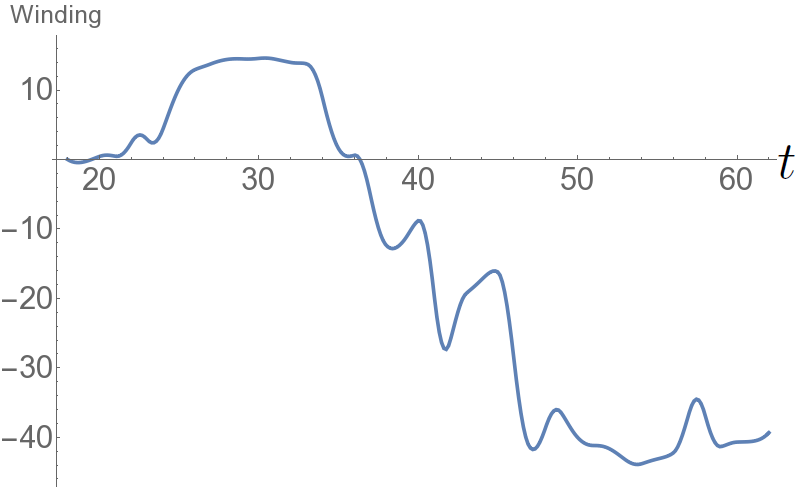}}
  \caption{\label{b2helcityinput}Photospheric helicity and winding  inputs for the emergence of the mixed helicity field. (a) the helicity input rate ${\rm d}{H}/{\rm d}{t}$. The input is both positive and negative. Also shown are lines at $t=25,42$ and $48$. (b) the net helicity input $H(t)$. (c) the winding rate ${\rm d}L/{\rm d}t$. (d) the total winding input $L(t)$, the sign of this input is generally opposite to that of the helicity.}  
\end{figure}
\begin{figure}
\subfloat[]{\includegraphics[width=6.5cm]{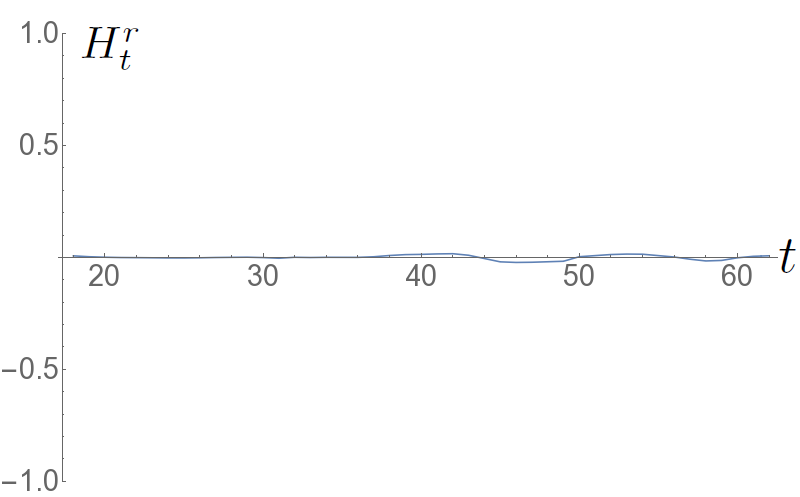}}\quad\subfloat[]{\includegraphics[width=6.5cm]{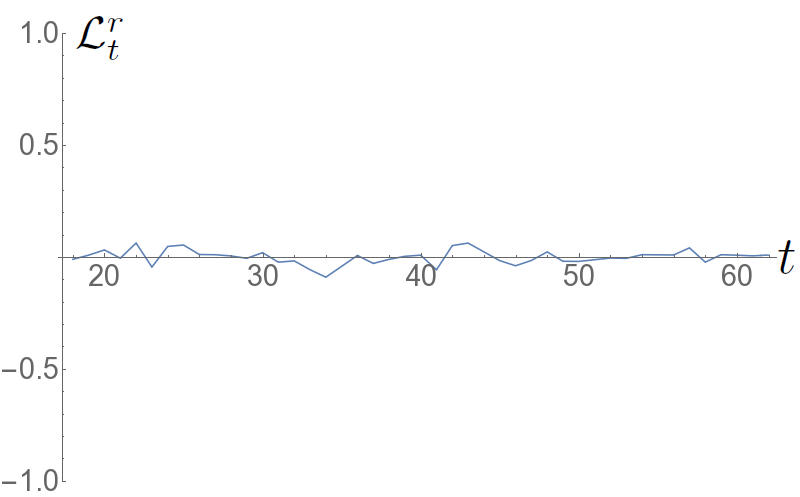}}\quad
\caption{\label{b2timeseries}Time series of the ratios (a) $H_t^r$ and (b) $L_t^r$ for the $n=2$ mixed helicity field emergence. } 
\end{figure}
Plots of the time series of the quantities ${\rm d}{H}/{\rm d}{t},\,H,\,{\rm d}L/{\rm d}t$ and $L$ are shown in Figure \ref{b2helcityinput}. The helicity rate ${\rm d}{H}/{\rm d}{t}$ (a) is both positive and negative and shows oscillatory behaviour (less temporally regular than for the twisted field). 
The net helicity input $H(t)$ (b) switches from negative to net positive. The winding input rate  ${\rm d} L/{\rm d}t$ (c) shows much more rapid variations. The net input $L$ (d) has a pattern qualitatively similar to the net helicity but with opposing sign.  The ratios $H_t^r$ and $L_t^r$ are found to be typically $1$-$2\%$ (Figure \ref{b2timeseries}), indicating the topological input is not far off neutral, {as to be expected}.

\begin{figure}
  \subfloat[]{\includegraphics[width=6cm]{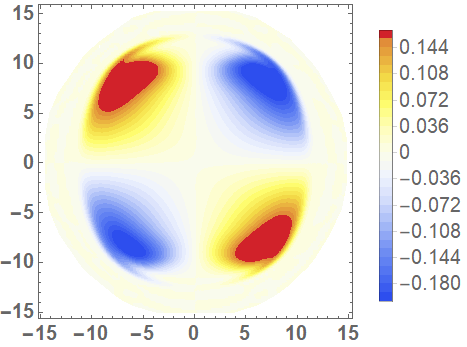}}\quad \subfloat[]{\includegraphics[width=6cm]{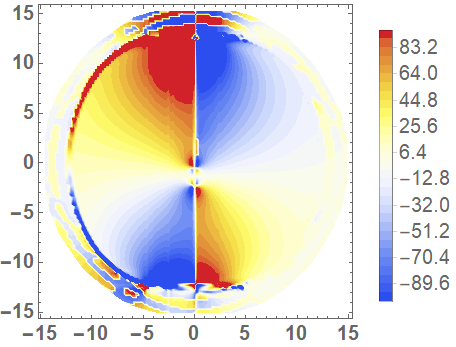}}\quad
  \subfloat[]{\includegraphics[width=6cm]{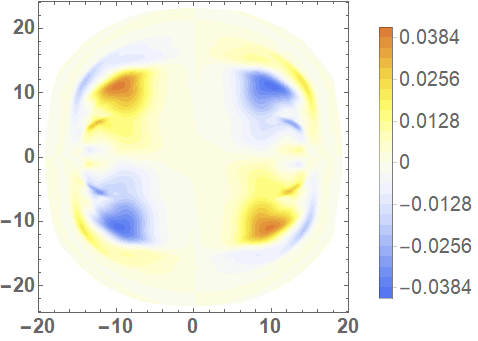}}\quad \subfloat[]{\includegraphics[width=6cm]{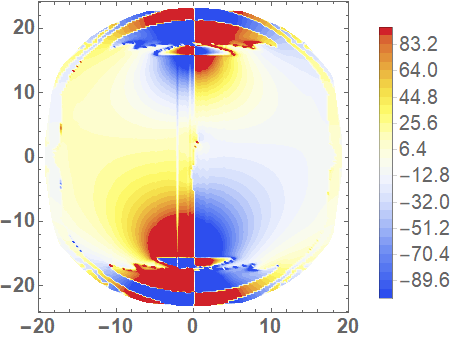}}\quad
  \subfloat[]{\includegraphics[width=6cm]{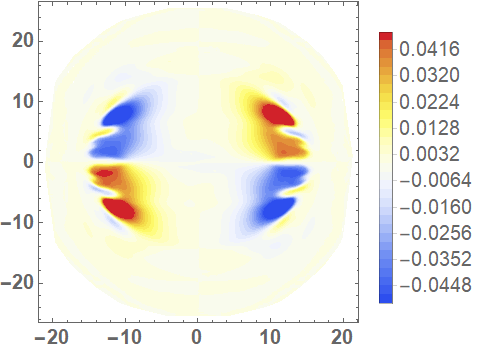}}\quad \subfloat[]{\includegraphics[width=6cm]{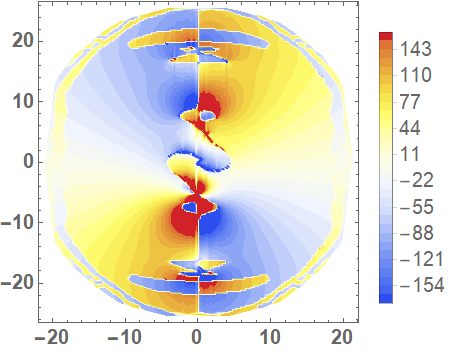}}
  \caption{\label{helwindplotsb2}Distributions of ${\rm d}{{\cal H}}/{\rm d}{t}$ and ${\rm d}{\cal L}/{\rm d}t$ which indicate the various characteristic stages of the distribution evolutions (indicated with vertical lines on Figure \ref{b2helcityinput}). (a) ${\rm d}{{\cal H}}/{\rm d}{t}$ and (b) ${\rm d}{\cal L}/{\rm d}t$ at $t=25$. In both cases there are two significant regions of positive input and two with negative input, a quadrupolar distribution. In (b) the straight PIL is visible along the $y$ axis. (c) ${\rm d}{{\cal H}}/{\rm d}{t}$ and (d) ${\rm d}{\cal L}/{\rm d}t$ at $t=42$. In (c) the four regions from (a) remain but are surrounded by thin strips of helicity input of opposing sign. In (d), by contrast, the sign of the four regions have swapped compared to (b) and there are additional sub regions of positive and negative input centered on the PIL.  (e) ${\rm d}{{\cal H}}/{\rm d}{t}$ and (f) ${\rm d}{\cal L}/{\rm d}t$ at $t=48$. The helicity input (e) still has four dominant domains but they have swapped sign from the distributions in (a) and (c). In (f) the flux region seen in the latter magnetograms (Figure \ref{braidemergence}) appears as a distorted PIL and there are strong concentrations of helicity either side of it.}
\end{figure}
The distributions ${\rm d}{\cal {H}}/{\rm d}{t}$ and ${\rm d}{\cal L}/{\rm d}t$ are shown in Figure \ref{helwindplotsb2}. The snapshots are chosen to display critical characteristics of these distributions which are present at various times of the emergence process. 

There are always four distinct regions of substantial positive and negative helicity (and winding) and these occur in pairs. These regions are generally of similar size and magnitude, which explains why the net helicity (and winding) input is much smaller than its absolute total. In practice, it is difficult to detect the imbalances in the maps of ${\rm d}{{\cal H}}/{\rm d}{t}$ that lead to the time series variations shown in Figure \ref{b2helcityinput}(a). By contrast, the winding distributions ${\rm d}{\cal L}/{\rm d}t$ capture more features related to emergence and submergence (we will return to this point later).  We ignore the distribution values around the edge of the magnetic domain as the field is very weak there.  The field line structure in the boundary current sheet is incoherent and, thus, any apparent winding input is not meaningful. 


The PIL is clear in all ${\rm d}{\cal L}/{\rm d}t$ distributions and its shape varies significantly as various new helicity structures appear (note that all additional structures  appear \emph{at} the PIL).


\subsection{Helicity and winding input: moving photosphere}
{
\begin{figure}
 \subfloat[]{\includegraphics[width=7cm]{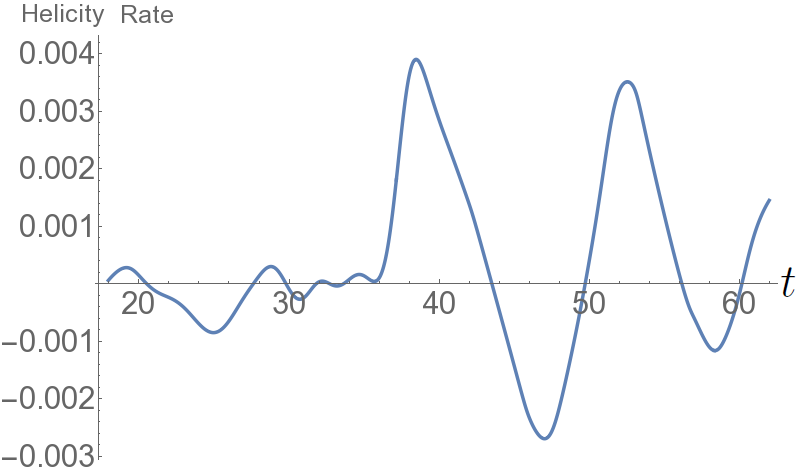}}\quad\subfloat[]{\includegraphics[width=7cm]{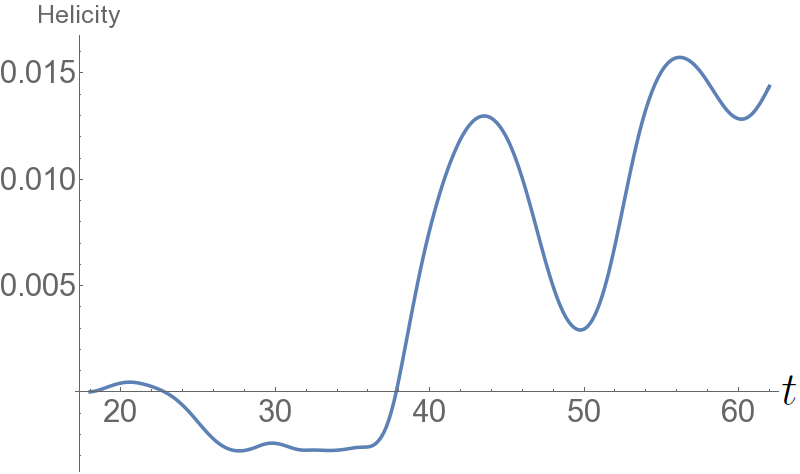}}\quad\subfloat[]{\includegraphics[width=7cm]{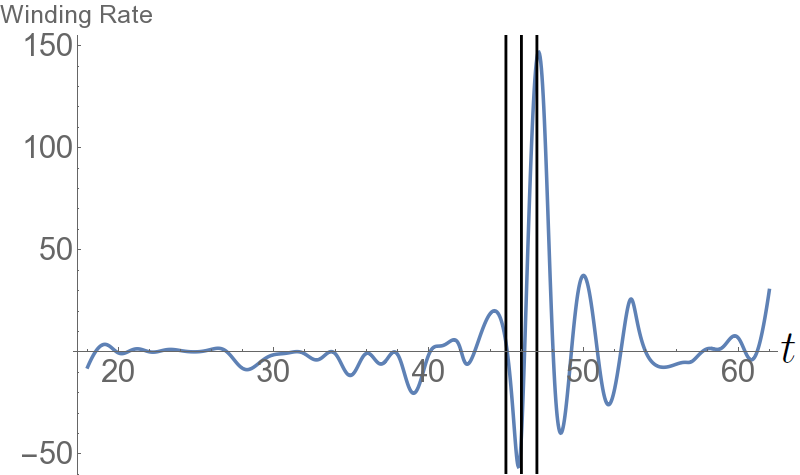}}\quad\subfloat[]{\includegraphics[width=7cm]{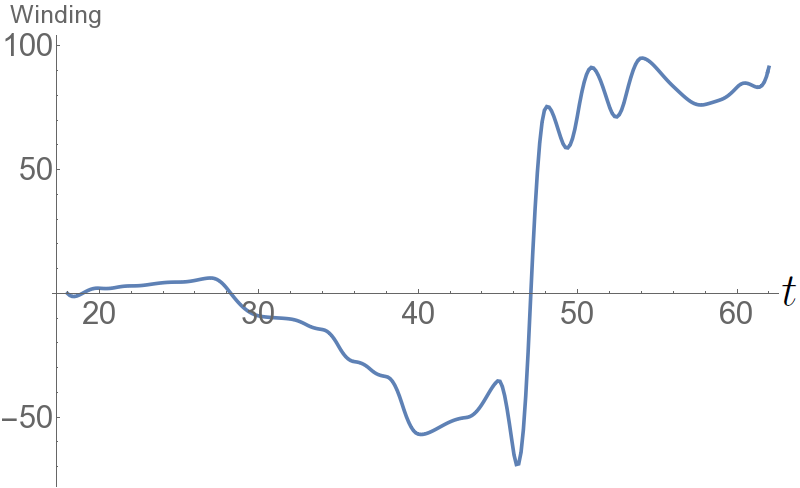}}
  \caption{\label{b2helcityinputnoneuc}Helicity and winding inputs for the emergence of the mixed helicity field, allowing for the varying photospheric geometry. (a) the helicity input rate ${\rm d}{H_v}/{\rm d}{t}$. (b) the net helicity input $H_v(t)$. (c) the winding rate ${\rm d}L_v/{\rm d}t$.  The vertical lines mark times $t=45,46,47$ at which the field is analyzed in detail in the text. (d) the total winding input $L_v(t)$.}  
\end{figure}
Time series of ${\rm d}{H_v}/{\rm d}{t}$, $H_v$, ${\rm d}L_v/{\rm d}t$ and $L_v$, which account for the varying photospheric geometry, are shown in Figure \ref{b2helcityinputnoneuc}. In order to prevent the winding measure being affected by any weak (unresolved) field, we utilize the modified definition of the function $\sigma_z$ (\ref{modsig}) with $\epsilon=0.0001$ for the ${\rm d}L_v/{\rm d}t$ and $L_v$ calculations. As was the case for the twisted field, the only significant change in the helicity time series compared to those shown in Figures \ref{b2helcityinput}(a) and (b) is in regard to the magnitudes. This difference could, again, be traced to differences in the vertical velocity distribution. The time series of the winding quantities ${\rm d}L_v/{\rm d}t$ and $L_v$, shown in  Figures \ref{b2helcityinputnoneuc}(c) and (d), differ significantly from those shown in Figures \ref{b2helcityinput}(c) and (d). We find that this is, in part, due to the varying photosphere $P_v$ and partly due to ignoring the contributions from significantly weak regions of the field. As discussed previously, the winding is more sensitive to large input from topologically complex field lines of very weak field strength. Hence, the calculation of the winding requires some care. 

The sign of the net input $L_v$ is largely dictated by a significant spike in the input ${\rm d}L_v/{\rm d}t$ at  $t=47$. After this, there are two large oscillations which are more balanced. The net inputs $H_v$ and $L_v$ now agree in terms of the net sign of their input over the simulation (as opposed to the quantities $H$ and $L$, emphasizing the sensitivity of $L$ or $L_v$).

\subsection{Field submergence and deformation}
As mentioned above, a clearer transition is revealed by the winding compared to the helicity. The key question to answer here is, what is causing this transition? Some indications can be found in the distributions of $\d\cl_v/\d t$ shown in Figure \ref{spikedists}. These distributions are at the times indicated by the parallel lines in Figure \ref{b2helcityinputnoneuc}(c). Figure \ref{spikedists}(a) displays $\d\cl_v/\d t$ when $\d L_v/\d t$ is changing from positive to negative. A small region of negative $\d\cl_v/\d t$ is visible at the centre (0,0). The region grows in (b), which corresponds to when $\d L_v/\d t$ takes its largest negative value. In (c), which corresponds to the large positive spike in $\d L_v/\d t$ and positive jump in $L_v$, the distribution changes to a clear antisymmetrical pattern. There is no longer the isolated patch of negative $\d\cl_v/\d t$ at the centre and the PIL is highly deformed.
\begin{figure}
 \subfloat[]{\includegraphics[width=7cm]{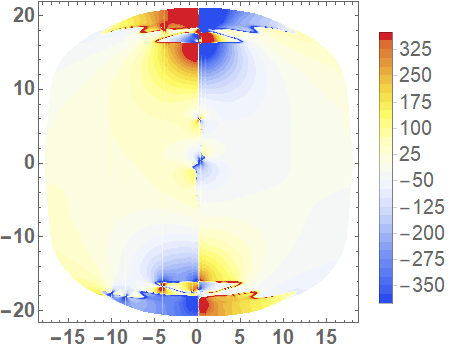}}\quad  \subfloat[]{\includegraphics[width=7cm]{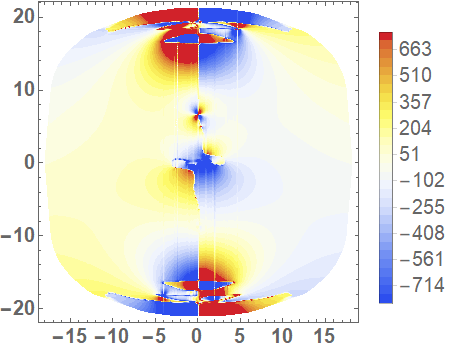}}\quad \subfloat[]{\includegraphics[width=7cm]{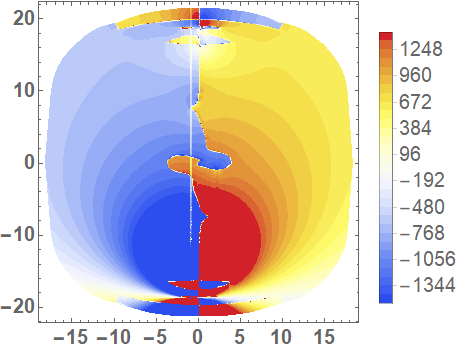}}
\caption{\label{spikedists}Distributions of the winding input density ${\rm d}\cl_v/{\rm d} t$ covering the period of the time series shown in Figure \ref{b2helcityinputnoneuc}(d) at which there is a strong spike in input. (a) $t=45$, before the spike. (b) $t=46$. (c) $t=47$ at the spike's peak.} 
\end{figure}
\begin{figure}
\centering
\includegraphics[width=7cm]{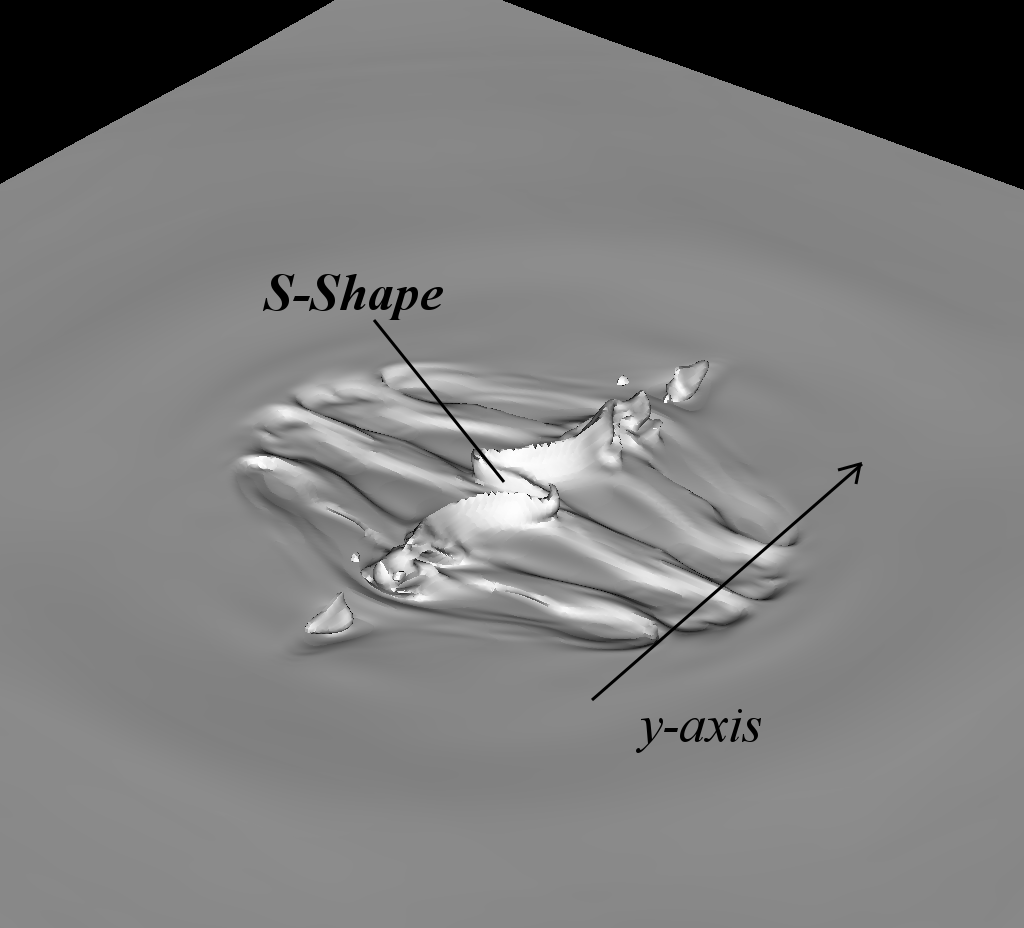}
\caption{\label{densitysig}A contour plot of density $\rho =0.2$, slightly above the $\rho=1$ photospheric level at $t=62$. There is a an $s$-shaped peak in the density profile at the centre of the domain which is parallel to the $y$-direction and has been established to correspond to a pooling of dense plasma in this location. This shape matches the morphology of line observed in the winding input densities ${\rm d}L/{\rm d}t$, shown in Figure \ref{helwindplotsb2}(f), and  ${\rm d}L_v/{\rm d}t$, shown in Figure \ref{spikedists}(c).}
\end{figure}
\begin{figure}
 \subfloat[]{\includegraphics[width=14cm]{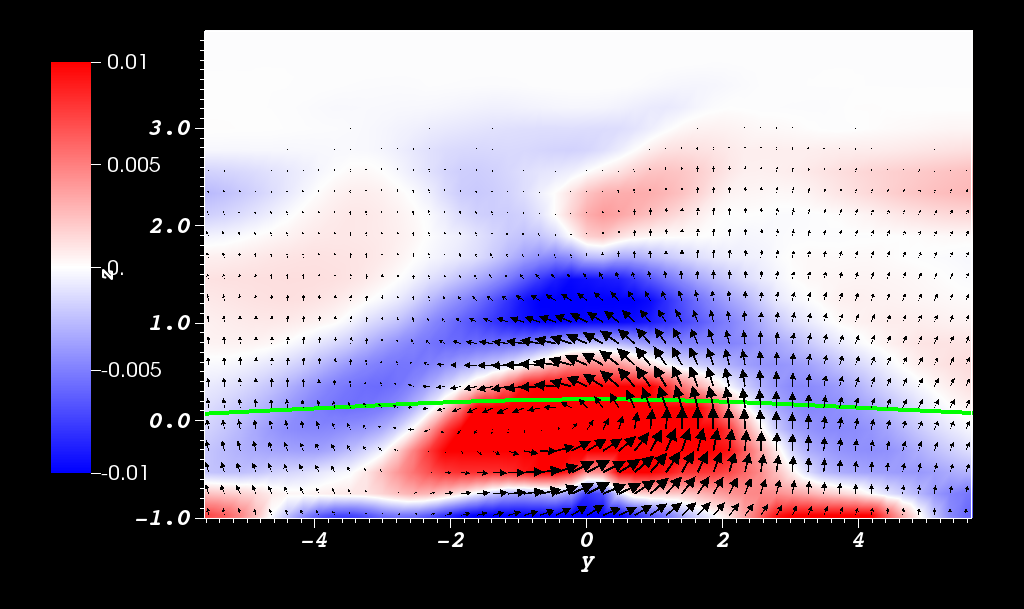}}\quad \subfloat[]{\includegraphics[width=14cm]{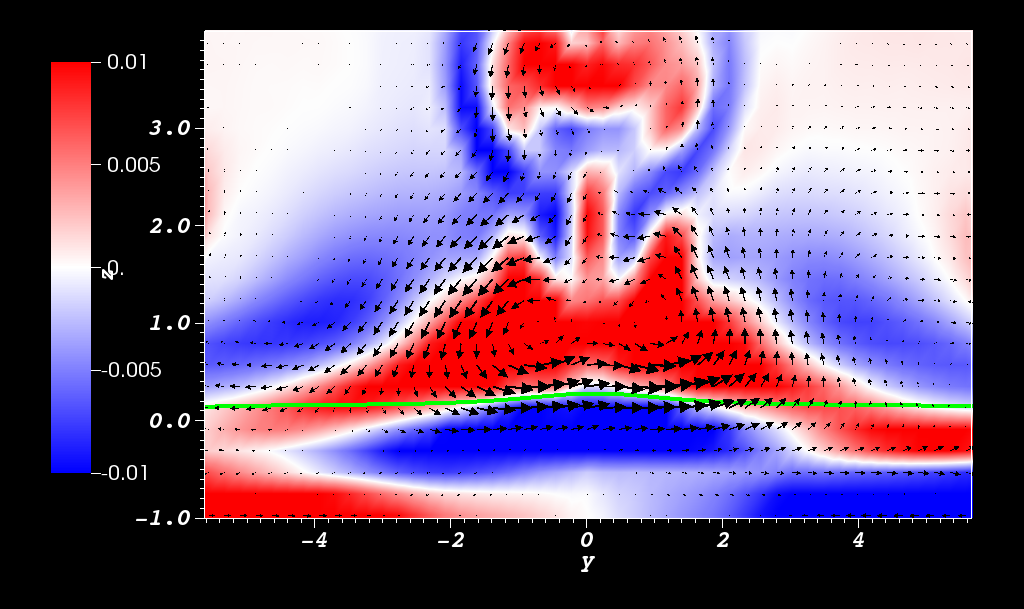}}
\caption{\label{fieldsliceb2a} Slices  in the $y$-$z$ plane, at the PIL at (a) $t=27$ and (b) $t=42$, revealing different phases of emergence. The slices display distributions of the out of plane component of the current density. Also shown, as a green line, the $\rho=1$ surface of the moving photosphere. (a) ($t=27$) there is a spiral structure, indicative of a locally twisted field just of the centre of the photospheric domain. It is part way through emerging through the $\rho=1$ photosphere line. (b) ($t=42$) the previous twist has split in two due to the deformation of the current structure in the atmosphere.}
\end{figure}
\begin{figure}
\subfloat[]{\includegraphics[width=14cm]{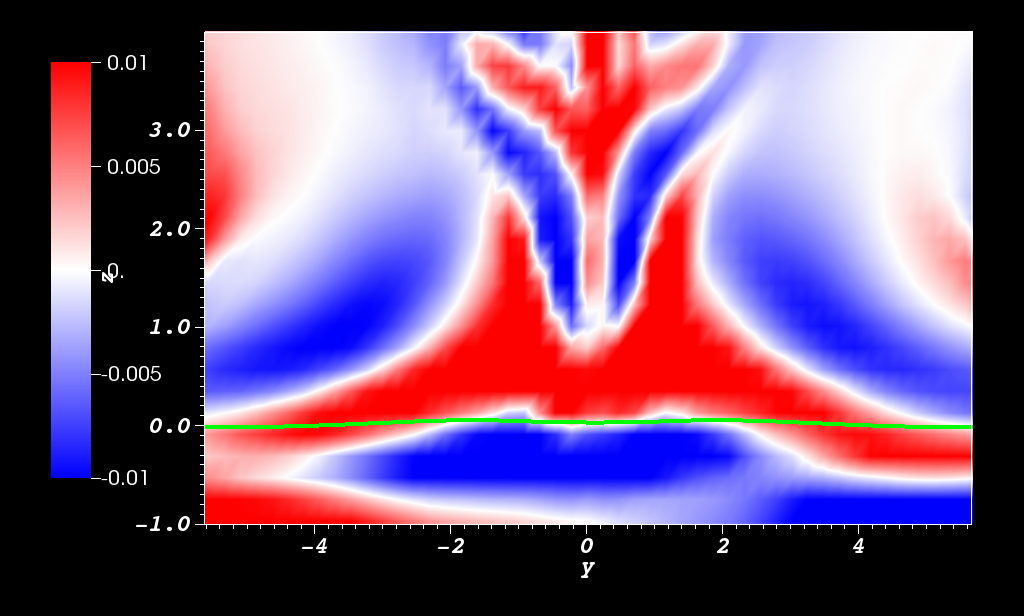}}\quad\subfloat[]{ \includegraphics[width=14cm]{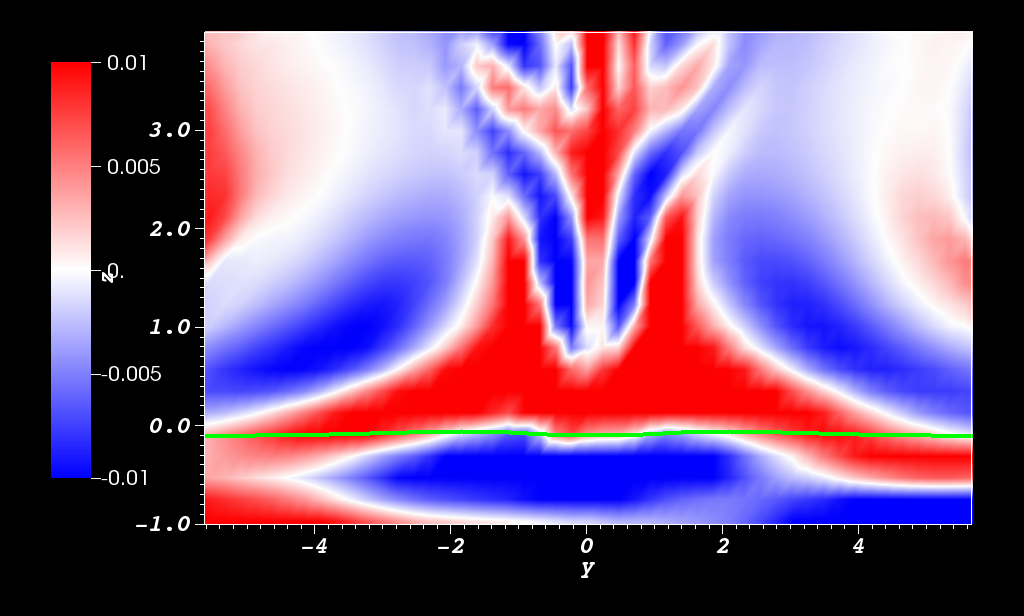}} 
\caption{\label{fieldsliceb2b}  Slices in the $y$-$z$ plane at the PIL at (a) $t=46$ and (b) $t=47$.  The slices display distributions of the out of plane component of the current density. Also shown, as a green line, the $\rho=1$ surface of the moving photosphere.}
\end{figure}

Since the emerging field has very little twist, it cannot easily support dense plasma in the atmosphere. This manifests itself in the atmosphere as the buckling of the central part of the emerging field, where dense plasma drains and restricts the magnetic field in the lower atmosphere. This process is visualized in Figure \ref{braidemergence} and the pooling of dense plasma which forms can be seen in Figure \ref{densitysig}. The patch of negative  $\d\cl_v/\d t$, described above, indicates the begining of motion leading to submergence, that is, the movement of helicity carrying field from the atmosphere down to the photospheric boundary. We now characterize various stages of the emergence up to and including this submergence event. 



Figure \ref{fieldsliceb2a}(a) displays the emergence of the mixed helicity field at the (relatively) early time of $t=27$. The slice at $x=0$ displays the $x$-component of current density, the projection of magnetic field arrows on the plane and the position of the moving photosphere. What is shown is a twisted structure begining to emerge into the atmosphere. Comparison with Figure  \ref{b2helcityinput} shows that the helicity and winding inputs are still relatively weak. Figure \ref{fieldsliceb2a}(b) displays a similar slice but for the much later time of $t=42$. Comparison with Figure  \ref{b2helcityinput} shows that this corresponds to the helicity approaching a local (positive) maximum and the winding near its most negative value. A current structure exits in the atmsophere in Figure \ref{fieldsliceb2a}(b) that was not present in (a). This current developed due to the buckling of the field due to dense plasma drainage, as described previously. This nonlinear deformation of the field has resulted in two `twist units', as opposed to the single unit displayed in (a). Where the two twist units meet, at $y=0$ and just above the photosphere, there is a small patch of negative current. This patch corresponds to the patch of negative winding displayed in Figures \ref{spikedists}(a) and (b), which reveal this patch growing at the later times of $t=45$ and $t=46$. 

As mentioned above, \ref{spikedists}(c) reveals that the patch of negative winding found in Figures \ref{spikedists}(a) and (b) disappears. Instead, the winding distribution reveals a highly deformed PIL and winding magnitudes double that of the previous time step. This behaviour is caused by submergence. Consider Figure \ref{fieldsliceb2b}. This figure shows $y$-$z$ slices (at $x=0$) of the $x$-component of the current density and the photospheric boundary at times (a) $t=46$ and (b) $t=47$, corresponding to the times of the winding distributions in Figures  \ref{spikedists}(b) and (c). In Figure \ref{fieldsliceb2b}(a) the structure of  postive current, created by dipped field in the atmosphere, is just glancing the photospheric boundary. In Figure \ref{fieldsliceb2b}(b), this structure has passed slightly beneath the photospheric boundary. Despite the appearance of Figures \ref{fieldsliceb2b}(a) and (b) being only marginally different, the submergence of a very small part of the field can have a large impact on the winding distributions (as well as the magnetograms). As submergence continues, the changing current structure leads to a large spike in the winding input (Figures \ref{b2helcityinputnoneuc}(c) and (d)).

It is interesting to note that the original emerging field (such as the twisted unit in Figure \ref{fieldsliceb2a}(a)) did not  lead to a spike in the winding input.  It was only after the emerged field above the photosphere changed and this new structure was submerged that an event was detected in the time series. 

Further confirmation that submergence dominates these helicity and winding time series is found, as we did for the twisted case, by examining the velocity flux. Time series of the the velocity flux and the $\d H/\d t$ are compared in Figure \ref{velb2timeseries}(a). After $t=40$, ${\cal V}_z<0$ for the period corresponding to the draining and submergence described above. We also note that there is a correlation between oscillations in the the helcity input time series and the velocity flux (Figure \ref{velb2timeseries}(b)).
\begin{figure}
  \subfloat[]{\includegraphics[width=7cm]{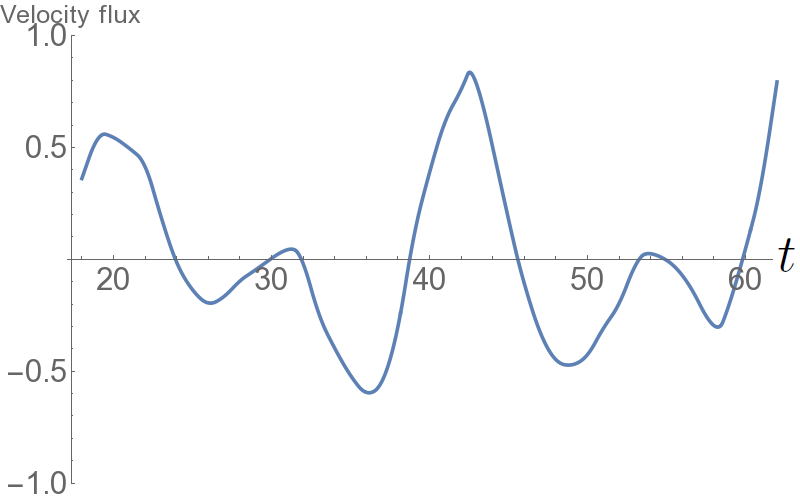}}\quad\subfloat[]{\includegraphics[width=7cm]{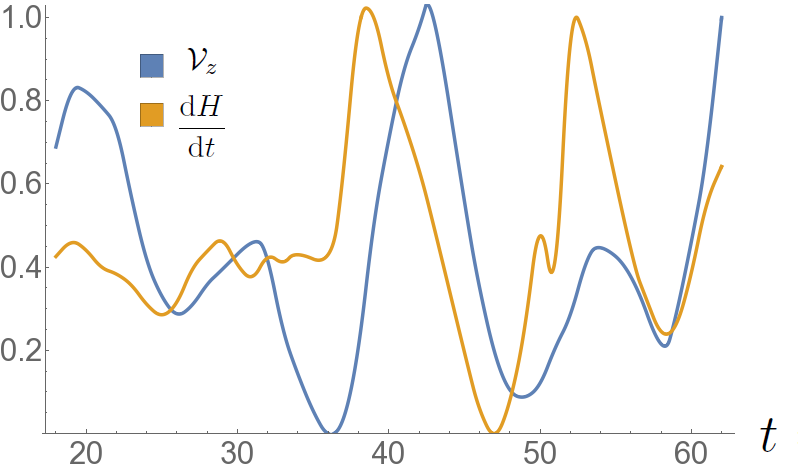}}
  \caption{\label{velb2timeseries}Plots of the velocity flux ${\cal V}_z$ and its correlation to the helicity input rate. (a) the temporal variation of ${\cal V}_z$. (b) a comparison of the temporal variation of the quantities ${\cal V}_z$ and the helicity input rate ${\rm d}{H}/{\rm d}{t}$ (both scaled to lie between $[0,1]$).}
\end{figure}

\section{Conclusions and discussion}
This article presents simulations of the emergence of magnetic flux tubes with varying internal geometries and topologies.  One simulation considers a twisted tube which has a single sign helicity (negative in this case). The other simulation focusses on a tube of mixed helicity. The two main quantities analysed are the (relative) helicity $H$ and the winding $L$, both found by considering their rate of change through the photosphere. We have performed calculations through a planar surface, representing the initial position of the photosphere in the simulations, as well as a surface of constant density which represents a (potentially) changing photosphere. Both quantities ($H$ and $L$) rely on the same baseline information, the changing entanglement of the points at which magnetic field lines pierce the photosphere, i.e. the evolving field topology {affected by both the in-plane fluid motion and the vertical motion of the field passing through the photosphere}. The major difference between $H$ and $L$ is the weighting of magnetic flux for the helicity input. The following observations emerge from the analysis:

\begin{enumerate}
\item{Large oscillations in the helicity input are consistently related to the reversal of plasma flow across the photosphere, occuring in both the single sign and mixed helicity emergence simulations. These reversals appear in the spatial distributions of the helicity and winding rates as the temporally varying sign of clear structures. These structures can be further correlated to the appearance of flux structures in the magnetogram distributions, leading to several features in the magnetograms being linked to submergence rather than emergence.}
\item{{The winding time series tends to be far more sensitive to the emergence (submergence) of field topology and can detect individual events of helicity/winding carrying structures intersecting the photospheric boundary. The helicity time series tends to give more gradual variations in input. It is sensitive to the emergence (submergence) of topologically complex field when this is linked to the transport of significant magnetic flux crossing the photospheric boundary.}}
\item{The ratios of the input of helicity and winding to their unsigned equivalents is far larger for the twisted emergence than for the mixed helicity emergence. This result was to be expected based on the initial conditions: the twisted tube has net helicity, the mixed helicity field does not.}
\item{It is important to take the changing geometry of the photosphere into account, particularly for the calculation of the winding. It is also important to remove the effect of weak or unresolved field, which can affect the winding calculation both quantitatively and qualitatively.}
\end{enumerate}

Based on our analysis of the simulations, we make the following suggestions which may aid or enhance the analysis of observational flux emergence data:
\begin{enumerate}
\item{We recommend that both the helicity \emph{and} winding input distributions and time series should be calculated. This task should be straight forward, in one sense, as the only main difference in the calculation of the winding compared to the helicity is to substitute an indicator function measuring the sign of $B_z$ instead  of $B_z$ itself. Since the winding is more sensitive to noise, however, suitable preprocessing of the observational data will be required. If possible, the varying geometry of the photosphere should be accounted for as it can affect both the magnitude and morphology of the input time series.}
  \item{We recommend that the behaviour of the helicity and winding rates and distributions should be correlated to the vertical velocity flux, where possible. This would aid in determining if a new structure in a magnetogram is the result of emergence or submergence, a task which would be difficult if reliable information is restricted only to photospheric magnetograms. On that note, we acknowledge that the $\d H/\d t$ and $\mathcal{V}_z$ time series for the twisted field simulation show a shift between anti-correlation and correlation (Figure \ref{velocityvariation}(d)), while for the mixed helicity simulation these quantities appear to be more coherently correlated (Figure \ref{velb2timeseries}(b)). Although we have not investigated the relationship between the time series in detail, that is beyond recognizing when magnetic field is emerging or submerging, it is an area worthy of future study and may reveal important characteristics of the emergence of different types of magnetic field.}
\item{We recommend a combined analysis of the winding input time series and distributional information. These can be used to detect events in which magnetic field structures with complex topology pass through the photosphere (from above or below). This approach provides more information than magnetogram distributions, which indicate new emergence/submergence events but no information regarding their topological content. However, care must be taken to discount events which have significantly weak magnetic field and may be spurious. Such events may have little significance for the overall evolution of the emerging magnetic field. This point is made clear in the difference between the winding series for the mixed helicity field. In the non-moving surface case the sign of the total helicity and winding inputs differ (Figure \ref{b2helcityinput} (b) and (d)). It is entirely plausible that this can occur for a field which has significant small-scale but well defined topology (something with a high winding measure) that occurs in a region of weak field ($B_z\ll 1$). Then the helicity assigns a very small weight to this topology (winding) due to the multiplication by a very small $B_z$. However, as discussed above, in simulations of emergence into an, initially, magnetically empty region above the photosphere, there is a current sheet surrounding the emerging flux rope. In practice this leads to a small region on the boundary of the emerged field where the field strength rapidly decreases to zero. The field line structure in this boundary layer is not resolved and, therefore, not to be trusted. It was found that by removing these contributions, the helicity and winding inputs have the same overall sign (see Figure \ref{b2helcityinputnoneuc} (b) and (d)). A significant part of this change was found to be due to the cut-off in field strength applied to the winding calculations. The size of $\epsilon$ (used in the cut-off) that needs to be prescribed will depend on the implementation of how the magnetic field is approximated numerically. In particular, it will depend on the numerical scheme (finite difference, finite element, etc.) and the grid resolution. Without rigorous theoretical results, a `trial and error' approach to $\epsilon$ selection is recommended. If the value is too low, the winding results will be significantly different for slightly higher values of $\epsilon$. If the value is too high, field that can be resolved numerically may be removed from the calculation.} 
\item{We recommend that both the helicity and winding ratios, $H_t^r$ and $L_t^r$, should be calculated. Again, these calculations should be straight forward as they use the same basic information as the helicity and winding inputs. The evidence from our study suggests that a consistently low value of these ratios is indicative of an emerging mixed helicity structure  where the internal twisting in the field is not dominated by one sign. Our results suggest that a ratio below $5$\% is indicative of a mixed helicity field. That being said, we have only considered a strongly twisted case. For emerging fields with lower twists, the ratio may also be small and this is something to be considered in future work. }
\end{enumerate}

 Apart from the points mentioned above, there are several directions in which to carry this work forward. So far, we have considered emerging regions before the formation of any eruptions, i.e. the motion we have studied is effectively ideal. How non-ideal effects (e.g. reconnection) change the helicity and winding signatures, including their interpretation, remains to be studied.  It is known that for high magnetic Reynolds number plasmas, dissipative losses to the (relative) helicity in the volume are small \citep{berger1984rigorous,sturrock2015sunspot,pariat2015testing}. This is true typically of dissipative helicity loss at the photospheric surface \citep{sturrock2015sunspot,pariat2015testing}. However, this does not necessarily imply that local reconnection rates are low (which depend on the electric current), just that the average losses are small. Thus, since there are small regions of relatively high current at the photosphere in these simulations, there may be some regions where local losses are significant. This, in turn, may lead to significant local changes in field topology.

 Since our focus in this study was not on the field's evolution above the photosphere, we did not make direct calculations of the relative helicity in the volume above the photosphere. This calculation has been performed in various flux emergence studies in which the changing structure of the field above the photosphere was of significant focus \citep[e.g.][]{sturrock2015sunspot,pariat2015testing,pariat2017relative}. As an example comparison, the helicity input in the twisted case of this manuscript has a linear total input with an oscillatory component.  This behaviour is in line with the results of \cite{sturrock2015sunspot} (allowing for the fact that the oscillations in our study are larger due to the the field's twisted core getting stuck at the photosphere), who perform the relative helicity calculation in the volume and confirm, as expected, that the total helicity input is approximately equal to the relative helicity in the volume. Theoretically, as long as the emerging field does not interact with the boundaries of the computational domain and the evolution is ideal, the helicity flux and direct calculations should give identical results. Errors in different numerical implementations and resistive effects can play a role in making these results diverge, so checking both calculations is, generally, a good idea.  This time, however, we feel that the above comparison with the results of \cite{sturrock2015sunspot} justifies our decision to postpone such calculation checks for a thorough study of the topology of the field above the photosphere.

 Further to this, previous analyses of helicity dissipation have only considered the effects of scalar diffusion. In flux emergence applications, a diffusion tensor, with components parallel and perpendicular to the magnetic field, represents a more accurate diffusion model for the region between the photosphere and the corona. Further still, diffusion perpendicular to the magnetic field can be orders of magnitude larger than that parallel to the field, in this region \citep[e.g.][]{arber2007emergence}. Therefore, more work needs to be done to investigate both the global and local dissipative effects on helicity and winding.


Another important aspect for further study, revealed by our results, is the sensitivity of  helicity and winding to changes at the photosphere. The present model should be updated to include a convectively unstable solar interior.

\section*{}
We acknowledge the computational resources provided by the EPSRC funded ARCHIE-WeSt High Performance Computer (www.archie-west.ac.uk), EPSRC grant no. EP/K000586/1.


\bibliographystyle{jpp}

\bibliography{braids}

\end{document}